\documentclass[11pt]{article}
\pdfoutput=1
\usepackage{amsmath,amssymb,epsfig,amsfonts}
\usepackage{graphicx}

\usepackage{subcaption}
\usepackage[usenames, dvipsnames]{color}
\usepackage{hyperref}
\usepackage[dvipsnames]{xcolor}
\usepackage{cite}
\usepackage{multirow}
\usepackage{slashed}
\usepackage{verbatim} 
\usepackage{enumitem}
\usepackage{rotating}
\usepackage{xcolor}
\usepackage{multirow}
\usepackage{bm}
\usepackage[normalem]{ulem}
\usepackage{cleveref}
\usepackage{booktabs} 
\usepackage{tikz-cd} 
\usepackage{empheq}

\usepackage{tikz}
\usepackage{diagbox}
\usetikzlibrary{positioning}
\usetikzlibrary{calc}
\usetikzlibrary{decorations.pathreplacing,calligraphy}

\usepackage{xstring}
\usetikzlibrary{decorations.pathmorphing} 
\usetikzlibrary{decorations.markings} 
\usetikzlibrary{arrows} 
\usetikzlibrary{shapes} 
\usetikzlibrary{matrix} 
\usetikzlibrary{positioning} 
\usepackage[english]{babel} 
\usepackage[autostyle]{csquotes}
\usepackage{pifont}
\usetikzlibrary{shapes.multipart}

\tikzset{->-/.style={decoration={
  markings,
  mark=at position .5 with {\arrow{>}}},postaction={decorate}}}

\newcommand{\bea}{\begin{eqnarray}}
\newcommand{\eea}{\end{eqnarray}}
\newcommand{\be}{\begin{equation}}
\newcommand{\ee}{\end{equation}}
\newcommand{\ba}{\begin{aligned}}
\newcommand{\ea}{\end{aligned}}
\newcommand{\bit}{\begin{itemize}}
\newcommand{\eit}{\end{itemize}}
\newcommand{\ben}{\begin{enumerate}}
\newcommand{\een}{\end{enumerate}}

\newcommand{\id}{\text{id}}
\newcommand{\dd}{\mathop{}\!\mathrm{d}}
\newcommand{\CS}{\text{CS}}
\newcommand{\BF}{\text{BF}}

\newcommand{\SymTFT}{\text{SymTFT}}
\newcommand{\Neu}{\text{Neu}}
\newcommand{\Dir}{\text{Dir}}
\newcommand{\Tr}{\text{Tr}}

\newcommand{\Bsym}{\mathfrak{B}^{\text{sym}}}
\newcommand{\Bphys}{\mathfrak{B}^{\text{phys}}}

\newcommand{\Hod}{\text{Hod}}
\newcommand{\SSB}{\text{SSB}}

\newcommand{\R}{{\mathbb R}}

\newcommand{\bR}{\mathbb{R}}

\newcommand{\cA}{\mathcal{A}}

\newcommand{\cD}{\mathcal{D}}

\newcommand{\cF}{\mathcal{F}}

\newcommand{\cJ}{\mathcal{J}}

\newcommand{\cL}{\mathcal{L}}
\newcommand{\cM}{\mathcal{M}}

\newcommand{\cO}{\mathcal{O}}
\newcommand{\cP}{\mathcal{P}}

\newcommand{\cR}{\mathcal{R}}

\newcommand{\cU}{\mathcal{U}}

\newcommand{\fg}{\mathfrak{g}}
\newcommand{\fh}{\mathfrak{h}}
\newcommand{\fm}{\mathfrak{m}}

\def\tr{\mathop{\mathrm{Tr}}\nolimits}

\newcommand{\ie}{\begin{equation}\begin{aligned}}
\newcommand{\fe}{\end{aligned}\end{equation}}

\newcommand{\sg}{\mathsf{g}}
\newcommand{\sR}{\mathsf{R}}
\newcommand{\df}{\coloneqq}

\protected\def\[#1\]{\begin{equation}\begin{split}#1\end{split}\end{equation}}

\usepackage{empheq}

\begin{document}

\baselineskip=18pt  
\numberwithin{equation}{section}  
\allowdisplaybreaks  

\thispagestyle{empty}

\begin{center}
{\Huge Symmetry TFTs for \\
\smallskip
Continuous  
Spacetime Symmetries
}

\vspace*{0.8cm}
Fabio Apruzzi$\,^1$, 
Nicola Dondi$\,^2$, 
Iñaki García Etxebarria$\,^3$, \\

Ho Tat Lam$\,^{4}$,   
Sakura Sch\"afer-Nameki$\,^5$

\vspace*{0.5cm} 

{\it $^1$ 
Dipartimento di Fisica e Astronomia “Galileo Galilei”, \\
Università di Padova,
Via Marzolo 8, 35131 Padova, Italy\\ 

$^1$  INFN, Sezione di Padova Via Marzolo 8, 35131 Padova, Italy
}

{\it $^2$ 
Abdus Salam International Centre for Theoretical Physics,\\
Strada Costiera 11, 34151, Trieste, Italy\\

$^2$ INFN, Sezione di Trieste, Via Valerio 2, I-34127 Trieste, Italy
}
 
 {\it $^3$ Department of Mathematical Sciences,
    Durham University,\\
    Durham, DH1 3LE, United Kingdom}

{\it $^4$ Center for Theoretical Physics — a Leinweber Institute,  \\
Massachusetts Institute of Technology, Cambridge, MA 02139, USA}

{\it $^4$ Leinweber Institute for Theoretical Physics, \\ 
Stanford University, Stanford, CA 94305, USA}

{\it $^4$ Department of Physics and Astronomy,
University of Southern California,\\ Los Angeles, CA 90089, USA}

 {\it $^5$  Mathematical Institute, University of Oxford, \\
 Andrew Wiles Building,  Woodstock Road, Oxford, OX2 6GG, UK}

 \end{center}
\vspace*{0.2cm}

\noindent
We propose a Symmetry Topological Field Theory (SymTFT) for continuous spacetime symmetries. For a $d$-dimensional theory, it is given by a $(d+1)$-dimensional BF-theory for
the spacetime symmetry group, and whenever $d$ is even, it can also include Chern-Simons couplings that 
encode conformal and gravitational anomalies. We study the boundary conditions for this SymTFT and describe the general setup to study symmetry breaking of spacetime symmetries. 
We then specialize to the conformal symmetry case and derive the dilaton action for conformal symmetry breaking. 
To further substantiate that our setup captures spacetime symmetries, we demonstrate that the  topological defects of the SymTFT realize the associated spacetime symmetry transformations.
Finally, we study the relation to gravity and holography. The
proposal classically coincides with two-dimensional Jackiw-Teitelboim gravity for $d=1$ as well as  the topological limit of four-dimensional gravity in the $d=3$ case.

\newpage

\tableofcontents

\section{Introduction}
\label{sec:Intro}

During the last decade, building mainly on the seminal work 
\cite{Gaiotto:2014kfa}, it has become clear that thinking of symmetries in terms of topological operators can be a very powerful
approach to understanding quantum field theory (QFT) in $d$ dimensions. 
 This new perspective 
elegantly unifies continuous and discrete symmetries, and
greatly extends the applicability of symmetry-based techniques by incorporating
symmetries acting on extended operators -- such as higher-form and higher-groups --  and allowing for symmetries
that compose according to algebraic structures more general than
ordinary groups. This more general point of view allows us to
reformulate multiple phenomena in QFT, which were previously
understood using ad-hoc techniques, in a unified manner using the
language of symmetry, and it has lead to impressive new insights. We
refer the reader to the reviews \cite{Schafer-Nameki:2023jdn, Brennan:2023mmt, Bhardwaj:2023kri, Shao:2023gho,Luo:2023ive,Iqbal:2024pee} for surveys of the field.

So far our understanding of symmetries as topological operators has
mostly focused on internal symmetries, namely those not acting on
spacetime itself.
It is clear that developing a formalism that
incorporates spacetime symmetries, in particular continuous ones,  in a way that is compatible with our
modern understanding of internal symmetries is one of the main open
questions in the field, and our goal in this paper is to address this
issue.

We will do so in the context of the {\bf Symmetry Topological Field
  Theory  (SymTFT)}, a topological field theory in $(d+1)$ dimensions, which encodes {the structure of topological defects and generalized charges of the associated symmetries} in $d$ dimensional field theories \cite{Ji:2019jhk, Pulmann:2019vrw, Gaiotto:2020iye,  Apruzzi:2021nmk, Freed:2022qnc}. 
The SymTFT approach has a number of advantages compared with a direct description in $d$ dimensions,
stemming from the fact that it separates questions about symmetries from questions about local dynamics of the field theory, which are
typically much harder to understand. 
Applied to internal (finite) symmetries,  for instance, it can be used to completely classify gapped (topological) phases with a given generalized symmetry, vastly extending the standard Landau paradigm \cite{Bhardwaj:2023fca}, phase transitions \cite{Chatterjee:2022tyg, Wen:2023otf, Bhardwaj:2023bbf,Bhardwaj:2024qrf} and quantifying anomalies of generalized symmetries \cite{Antinucci:2025fjp} -- to name a few applications. Crucially, it encodes the charges under generalized symmetries 
\cite{Bhardwaj:2023wzd, Bartsch:2022ytj, Bhardwaj:2023ayw, Bartsch:2023wvv}. 
{An important advantage compared to other approaches to generalized symmetries, which
will play a key role in our analysis, is that in situations where the $d$-dimensional field theory at hand has a holographic dual, the
SymTFT is closely connected to this bulk gravitational theory}. This connection
is by now well understood for internal symmetries \cite{Witten:1998wy, Apruzzi:2021phx, Apruzzi:2022rei, GarciaEtxebarria:2022vzq, Antinucci:2022vyk}, and in this paper
we will explain how the correspondence extends to  spacetime
symmetries. It should be emphasized that our proposal is general, and
we will give evidence that it works also for theories with no
tractable holographic dual, but the holographic case is an excellent testing ground for our ideas.

For finite spacetime symmetries, we should note that the SymTFT approach has been applied in \cite{Pace:2024acq, Pace:2025hpb, Antinucci:2025fjp}. In all these setups, however, the bulk TQFT is enriched by the finite spacetime symmetry, and this is quite different from the setting we consider here, where we flat-gauge continuous spacetime symmetries.

\paragraph{Proposal for Spacetime SymTFT. }
If one has a $d$-dimensional theory with an anomaly-free internal
symmetry group $G_{\text{internal}}$, which might be finite or
continuous, the SymTFT is a $(d+1)$-dimensional theory of flat
$G_{\text{internal}}$ connections. For finite groups, this is a vanilla gauge theory for $G_{\text{internal}}$, whereas for continuous $G_{\text{internal}}$, it is a BF-theory. The basic observation that we make in this
paper is that one can drop the adjective ``internal'':~the SymTFT for spacetime symmetry group $G_{\text{spacetime}}$ is
also a BF-theory for $G_{\text{spacetime}}$, except in odd bulk dimensions (or $d$ even), where there is an additional CS-term for $G_{\text{spacetime}}$.
This is akin to adding possible anomalies as interaction terms in the SymTFT for internal symmetries. 
Our main working example will be 
\be
G_{\text{spacetime}}=SO(d+1,1)\,,
\ee
the Euclidean conformal group in $d$ dimensions. Anomalies complicate the previous
statements somewhat, in a way that is well understood in the case of
internal symmetries; we will explain below how to incorporate
conformal and gravitational anomalies. 

So our proposal is in a sense a very natural guess, but it presents
two related basic conceptual puzzles:
\begin{enumerate}
\item[Q1.] How do we relate the action of extended operators in the  SymTFT
  to the expected action of $G_{\text{spacetime}}$? As a particularly
  vexing example, how can topological operators in the SymTFT implement
  translations of operators in the $d$-dimensional QFT?
\item[Q2.] Whenever the $d$-dimensional theory has a holographic dual, what is the relation of the SymTFT to the gravitational sector of the holographic dual?
\end{enumerate}
We will answer both of these questions.

\paragraph{Non-Abelian BF+CS Theory and Boundary Conditions.}
An essential technical tool that we require to study the properties of this SymTFT is a detailed formulation of the topological defects of non-abelian BF (+ CS) theories. We rely on various results starting with Horowitz \cite{Horowitz:1989ng} and more recently the analysis of topological defects by  Cattaneo and Rossi \cite{Cattaneo:2000mc, Cattaneo:2002tk}. Related continuous SymTFTs (although for internal symmetries, abelian and non-abelian) were recently constructed in \cite{Antinucci:2024zjp, Apruzzi:2024htg, Brennan:2024fgj, Bonetti:2024cjk, Jia:2025jmn}. 
We show that the SymTFT results in topological defects which have the correct braiding relations. Furthermore, imposing Dirichlet boundary conditions (BCs) results in precisely the spacetime symmetry generators, as expected from a SymTFT. 

Another crucial input into the SymTFT framework is the set of boundary conditions (BCs). To our knowledge, a comprehensive analysis of BCs for non-abelian BF theories does not exist so far in the literature -- including for compact groups. We discuss several BCs, starting with the canonical, Dirichlet one, which is a gapped (topological) BC, which when placed as the symmetry boundary of the SymTFT, gives rise to the input symmetry. For our purposes this is the spacetime symmetry group $G_{\text{spacetime}}$. Starting from this, we can consider BCs that are obtained by gauging a non-anomalous subgroup of the spacetime symmetry group, that are (partial) Neumann BCs. These will be used as physical BCs in the context of the SymTFT. The results on this can be equally applied to compact groups and should have utility beyond the applications we consider here. 

{However, for continuous groups, spontaneous symmetry breaking results in gapless Goldstone modes. We thus have to consider modified Neumann BCs, which are gapless and give rise to effective theory of the Goldstone bosons after compactification of the SymTFT sandwich. Essentially this corresponds to adding a non-topological term to the partial Neumann BC, that is leading order in derivatives. This will be discussed in section \ref{sec:GaplessBC}.
}

\paragraph{How to move a point.} 
The Dirichlet boundary condition can be seen to implement the spacetime symmetry as follows:~
From the braiding of the topological defects in the SymTFT, and the induced action of symmetry generators on Dirichlet boundary conditions, we will be able to infer the action of symmetry generators on local operators, and show how they ``move (insertion) points", answering the first question above. This
shows that the SymTFT that we propose satisfies the main basic requirement
 for being the SymTFT for spacetime symmetries. 

\paragraph{Symmetry Breaking Phases from the SymTFT.}
One of the central utilities of SymTFTs is that they allow a study of symmetric ``phases". Applied to finite internal symmetries, they have successfully been applied to classification in 1+1d and 2+1d gapped and to some extent gapless phases, most importantly extending the classification beyond group-like symmetries to a categorical Landau paradigm \cite{Bhardwaj:2023fca, Bhardwaj:2023idu, Bhardwaj:2024kvy, Pace:2025hpb, Chatterjee:2024ych, Bhardwaj:2024wlr, Warman:2024lir, Aksoy:2025rmg, Bottini:2025hri, Lu:2025rwd, Bhardwaj:2024qiv, Xu:2024pwd, Bullimore:2024khm,  Bhardwaj:2025piv,  Inamura:2025cum,Chatterjee:2022tyg, Bhardwaj:2023bbf, Wen:2023otf, Bhardwaj:2024qrf,  Wen:2024qsg,  Bhardwaj:2025jtf, Wen:2025thg}.

The study of continuous symmetries and their breaking patterns is of course conceptually different, as we generically expect the appearance of a Goldstone mode. 
In the present context, we will apply the SymTFT to determine the symmetry breaking from e.g.~conformal symmetry to {Poincar\'e} symmetry, and derive the dilaton effective action. In the SymTFT, this arises by studying a boundary condition that breaks the spacetime symmetry (e.g.~the conformal symmetry for the dilaton action). A careful analysis of the compactification of the SymTFT, with one boundary realizing the conformal symmetry, and the other the spontaneous symmetry breaking, then yields the expected EFTs
for conformal symmetry breaking.

\paragraph{Brief Review of Spacetime Symmetry Breaking.}
Before outlining the characterization of symmetry breaking of spacetime symmetries in the SymTFT, it is useful to 
 briefly review some well-known results on this topic.
In contrast to discrete symmetries, breaking a continuous symmetry leads to Nambu-Goldstone (NG) modes that govern the low energy EFT. When spacetime symmetries are involved in the breaking pattern, a series of subtleties arise. For example, the number of NG modes effectively present at the low energy might be less than the number of symmetry generators broken \cite{Low:2001bw, Watanabe:2019xul}. This phenomenon is the so-called inverse Higgs effect \cite{Brauner:2014aha}, where redundant NG modes get integrated out. 

One case of relevance is when breaking patterns mix internal and spacetime symmetries. These have been used to provide a classification of various gapless phases \cite{Nicolis:2015sra} of relevance to condensed matter physics. {Here we will consider spacetime symmetries only, leaving this generalization for future work.} A systematic way to derive EFTs with these non-trivial breaking pattern is via the so-called coset construction \cite{Coleman:1969sm, Callan:1969sn} extended to spacetime symmetries \cite{ogievetsky1974nonlinear}, see also \cite{Delacretaz:2014oxa,Monin:2016jmo}. This prescription arises naturally in our SymTFT setup.

A paradigmatic instance of spacetime symmetry breaking is spontaneous breaking of conformal symmetry \cite{Schwimmer:2010za, Komargodski:2011vj}. In this case, the dilaton is the only NG boson present in the effective description. 
Its dynamics is partially governed by conformal anomalies, analogous to the WZW term included in pion Lagrangian to match chiral anomalies. These anomalies have a long history \cite{Duff:1993wm} and behave quite differently from ordinary internal symmetry anomalies. In 3+1d CFTs, there are two genuine anomalies, the $c$-anomaly (Type-B) and the the $a$-anomaly (Type-A). The latter is known to satisfy RG monotonicity theorems similarly to the $d=2$ $c$-anomaly \cite{Zamolodchikov:1986gt}. 

\paragraph{SymTFT Realization of Spacetime Symmetry Breaking.}
We illustrate the utility of the spacetime SymTFT by applying it precisely to this framework of conformal symmetry breaking. 
The starting point is the SymTFT for the conformal group, with the gapped symmetry boundary condition chosen to be Dirichlet. We construct the topological defects on this symmetry boundary explicitly. 
On the physical boundary, we place a partial Neumann bondary condition that breaks the symmetry from conformal to the Poincar\'e group. The SymTFT sandwich thus constructed, gives rise to the dilaton action after compactification. In odd $d$ dimensions, the SymTFT is simply the BF-theory for the conformal group and there is no anomaly. Whereas in even $d$, we include the CS-terms which we show, in $d=2,4$ give rise to the Type-A anomalies.

\paragraph{SymTFT and Gravity.}
To address the second question above,  we will show that the SymTFT we propose is indeed in certain cases a well-defined topological limit of gravity. In 2d gravity, the SymTFT for the conformal group is in fact JT gravity in first order BF-formulation. In higher dimensions, including in 3d, some more care needs to be taken and we discuss this in section \ref{sec:gravity}.
As a particularly interesting case, for 4d gravity with negative cosmological constant, we find that the BF-theory we propose is the \mbox{$G_N\to 0$} limit of a first-order formulation of general relativity. In higher dimensions, we expect similarly that the $G_N\to 0$ limit reduces gravity to the SymTFT, though we do not know a suitable higher dimensional analogue of the first-order formulation we use that would make generalisation of our arguments straightforward. One possible approach, which we will not explore here, could perhaps be to generalize the AKSZ formulation of gravity in \cite{Borsten:2024alh, Borsten:2025pvq}.  More broadly, we refer the reader to  \cite{Celada:2016jdt,Krasnov_2020} for reviews on BF-formulations of gravity.\\

We will spend the remainder of this introduction summarizing briefly the standard SymTFT construction for  internal symmetries and then give an overview of the SymTFTs for spacetime symmetries including boundary conditions relevant for the symmetry breaking. 

\paragraph{Recap:~SymTFT for Internal Symmetries.}
Let us recall the by now standard lore of classification of phases via the SymTFT (as it is applied to internal symmetries):~given a global symmetry $G$ (we focus here on groups but any fusion higher category is equally admissible), acting on a physical theory in $d$ spacetime dimensions, we gauge  the symmetry in $(d+1)$ dimensions, coupling it to flat $G$-background fields. This is the SymTFT, which has a BF-term for the gauge field of $G$, but there can be other topological couplings that capture anomalies etc. In our case, we will often have CS-terms. 

The most important aspects of the SymTFT are its topological defects:~they furnish both the symmetry $G$ as well as the charges under the symmetry.  For a BF-theory for a 0-form symmetry, these are topological defects of dimension $d-1$ and $1$, respectively
\be
U_{1}^\alpha(\Sigma_{1}) \quad \text{and} \quad U_{d-1}^a(\Sigma_{d-1}) \,,
\ee
defined on suitably dimensional subspaces $\Sigma$,
where $a$ and $\alpha$ specify some group-theoretical (or representation-theoretical) data. 
Note that crucially $U_{d-1}$ and $U_1$ link non-trivially in $(d+1)$ dimensions, which corresponds to the action of the symmetry on the (generalized) charges.

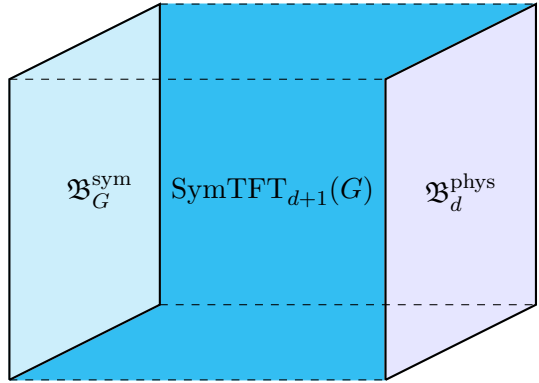
\begin{figure}
$$
\begin{tikzpicture}
\scalebox{1}{
\draw [cyan, fill= cyan, opacity =0.8]
(0,0) -- (0,4) -- (2,5) -- (7,5) -- (7,1) -- (5,0)--(0,0);
\draw [white, thick, fill=white,opacity=1]
(0,0) -- (0,4) -- (2,5) -- (2,1) -- (0,0);
\begin{scope}[shift={(5,0)}]
\draw [white, thick, fill=white,opacity=1]
(0,0) -- (0,4) -- (2,5) -- (2,1) -- (0,0);
\end{scope}
\begin{scope}[shift={(0,0)}]
\draw [cyan, thick, fill=cyan, opacity= 0.2]
(0,0) -- (0, 4) -- (2, 5) -- (2,1) -- (0,0);
\draw [black, thick]
(0,0) -- (0, 4) -- (2, 5) -- (2,1) -- (0,0);
\node at (1.2,2.5) {$\Bsym_{G}$};
\end{scope}
 \node at (3.5, 2.5) {$\SymTFT_{d+1}(G)$};
\begin{scope}[shift={(5,0)}]
\draw [black, thick, fill= blue,opacity=0.1] 
(0,0) -- (0, 4) -- (2, 5) -- (2,1) -- (0,0);
\draw [black, thick]
(0,0) -- (0, 4) -- (2, 5) -- (2,1) -- (0,0);
\node at (1,2.5) {$\Bphys_d$};
\end{scope}
\node at (3.3, 2.5) {};
\draw[dashed] (0,0) -- (5,0);
\draw[dashed] (0,4) -- (5,4);
\draw[dashed] (2,5) -- (7,5);
\draw[dashed] (2,1) -- (7,1);  
}
\end{tikzpicture}
$$
\caption{SymTFT setup with symmetry and physical BCs. We will consider $G$ to be a (continuous) spacetime symmetry, and $\Bsym_G$ the symmetry boundary that is a gapped boundary condition, on which the symmetry defects form the 0-form symmetry group $G$. The setup is applicable to any continous, abelian or non-abelian group $G$, but the examples we consider will be spacetime symmetries.
\label{fig:monkeydoodle}}
\end{figure}

The $(d+1)$-dimensional SymTFT is compactified on an interval with two sets of boundary conditions (BCs), see figure \ref{fig:monkeydoodle}:~the symmetry boundary, $\Bsym_G$, which is a gapped/topological boundary condition (BC), and the physical boundary $\Bphys$, which may or may not be gapped, depending on whether the initial theory was topological or not. This setup is referred to as the {\bf SymTFT sandwich} \cite{Ji:2019jhk, Gaiotto:2020iye, Apruzzi:2021nmk, Freed:2022qnc}. If one only considers the SymTFT with the symmetry boundary, which is useful for various computation, this is referred to as the {\bf SymTFT quiche}. 

One of the important tasks given a SymTFT is to determine its gapped BCs as they specify which symmetries can be realized. A canonical choice of gapped BC gives rise to the symmetry $G$ that we started with, and we call this the Dirichlet BC. 
Two symmetries that are both realizable on gapped BCs of a given SymTFT are related by topological manipulations (such as flat gauging). 

The physical boundary can be chosen to be topological -- if one is interested in symmetric gapped phase --  or non-topological. E.g.~chosing the same gapped Dirichlet boundary as the physical boundary gives rise to a spontaneous symmetry breaking (SSB). 
More generally, choosing the physical boundary condition to be  a partial Neumann boundary condition for the symmetry that remains intact, gives the associated SSB phase.
As we will discuss in depth later on, the setup is quite different for continuous symmetries, and spontaneous symmetry breaking of a continuous symmetry will always generate a Goldstone mode so that the relevant physical boundary conditions should be gapless ones. \footnote{SSB does not exist in $d=2$ QFTs if only internal symmetries are involved \cite{Mermin:1966fe,Hohenberg:1967zz,Coleman:1973ci}. Exceptions exists involving spacetime symmetries \cite{Komargodski:2021zzy}. Our procedure constructs non-linear sigma models with target $G/H$ based on symmetry, which can be well defined in the infinite volume limit or not.
}

\paragraph{Plan.}
The plan of the paper is as follows:~
Section \ref{sec:SymTFTST} provides an in depth analysis of BF+CS-theories for continuous non-abelian groups, their topological defects and boundary conditions. This is applicable to compact and non-compact continuous symmetry groups, and will have utility beyond the spacetime symmetry application.
The fundamental background  on BF-theories for continuous (not necessarily compact) symmetries is discussed in section~\ref{sec:BF-review}. The main SymTFT proposal for spacetime symmetries is then presented in section~\ref{sec:SymTFTProp}. 
Dirichlet and partial Neumann BCs are then discussed for non-abelian BF-theories in section \ref{sec:BCs} and BF+CS-theories in \ref{sec:BCsCS}. As this part of the analyis is also new for compact groups, we give an example of how the SymTFT sandwich is constructed when describing SSB for compact groups and how it results in the theory for the Goldstone boson in section \ref{sec:CompEx}. 
Our main application to spacetime symmetries is presented subsequently. 
We start with the conformal symmetry and study its  breaking in section \ref{sec:conformal-anomaly}. In particular we consider the cases of 3d and 5d bulk and 2d and 4d conformal anomalies. 
We discuss the action of topological defects associated to spacetime symmetries
in section~\ref{sec:moving-points}, thereby answering question Q1. Finally, we answer Q2 in section~\ref{sec:gravity}. Some future  applications of this framework are discussed in section \ref{sec:Yipee}. Various appendices summarize conventions and technical details.

\paragraph{Notation.} We indicate form degree as $\omega_p$ for $p$-forms, with the exception of one-form gauge connections $A$ and two-form curvatures $F$. Capital letters $A, B_{d-1}$ are used for dynamical fields while calligraphic letters, $\mathcal{A}, \mathcal{F}$, for backgrounds. Gauge transformations by $g \in G$ on connection and fields are indicated as $A \mapsto A^{(g)}$. For split algebras $\fg = \fh \oplus \fm$ we introduce projectors $\mathbb{P}_\fh, \mathbb{P}_\fm$ and leave their action implicit $\mathbb{P}_{\fh}(A) \equiv A_{\fh}$ etc.. For dual algebra split $\fg^* = \fh^* \oplus \fm^*$ we indicate $\mathbb{P}^{\fh^*}(B_{p}) \equiv B_p^{\fh^*}$ etc. Projections into sub-spaces are always taken \emph{after} gauge transformations, namely $A_{\fh}^{(g)} \equiv \mathbb{P}_{\fh}(A^{(g)})$ and similarly for $\mathfrak{m}$.

\section{SymTFT for Continuous Symmetries}
\label{sec:SymTFTST}

\subsection{Continuous Non-Abelian BF-theory} 
\label{sec:BF-review}

In this initial section, we will discuss the basics of non-abelian BF-theory for a continuous group $G$. 
We follow the exposition in \cite{Cattaneo:2000mc, Cattaneo:2002tk}, and assume that $G$ is compact, though many results will carry through to the non-compact case with some care.
The BF-theory will be defined in $(d+1)$ spacetime dimensions, so that the physical theory, obtained after compactification of the  SymTFT interval, is $d$-dimensional. 

\subsubsection{Lightning Review of Non-Abelian BF-Theories}
\label{sec:NA-BF-review}

The BF-action on a $(d+1$)-dimensional manifold $M_{d+1}$ with gauge group $G$ is given by the functional \footnote{We assume trivial $G$-bundles $P$. The treatment can be extended by taking forms valued in $\text{Ad}\, P$, the adjoint bundle of the $G$-bundle $P$, and $\text{Ad}^*\, P$, the co-adjoint bundle of the $G$-bundle $P$. } 
\be\label{eq:BF_action}
S_{\text{BF}} = \frac{i}{2\pi} \int_{M_{d+1}} \langle \,B_{d-1} , F \rangle \,, 
\ee
where 
\be 
\ba
B_{d-1} &\in \Omega^{d-1}(M_{d+1},\mathfrak{g}^* )\cr 
F \df \dd A + A\wedge A &\in \Omega^{2}(M_{d+1}, \mathfrak{g}) \,.
\ea
\ee
We take $\langle \, , \, \rangle$ to be the canonical 
inner product between $\mathfrak{g}$ and $\mathfrak{g}^*$, where $\fg$ is the Lie algebra of $G$ and $\mathfrak{g}^*$ is its dual. Later we will also refer to this product as $\langle\,,\,\rangle_{\BF}$ to distinguish it from the CS-form. 
This is a minimal choice corresponding to so-called ``canonical" BF-theories \cite{Cattaneo:2000mc, Cattaneo:2002tk}. If a non-degenerate Ad-invariant inner product exists on $\mathfrak{g}$, we can define a BF-theory where $B_{d-1}$ takes values in $\mathfrak{g}$ as well by using the inner product to construct an isomorphism between $\fg$ and $\fg^*$.

The equations of motion in the absence of boundaries are 
\be\label{eq:BF_EOM}
\ba
F = 0\,,\quad \dd_A B_{d-1} \coloneqq \dd B_{d-1} + \text{ad}_{A}^* B_{d-1} = 0\,.
\ea
\ee
Here, $\text{ad}_A^*:\mathfrak{g}^*\rightarrow \mathfrak{g}^*$ denotes the coadjoint action of $A$, defined as the Hermitian conjugate of the usual adjoint action of $A$, $\text{ad}_A:\mathfrak{g}\rightarrow \mathfrak{g}$. For a $k$-form $\omega\in\Omega^k(M_{d+1},\mathfrak{g})$, the adjoint action is given by
\ie
\text{ad}_A \omega = A\wedge \omega-(-1)^k\omega\wedge A\,,
\fe
and the coadjoint action satisfies
\ie
\langle \text{ad}_A\omega, \tilde\omega\rangle =-(-1)^k \langle  \omega, \text{ad}_A^*\tilde \omega\rangle\,.
\fe
The solutions to the equations of motion are  flat connections and covariantly-closed $(d-1)$-forms. 
The theory has a gauge symmetry
\be\label{Gtildi}
{\mathcal{G}} =   \Omega^{d-2}(M_{d+1},\mathfrak{g}^*) \rtimes_{\text{Ad}}\Omega^0(M_{d+1}, G)\,,
\ee
which
acts as \footnote{The coadjoint action of $g\in G$ on $B\in\mathfrak{g}^*$ is defined as the Hermitian conjugate of the adjoint action of $g$ on $A\in\mathfrak{g}$ with respect to the canonical inner product:~$\langle A, \text{Ad}_g^* B\rangle=\langle \text{Ad}_{g^{-1}} A,  B\rangle=\langle g^{-1}Ag,  B\rangle$. Here, we denote it as $\text{Ad}_g^* B=g^{-1} B g$. This notation makes sense if we have a non-degenerate inner product on $\mathfrak{g}$, such as trace. } 
\be
\ba
\label{eq:gauge_trans}
A &\ \mapsto\  A^{(g)} = g^{-1} A  g+ g^{-1} \dd g\, , \\ 
B_{d-1} &\ \mapsto \ B_{d-1}^{(g,\sigma)} = g^{-1} B_{d-1} g + \dd_{A^{(g)}} \sigma_{d-2} \, ,
\ea
\ee
with $g\in \Omega^0(M_{d+1}, G)$ and $\sigma_{d-2} \in \Omega^{d-2}(M_{d+1}, \mathfrak{g}^*)$.
Infinitesimally these read \footnote{We always take algebra generators to be anti-hermitian.} 
\begin{equation}
\delta_{(\epsilon)} A = \dd_A \epsilon , \quad \delta_{(\epsilon,\sigma)} B_{d-1} = \dd_A \sigma_{d-2} +  \text{ad}_{\epsilon}^*\, B_{d-1}\,.
\end{equation}
This gauge invariance is tightly related to the topological nature of the BF-theory \cite{Birmingham:1991ty}.  Diffeomorphisms generated by vector fields $\xi$ on $M_{d+1}$ \footnote{See appendix~\ref{sec:App_finite_transform} for a discussion of the extension to finite transformations.} as
\[
  \label{eq:inf-diffs-as-gauge}
\mathcal{L}_{\xi} A &= \iota_{\xi} (F) + \dd_A (\iota_{\xi} A)\,\\
\mathcal{L}_{\xi} B_{d-1} &= \iota_{\xi} (\dd_A B_{d-1}) + \dd_A (\iota_{\xi} B_{d-1}) + \text{ad}^*_{\iota_{\xi} A} \, B_{d-1} ,
\]
where $\mathcal{L}_{\xi}$ is the Lie derivative with respect to ${\xi}$ and $\iota_{\xi}$ is the interior product with ${\xi}$.
On shell, these are equivalent to an infinitesimal gauge transformation with parameters $\epsilon = \iota_{\xi} A, \, \sigma_{d-2} = \iota_{\xi} B_{d-1}$ \footnote{There is a small subtlety arising when we consider non-trivial principal $G$-bundles $P$. Gauge transformation parameters are elements of $\Omega^0(M, \text{Ad}\, P)$ with $\text{Ad}\,P$ the adjoint bundle of the $G$-bundle $P$, while $\iota_{\xi} A$ lives in the image under $\iota_{\xi}$ of $\text{Conn}(P,G)$, the space of connection of the $G$-bundle $P$. Working locally on opens $U \subset M_{d+1}$ (or for $P$ trivial $G$-bundle) there is no distinction between the two sets.}. This relation between gauge transformations and diffeomorphisms will play a key role in section~\ref{sec:moving-points}, where we discuss the interpretation of translations and rotations in our formalism.

\subsubsection{Topological Defects} 

In this section, we review the construction of topological operators in the BF-theory, as originally studied  in \cite{Cattaneo:2002tk}. The same construction has recently appeared in more detail in \cite{Jia:2025jmn} in applications to flavor symmetries. 

In the above BF-theory, we can define two sets of topological defects, that arise from the holonomies of $A$ and of $B_{d-1}$, respectively. The topological Wilson lines along a curve $\gamma$ are holonomies of $A$
\begin{equation}\label{sneez1}
\mathcal{U}_1^{\mathcal{R}}[\gamma] \coloneqq \Tr_{\cR}  \left(\mathcal{P}\, \exp\left\{  \oint_\gamma A \right\} \right)\,.
\end{equation}
These are labeled by representations $\mathcal{R}$ of the gauge group $G$ and in general can be decomposed into a direct sum of simple lines, each labeled by an irreducible representation (irrep). 

In addition, on a $(d-1)$-dimensional surface $\Sigma_{d-1} \subset M_{d+1}$ one can define topological defects that depend on the group conjugacy classes  $[g(X)=e^X]$  for $X \in \fg$ 
\be\label{holycow}
\mathcal{U}^{[g(X)]}_{d-1}[\Sigma_{d-1}]  \coloneqq \int \mathcal{D} \alpha_0 \mathcal{D} \beta_{d-2}\, \text{exp}\left\{ i  \oint_{\Sigma_{d-1}} \langle \alpha_0 , \dd_{A} \beta_{d-2}  + B_{d-1} \rangle   \right\}\,,
\ee
where 
\be\ba 
\alpha_0 &\in \Omega^0(\Sigma_{d-1}, [X] ) \,,\cr 
\beta_{d-2} &\in \Omega^{d-2}(\Sigma_{d-1} , \mathfrak{g}^*)\,.
\ea\ee
It is evident that $\mathcal{U}^{[e^X]}_{d-1}[\Sigma_{d-1}]$ is
gauge invariant, where the auxiliary fields transform as 
\be
\ba
\alpha_0 &\mapsto \alpha_0^{(g)} = g^{-1} \alpha_0 g \cr 
\beta_{d-2} &\mapsto \beta_{d-2}^{(g,\sigma)} = g^{-1} \beta_{d-2} g - \sigma_{d-2} \, .
\ea
\ee
We will show below that the operators~\eqref{holycow} are topological. 

More generally, the topological defects\footnote{{Strictly speaking these are part of a braided higher category, where we should think of the $d-1$ dimensional defects as objects and the lines as higher morphisms.}} are labeled by a conjugacy class $[g]$ with $g\in G$ and an irrep $\cR_{H_g}$ of the centralizer $H_g$ of any representative in $[g]$:
\be\label{combo}
\begin{aligned}
&\cU^{([e^X], \cR_{H_g})}[\Sigma_{d-1},\gamma\subset \Sigma_{d-1}] \coloneqq
\\
& \int \mathcal{D} U \mathcal{D} \beta_{d-1}\, \text{exp}\left\{ i  \oint_{\Sigma_{d-1}} \langle UXU^{-1} , \dd_{A} \beta_{d-2}  + B_{d-1} \rangle   \right\} \text{Tr}_{\cR_{H_g}}\!\left( P\exp\left\{\oint_{\gamma} A^{(U)}\right\}\right)\,,
\end{aligned}
\ee
where $U\in \Omega^0(\Sigma_{d-1}, G)$.
On $\Sigma_{d-1}$, there is an additional $H_g$ gauge symmetry, which together with the bulk gauge symmetry acts as
\ie
U\mapsto g^{-1}Uh\,,
\qquad
A^{(U)}\mapsto h^{-1} A^{(U)} h+h^{-1}\dd h~,
\fe
with $h\in \Omega^0(\Sigma_{d-1},H_g)$. Since $A^{(U)}$ transforms as an $H_g$-gauge field, we can use it to build an $H_g$ Wilson line. 
These more general operators can be interpreted as decorating the surface defect $\cU_{d-1}^{[g]}(\Sigma_{d-1})$ with an $H_g$-Wilson line on $\gamma\subset\Sigma_{d-1}$. Note that this Wilson line is generally stuck on $\Sigma_{d-1}$ and cannot move off it. 
The defects \eqref{sneez1} and \eqref{holycow} correspond to choosing $[g]=[\id]$, for which the centralizer is $G$, and $([g],1)$ (i.e.~the trivial irrep), respectively. 
The most general topological defect has in addition condensation defects of these lines stacked on top of the surfaces as in (\ref{combo}).

This is very similar to the structure of the topological defects in  BF-theory (or Dijkgraaf-Witten theory) with finite groups $G$, which is the SymTFT of the finite $G$ 0-form symmetry: for instance for 2+1d theories, the SymTFT has topological defects given by surfaces labeled by $[g]$ with $g\in G$ and on these surfaces, there are lines in irreps of the centralizer $H_g$ of $g$ \cite{Kong:2019brm, Bhardwaj:2025piv}.

\paragraph{Proof that $\cU_{d-1}$ is topological.} 
To show that this operator $\cU_{d-1}$ in (\ref{holycow}) is topological, assume first that 
$\alpha_0 \in \Omega^0(\Sigma_{d-1}, \mathfrak{g})$.
First  integrate out $\beta_{d-2}$, which localizes on $\alpha_0$ configurations which are covariantly constant 
\be
\dd_A \alpha_0 = 0\,.
\ee 
These solutions are in one-to-one correspondence with elements $X \in \mathfrak{g}$ which are invariant under the holonomy group of $A|_{\Sigma_{d-1}}$. They can be constructed from a reference point $x_0 \in \Sigma_{d-2}$ where $\alpha^X_0(x_0) = X$ by  applying parallel transport with $A$ throughout $\Sigma_{d-1}$.  Invariance under the holonomy group of the connection guarantees these solutions to be single-valued. More explicitly, they can be written as
\begin{equation}\label{eq:CovConstSol}
\alpha^X_0( x \, | \, x_0 , A) = \mathcal{U}_1[\gamma_{[x,x_0]}]^{-1}\, X \,\mathcal{U}_1[\gamma_{[x,x_0]}]\,,
\end{equation}
where $\cU_1[\gamma_{[x,x_0]}]\in G$ denotes the holonomy of the connection along $\gamma_{[x,x_0]}$ and the dependence on the specific path $\gamma_{[x,x_0]}$ chosen is immaterial due to the flatness  $F = 0$ from \eqref{eq:BF_EOM}\footnote{We assume trivial topology for the surface $\Sigma_{d-1}$.
}.  The surface operator can then be equivalently written as
\begin{equation}\label{eq:TopOp2}
\begin{aligned}
\mathcal{U}_{d-1}[\Sigma_{d-1}] &= \int_{G} \dd g(X) \,\exp\left\{ i \oint_{\Sigma_{d-1}}   \langle \alpha_0^X(x \, | \, x_0 ,A ) , B_{d-1} \rangle \right\} \,,
\end{aligned}
\end{equation}
where $g(X)= e^{X}$.
In this simplified expression, $\dd_AB_{d-1}=0$ as follows from \eqref{eq:BF_EOM}
\ie
\label{eq:closed-exponent}
\dd \langle \alpha_0^X(x \, | \, x_0 ,A ) , B_{d-1} \rangle =\langle X, \dd_A B_{d-1}\rangle=0 \,.
\fe
This implies that the integrand of \eqref{eq:TopOp2} is itself topological, i.e.~invariant under continuous deformation of $\Sigma_{d-1}$, and therefore so is the surface operator $\mathcal{U}[\Sigma_{d-1}]$.

In general, $\mathcal{U}[\Sigma_{d-1}]$ is reducible and it splits into multiple topological and gauge-invariant $(d-1)$-dimensional surface operators.
The integrand of \eqref{eq:TopOp2} is invariant under $B_{d-1} \rightarrow B_{d-1} + \dd_A \sigma_{d-2}$ for each fixed $X \in \mathfrak{g}$, but not invariant under $G$ gauge transformations. In fact, the solutions we found are only gauge covariant, and also depend on the reference point $x_0$:~
\begin{align}
\alpha_0^X(x \, | \, x_0, A^{(g)} ) &= g(x)\, \alpha_0^{X'} (x \, | \, x_0, A )\,  g(x)^{-1} & & \text{\ for }  X' = g(x_0)^{-1} X g(x_0) \, ,\nonumber\\
\alpha_0^X(x \, | \, x_0' , A) &= \alpha_0^{X} (x \, | \, x_0', A ) & &  \text{\ for } X' = \mathcal{U}_1[\gamma_{[x_0',x_0]}]^{-1} X \mathcal{U}_1[\gamma_{[x_0',x_0]}] \,.
\end{align}
From these follows that the integrand of \eqref{eq:TopOp2} alone cannot define a gauge invariant topological operator, labeled by Lie algebra elements $X\in\mathfrak{g}$. 
However, it can be made gauge invariant if integrated over the conjugacy class $[g(X)]$ of $X\in\mathfrak{g}$ and the simple components are labeled by conjugacy classes of the algebra $\mathfrak{g}$:
\begin{equation}
\mathcal{U}_{d-1}^{[g(X)]}[\Sigma_{d-1}] = \int_{[X]} \dd X \, \exp\left\{ i \oint_{\Sigma_{d-1}} \langle \alpha_0^X( x \, | \, x_0, A) , B_{d-1} \rangle  \right\} \, .
\end{equation}
Reintroducing $\beta_{d-2}$,  these operators are precisely \eqref{holycow}. 

\subsubsection{Linking of Topological Defects}

A crucial property of the topological defects is their linking (mathematically this is encoded in the braided structure of the category of defects). This determines for instance whether two defects can end on the same gapped boundary condition, and furthermore encode the charges under the symmetries. 

We will focus on the non-trivial linking for the defects  $\mathcal{U}_{d-1}^{[X]}[\Sigma_{d-1}]$ and $\mathcal{U}_1^{\mathcal{R}}[\gamma]$, which is due to the BF-term. The more general defects labeled by $([g], \cR)$ also have non-trivial linking which we do not consider here. 
The insertion of the surface operator introduces a source for the curvature as follows:
\begin{equation}
F(x) + \alpha_0^X (x\, | \, x_0, A) \delta^{(2)}(x \in \Sigma_{d-1} ) = 0 \, .
\end{equation}
The linking factor is the expectation value of the Wilson line $\mathcal{U}_1^{\mathcal{R}}[\gamma]$ for this solution. This can be evaluated using the non-abelian version of Stokes' theorem:
\begin{equation}
\tr_{\mathcal{R}} \, \mathcal{P} \exp\left\{\oint_\gamma A \right\} = \tr_{\mathcal{R}} \mathcal{P}_\gamma \, \exp\left\{ \int_{\Sigma_2 \, | \, \partial \Sigma_2 = \gamma} \mathcal{U}_1[\gamma_{[x,\bar{x}]}]^{-1} F(x)\, \mathcal{U}_1[\gamma_{[x,\bar{x}]}]  \right\}\,,
\end{equation}
where $\bar{x}\in \gamma$ is a base point for the loop $\gamma$, $\gamma_{[x,\bar{x}]}$ an arbitrary path connecting the base point and $x \in \Sigma_2$ and $\mathcal{P}_\gamma$ is path ordering along the boundary $\partial \Sigma_2 = \gamma$ (for more details see \cite{schreiber2011,Alvarez:1997ma}). The linking can then be determined to be 
\begin{equation}
\langle \mathcal{U}^{[X]}_{d-1}[\Sigma_{d-1}] \mathcal{U}_1^{\mathcal{R}}[\gamma] \rangle = \tr_{\mathcal{R}} \left[ e^{- X \text{Link}(\Sigma_{d-1},\gamma)} \right] \langle \mathcal{U}^{[X]}_{d-1}[\Sigma_{d-1}]  \rangle\,,
\end{equation}
where $\text{Link}(\Sigma_{d-1},\gamma)$ is the topological linking between $\gamma$ and $\Sigma_{d-1}$. 
Recall that our generators are anti-hermitian, so this linking is a phase for abelian compact groups. 
Note also, that for non-abelian group, the character can vanish for general $\mathcal{R}$ and $[X]$ so these topological operators are generally non-invertible. For non-compact group, $\mathcal{R}$ is generally infinite dimensional. In this case, the character might require more care and regularization. One example is discussed in the context of the Virasoro TQFT  in \cite{Collier:2023fwi}.

\subsubsection{Gapped Boundary Conditions}

In the context of the SymTFT approach, we are interested in BF-theories with boundaries. We will study these in detail in section \ref{sec:SymTFTST}. 
Here, let us summarize a few salient points about the variation of \eqref{eq:BF_action} in the presence of boundaries, which produces additional terms given by 
\begin{equation}
\delta S_{\rm BF}\vert_{\partial M_{d+1}} = \frac{i}{2\pi} \int_{\partial M_{d+1}} \langle \delta A , B_{d-1} \rangle\,.
\end{equation}
Gauge invariance under ${\mathcal{G}}$ as defined in (\ref{Gtildi}) is spoiled by a boundary term
\begin{equation}
S_{\rm BF}[A^{(g)} , B^{(g,\sigma)}] - S_{\rm BF}[A,B] = \frac{i}{2\pi}\int_{\partial M_{d+1}} \langle \sigma_{d-2}, g^{-1} F g \rangle\, .
\end{equation}
A consistent boundary conditions must have $\delta S_{\text{BF}}\vert_{\partial M_{d+1}}=0$ and can explicitly break the ${\mathcal{G}}$-gauge symmetry, which can be restored if desired  by introducing a Stückelberg field. 

The topological operator \eqref{holycow} splits further when inserted on a gapped boundary. For example, for a boundary with Dirichlet boundary condition of the form \footnote{An inhomogeneous boundary condition for $A$ has to be flat in order to be compatible with the bulk equation of motion. We also assume the boundary not to have any non-trivial cycles.}
\be\label{eq:~Dirich}
A|_{\partial M_{d+1}} \equiv \mathcal{A} = h^{-1} \dd h
\ee
with $h$ fixed, the solutions of $\dd_{\mathcal{A}} \alpha_0 = 0$ are simply $\alpha_0^X(x \vert x_0,\mathcal{A}) = h^{-1}(x) h(x_0)\, X\, h^{-1}(x_0) h(x) $. Thus, $\mathcal{U}_{d-1}[\Sigma_{d-1}]$ splits into gauge-invariant topological operators of the form
\begin{equation}
\mathcal{U}_{d-1}^{g(X)}[\Sigma_{d-1}] =  \exp\left\{ i \int_{\Sigma_{d-1}} \langle h^{-1}(x) h(x_0)\, X\, h^{-1}(x_0) h(x) , B \rangle  \right\}\, .
\end{equation}
In this case, we are not required to impose gauge invariance under $G$, since these are explicitly broken by the boundary condition. We can restore the $G$ gauge symmetry by introducing a Stückelberg field $U\in\Omega(\partial M_{d+1},G)$, which transforms as $U\rightarrow U g$. This amounts to replacing $h\rightarrow h U$ in all the expressions above. 
 These operators are labeled by elements of the entire algebra, and generate a non-abelian zero-form symmetry $G^{(0)}$ on the boundary theory. 

We will discuss other BCs in subsequent sections, in particular how flat gauging the symmetry results in partly Neumann BCs.

\subsection{SymTFT as Non-Abelian BF- and CS-Theory}
\label{sec:SymTFTProp}

We now propose the SymTFT for continuous spacetime symmetries. Consider the spacetime symmetry group $G$, e.g.~the Poincar\'e or Conformal Groups in $d$ spacetime dimensions. Then we show that the SymTFT is given by a combination of a (non-abelian) $G$-BF-theory and when $(d+1)$ is odd, additional terms that capture anomalies, given in terms of the CS-theory for $G$. 

Concretely, the SymTFT for a spacetime symmetry group $G$ is 
\be\label{SymTFTBFCS}
\ba
S_{\SymTFT} &= S_{\BF} + S_{\CS} \cr 
&= {i\over 2 \pi} \int_{M_{d+1}} \langle B_{d-1} , F \rangle_{\BF}  + \frac{i k}{(2\pi)^{n}{(n+1)!}} \CS_{d+1= 2n+1}(A) \,,
\ea
\ee
The details of the CS-term are provided in general odd dimension in appendix \ref{sec:CS_conventions}. Concretely for $d+1= 3,5$ they are 
\be\label{CS35}
\ba
S_{\CS}^{(3)} &= \frac{i k}{2(2\pi)} \int_{M_3} \left[\left\langle A, F \right\rangle_{\CS} - \frac{1}{3}\left\langle A, A \wedge A \right\rangle_{\CS}\right] \cr 
S_{\CS}^{(5)} &= \frac{i k}{6(2\pi)^2} \int_{M_5} \left[\left\langle A, F,F\right\rangle_{\CS} - \frac{1}{2} \left\langle A, A \wedge A , F\right\rangle_{\CS} + \frac{1}{10} \left\langle A, A \wedge A , A \wedge A \right\rangle_{\CS}\right] \,.
\ea
\ee
The multilinear bracket $\langle ... \rangle_{\CS}$ is defined in (\ref{eq:CS_innerproduct}), and is to be distinguished from the BF one, that we introduced in (\ref{eq:BF_action}).
 We should make a few comments before studying the important question of boundary conditions for this SymTFT. 
If $G$ is non-compact, this  requires some modification compared to the compact, non-abelian $G$ BF-theories studied in the last subsection. We will discuss the related subtleties in section \ref{sec:conformal-anomaly} when applying the formalism to the conformal group.

\subsection{Dirichlet and Partial Neumann BCs for the BF-theory}
\label{sec:BCs}

In this section we will determine some gapped boundary conditions (BCs) for the non-abelian BF-theory. We will present first the standard description, which generically explicitly breaks gauge transformations at the boundary. 
In this picture, some topological operators can end on such boundary and the endpoints (interpreted as generalized charges) will transform under the broken gauge symmetries, which then must be interpreted as global symmetries.  

The second formulation will restore full gauge invariance by introducing St\"uckelberg fields on the boundary. In this case, topological operators which in the first formulation were allowed to end on the boundary, now they must end on operators built out of St\"uckelberg fields. However, the symmetry action on these endpoints, gives rise to the same generalized charge, as is determined by the bulk linking (independently of the St\"uckelbergs).
This formulation is useful as it will allow us to perform the SymTFT sandwich compactification to $d$-dimensions more straightforwardly.

\subsubsection{Boundary Conditions for Non-Abelian BF-theories}

We now discuss gapped/gapless boundary conditions for the SymTFT. Our starting point will be the Dirichlet boundary condition, which realizes the original $G$ global symmetry. We obtain the other BCs by (flat) gauging a subgroup $H$, these are the partial Neumann boundary conditions. 

\paragraph{Dirichlet BC.}
The Dirichlet boundary condition of the type $A|_{\partial M_{d+1}} = \mathcal{A}$ where $\mathcal{A}$ is a fixed flat connection can be imposed with the action 
\begin{empheq}[box=\fbox]{align}
\Dir(G):\qquad S_{\rm bdry} = - \frac{i}{2\pi} \int_{\partial M_{d+1}} \left\langle A- \mathcal{A} , B_{d-1}\right\rangle_{\BF} \, .
\end{empheq}
This implies the standard condition for Dirichlet boundaries where the gauge field is fixed to a particular value
\be
A= \cA \,.
\ee
Consistency with the bulk equation of motion simply constraint the auxiliary field boundary value $B|_{\partial M_{d+1}}$ to be $\dd_{\mathcal{A}}$-closed. 

In the SymTFT we will use this Dirichlet BC as the symmetry boundary $\Bsym_G$ for the symmetry $G$. We show in section  \ref{app:Charges} that on this boundary the topological defects that cannot end, but are confined give rise to the generators of $G$.

\paragraph{Partial Neumann BCs.}
To obtain a partial Neumann boundary conditions,  consider the generic case in which the algebra can be split as 
\be
\mathfrak{g} = \mathfrak{h} \oplus \mathfrak{m}\,,
\ee
where $\mathfrak{h}$ is a subalgebra generating the Lie algebra associated to the subgroup $H < G$. Given such a split, we can define projectors on the algebra subspaces $\mathbb{P}_{\mathfrak{h}}$, $\mathbb{P}_{\mathfrak{m}}$ and the dual subspace $\mathbb{P}^{\mathfrak{h}^*}$, $ \mathbb{P}^{\mathfrak{m}^*}$.
In the following we will consider the instance when $\mathfrak{g}$ and $\mathfrak{h}$ form a reductive pair, i.e.~$G/H$ is a reductive coset, i.e.~
\be\label{eq:reductive}
[\mathfrak{h}, \mathfrak{h}] \subseteq \mathfrak{h}, \quad [\mathfrak{h}, \mathfrak{m}] \subseteq \mathfrak{m} \,.
\ee
This simplifies some of the analysis, in the following, although many arguments can be carried through for the non-reductive cases as well\footnote{See \cite{Son:2008ye, Balasubramanian:2008dm, Schafer-Nameki:2009dsc} for instances of non-reductive cosets.}.

The boundary condition that in BF-theory imposes Neumann BCs for $\mathfrak{h}$-components of $A$ is
\begin{empheq}[box=\fbox]{align}\label{eq:Neumann}
\Neu(G, H):\qquad S_{\rm bdry} = - \frac{i}{2\pi} \int_{\partial M_{d+1}} \langle  A_{\mathfrak{m}}, B_{d-1} \rangle_{\BF}\,,
\end{empheq}
where $ A_{\mathfrak{m}}=\mathbb{P}_{\mathfrak{m}}(A)$ and  the inner product will select out the component of $B$ along $\fm, \fm^*$, i.e.~$ B_{d-1}^{\mathfrak{m}^*}=\mathbb{P}_{\mathfrak{m}^*}(B_{d-1})$ \footnote{The space $\fh^* , \fm^*$ are simply defined as the spaces of functional which vanish on $\fm, \fh \subset \fg$ respectively. The canonical inner product, being defined as $\langle X, Y^* \rangle_{\BF} \coloneqq Y^*(X)$ automatically satisfies $\langle \fh, \fm^* \rangle_{\BF} = \langle \fm, \fh^* \rangle_{\BF} = 0$.}. 

In the SymTFT considerations, this partial Neumann BC will be used in terms of the physical boundary. The sandwich then corresponds to the spontaneous symmetry breaking from $G$ to the subgroup $H$.

\paragraph{Equations of motions for the partial Neumann BC.}
Let us work out explicitly the solution of the variational problem given by bulk and boundary with this action. The joint bulk and boundary action variation gives
\begin{equation}
    \frac{i}{2\pi} \int_{\partial M_{d+1}}\left(\langle \delta A_\fh , B^{\fh^*}_{d-1} \rangle + \langle \delta A_\fm , B^{\fm^*}_{d-1} \rangle - \langle \delta A_\fm, B^{\fm^*}_{d-1} \rangle - \langle  A_\fm , \delta B^{\fm^*}_{d-1} \rangle\right)=0  \,,
\end{equation}
More generally we have 
\begin{equation}\label{eq:BC_Neumann}
 B_{d-1}^{\mathfrak{h}^*} |_{\partial M_{d+1}} = 0 \,, \qquad A_{\mathfrak{m}} |_{\partial M_{d+1}} = 0 \, .
\end{equation}
Consistency with the bulk equation of motion require $A_\mathfrak{h}|_{\partial M_{d+1}}$ to be a flat $H$-connection. The other equation of motion instead is
\be\label{Bmhm}
0 = \dd B_{d-1}^{\mathfrak{m}^*}+ \text{ad}_{A_{\mathfrak{h}}}^* B_{d-1}^{\mathfrak{m}^*} \, .
\ee
To determine wether this condition is consistent with \eqref{eq:BC_Neumann}, pick dual basis 
$\fm = \text{span} \{M_i\}$, 
$\fh = \text{span} \{H_a \}$ and likewise for the duals,
 such that $ \langle M_i , M_j^* \rangle_{\BF} = \delta_{ij}, \, \langle H_a, H_b^* \rangle_{\BF} = \delta_{ab}$ and zero otherwise. Then this equation projected on $\fm, \fh$ gives
\be
\ba
0 &= \dd B_{d-1}^{\fm^*, i} - A_\fh^a \wedge B^{\fm^*, j}_{d-1}\langle [H_a, M_i], M_j \rangle_{BF} \\
0 &= A_\fh^a \wedge B^{\fm^* j}_{d-1} \langle [H_a, H_b] , M_j \rangle_{\BF}\,.
\ea\ee
Using the orthogonality of $\mathfrak{h}$ and $\mathfrak{m}^*$ etc, the second equation vanishes automatically, and the first only gets contributions when $[H_a, M_i] \in \mathfrak{m}$.

\paragraph{Partial Neumann from Gauging.}
\label{sec:gaugings}
We can get the partial Neumann BC also from the Dirichlet one by a partial flat gauging of $H$. Consider the Dirichlet quiche configuration on $M_d \times \mathbb{R}^+$ with $\partial M_d =\emptyset$, corresponding to the path integral 
\begin{equation}
\mathcal{Z}_{\Dir(G)}[\mathcal{A}] = \int \frac{\mathcal{D} A \mathcal{D} B_{d-1}}{\text{Vol}\, \mathcal{G}}\, \exp\left\{  \frac{i}{2\pi} \int_{M_d \times \mathbb{R}^+} \langle F, B_{d-1} \rangle_{\rm BF} - \frac{i}{2\pi} \int_{M_d} \langle A - \mathcal{A}, B_{d-1} \rangle_{\rm BF}  \right\} \, .
\end{equation}
This is well defined on gauge equivalence classes
\begin{equation}
\mathcal{Z}_{\Dir}[g^{-1} \mathcal{A} g + g^{-1} \dd g] = \mathcal{Z}_{\Dir}[\mathcal{A}]\, ,
\end{equation}
and as such we can perform (partial) gaugings. To get partial Neumann conditions for a subgroup $H < G$ with algebra split $\mathfrak{g} = \mathfrak{h} \oplus \mathfrak{m}$, we can start for a Dirichlet boundary condition which takes values in $\mathfrak{h}$, namely $\mathcal{A}^{\mathfrak{h}} \in \Omega^1(M_d,\mathfrak{h})$. We then perform gauging as follows:
\begin{equation}
\mathcal{Z}_{\Neu(G,H)}[\mathcal{B}_{d-2}] = \int \frac{\mathcal{D} \mathcal{A}^\mathfrak{h}}{\text{Vol} \mathcal{H}}\, \exp\left\{- \frac{i}{2\pi} \int_{M_d} \langle \mathcal{A}^{\mathfrak{h}} , \dd_{\mathcal{A}^\mathfrak{h}} \mathcal{B}_{d-2}^{\mathfrak{h}^*} \rangle \right\} \mathcal{Z}_{\Dir(G)}[\mathcal{A}^\mathfrak{h}] \, ,
\end{equation}
where now the background field $\mathcal{A}^{\mathfrak{h}}$ is regarded as dynamical, and $\mathcal{B}_{d-2}^{\mathfrak{h}^*}$ is the dual field. From the resulting quiche configuration, one can integrate out $\mathcal{A}^{\mathfrak{h}}$ via its equation of motions. One then gets 
\begin{equation}
\label{NeuGHFromGauging}
\ba
\mathcal{Z}_{\Neu(G,H)}[\mathcal{B}_{d-2}^{\mathfrak{h}^*}] & = \int \frac{\mathcal{D} A \mathcal{D} B_{d-1}} {\text{Vol}\, \mathcal{G}}\times\cr 
& \times \exp\left\{  \frac{i}{2\pi} \int_{M_d \times \mathbb{R}^+} \langle F, B_{d-1} \rangle_{\rm BF} - \frac{i}{2\pi} \int_{M_d}  \left( \langle A^{\mathfrak{m}} , B_{d-1}^{\mathfrak{m}^*} \rangle_{\rm BF} 
+ \langle A^{\mathfrak{h}}, \dd_{A^{\mathfrak{h}}} \mathcal{B}_{d-2}^{\mathfrak{h}^*} \rangle \right)  \right\} \,.
\ea
\end{equation}
This results in the same equations of motion as those we obtained in (\ref{eq:Neumann}), modulo the $\mathcal{B}_{d-2}^{\mathfrak{h}^*}$ background for the dual symmetry.

\subsubsection{Gapped Boundary Conditions with St\"uckelberg Fields}

When considering the SymTFT we will use the $\Dir (G)$ BC in order to fix the symmetry boundary. The topological defects on it are precisely generators of the symmetry group $G$. Likewise we will consider the partial Neumann $\Neu (G,H)$ as physical boundary, to describe the SSB from $G$ to $H$. 

However, in the compactification, i.e.~the actual dimensional reduction to $d$ dimensions, it is useful to retain the full (remaining) gauge invariance of the system. This is achieved by considering  gauge-invariant versions of the Dirichlet and partial Neumann boundary conditions. 

We will now construct such boundary actions that are invariant under the global component of (\ref{eq:gauge_trans}). To restore gauge invariance, it is sufficient to introduce St\"uckelberg fields $U :~\partial M_{d+1} \rightarrow G$ and $\lambda_{d-2}$ transforming as 
\be\label{eq:SuckGaugeTransf}
U \mapsto g^{-1} U\,,\quad\lambda_{d-2} \rightarrow g^{-1}\lambda_{d-2} g+\sigma\,.
\ee
The fully gauge invariant version of the $\Dir(G)$ BC is then 
\begin{empheq}[box=\fbox]{align}\label{DirichletEven}
D(G):~\quad S_{\rm bdry} =& - \frac{i}{2\pi} \int_{\partial M_{d+1}} \left\langle A - \cA^{(U^{-1})} , B_{d-1} -\dd_A\lambda_{d-2}\right\rangle_{\BF} \nonumber\\
& -\frac{i}{2\pi}\int_{\partial M_{d+1}} \langle \lambda_{d-2}, F  \rangle_{\BF}, \quad \text{with} \quad \mathcal{F} = 0 \,. 
\end{empheq}
Here 
\be
\cA^{(U)}=U^{-1} \cA U+U^{-1}dU\,.
\ee
The fully Neumann boundary condition can be instead realized simply by
\begin{equation}
\boxed{
N(G) :~\quad S_{\rm bdry} = -\frac{i}{2\pi}\int_{\partial M_{d+1}} \langle \lambda_{d-2}, F_A  \rangle\,.
}
\end{equation}

The partial Neumann boundary condition  can again be obtained from the partial gauging from Dirichlet, i.e.~analogous to (\ref{NeuGHFromGauging}), applied to $D(G)$ instead of $\Dir (G)$. 
We can integrate out $A^\mathfrak{h}$ and this remains fully gauge-invariant, if $\mathfrak{g}=\mathfrak{h} \oplus \mathfrak{m}$ are such that $(\mathfrak{g}, \mathfrak{h})$ correspond to a reductive coset. As stated earlier, we make the simplifying assumption of reductiveness, although it is not strictly necessary -- we can also obtain the gauge-invariant BC analogous to (\ref{NeuGHFromGauging}). 
Once we  restore full gauge invariance with the  St\"uckelberg fields the partial Neumann BC becomes 
\begin{equation}\label{eq:NeumannBC}
\boxed{
N(G,H):~\quad S_{\rm bdry} =
- \frac{i}{2\pi} \int_{\partial M_{d+1}} \langle  A_{\mathfrak{m}}, B_{d-1} - \dd_A \lambda_{d-2} \rangle_{\BF} - {i\over 2 \pi} \int \langle \lambda_{d-2}, F_A \rangle \,.
}
\end{equation}
This boundary condition is gauge invariant under $G,H$ implemented as
\begin{align}
H:~&\quad U \mapsto h^{-1} U h, \quad A \mapsto A^{(h)}, \quad B \mapsto B^{(h)} \nonumber\\
G:~&\quad U \mapsto g^{-1} U, \quad A \mapsto A^{(g)}, \quad B \mapsto B^{(g)}\,.
\end{align}

\subsubsection{{Gapless Boundary Conditions}}
\label{sec:GaplessBC}

When considering SSBs for continuous symmetries from $G$ to a subgroup $H$, it is important to also characterize gapless BCs, which incorporate the Goldstone bosons of the symmetry breaking.
We can consider BF-theories {with a non-canonical pairing, which for instance appears} whenever $\fg$ admits a non-degenerate Killing form, so that $B_{d-1}$ can be taken to be valued in $\fg$. We refrain from providing a complete classification, focusing instead on the ones that we will use in the next section to construct SSB phases. 
 We will refer to these as {\bf modified (partial) Neumann BCs}, $N^*(G,H)$. 
Note that we again refer to the decomposition $\mathfrak{g} = \mathfrak{h} \oplus \mathfrak{m}$. 
The action differs from the gapped Neumann BC as follows:  
\be 
\ba \label{eq:modNeu}
S_{N^*(G,H)}=&   - \frac{i}{2\pi} \int_{\partial M_{d+1}} \left\langle  A_\mathfrak{m}  , B_{d-1}\right\rangle_{\kappa}  +{f^2\over 2} \left \langle B_{d-1} , * B_{d-1} \right \rangle_{\kappa}  \,,
\ea
\ee
where $\langle \; , \; \rangle_{\kappa}$ is the quadratic pairing with the Killing form $\kappa_{ab}={\rm Tr}(T_aT_b)$ and the generators of the Lie algebra are $T_a,T_b\in \mathfrak{g}$. {The algebra generators can always be chosen in such a way that the split $\fg = \fm \oplus \fh$ is orthogonal with respect to the Killing form if this is non-degenerate.}
This modified Neumann BC $N^*(G,H)$ includes the singleton term $\langle B_{d-1}, * B_{d-1}\rangle_{\kappa}$ with dimensionful coupling $f^2$ and preserves the  $H$-gauge symmetry. This term is the  leading order non-topological term in the fields that we have in the bulk \cite{Maldacena:2001ss}. When applied in the SymTFT sandwich, it will give rise to the action -- in particular the kinetic term -- of the Goldstone boson  that captures the SSB. 

Note that the above strictly applies to internal symmetries, because in the $*$ operation we are choosing a particular metric. In the application to  spacetime symmetries, however, we will not commit to any particular choice of the metric and therefore we will define a metric independent Hodge dual operation $\Hod$. 
This will be discussed in detail in section \ref{sec:Sandwichodd} and appendix \ref{app:Hodge}.

\subsection{Boundary Conditions for BF+CS-Theory}
\label{sec:BCsCS}

When extending the system with a CS term in the bulk with dimension $d+1=2n+1$, the bulk equation of motion are still
\begin{equation}
F = \dd_A B = 0 \, .
\end{equation}
However, new terms are present in the boundary variations and in general will obstruct the existence of some boundary conditions. 

\paragraph{Three-Dimensional Bulk.} In $d+1=3$ the full boundary variation reads
\begin{align}
\delta S|_{\partial M_{3}} = \frac{i}{2\pi} \int_{\partial M_3} \left\{  \langle \delta A , B_1 \rangle_{\rm BF} + \frac{k}{2}  \langle \delta A, A \rangle_{\rm CS}   \right\} \, .
\end{align}
To ensure a good variational principle, one has to modify the Dirichlet boundary conditions as follows (this is thus motivated in a similar way to the improvement terms required in making the standard holographic variational principle well-defined in which case one adds the Gibbons-Hawking-York terms):
\begin{align}
D_k(G)^{(\text{3d})}:~\quad S_{\rm bdry} &= - \frac{i}{2\pi} \int_{\partial M_{3}} \Bigg\{ \langle A -\cA , B_1\rangle_{\rm BF} + \frac{k}{2} \langle A, \cA \rangle_{\rm CS} \Bigg\} \, .
\end{align}
To restore full $G$-gauge invariance of the Dirichlet BC action we introduce St\"uckelberg fields as in the previous section. There is a gauge variation localized on $\partial M_3$ coming from the bulk Chern-Simons functional. To cancel that, one needs to introduce a specific topological action for $U$ coupled to $A|_{\partial M_3}$. How to derive such action in arbitrary dimension is outlined in \ref{sec:CS_conventions}. In $d=3$, one extends $U$ to an arbitrary three-manifold $X_3$ with $\partial X_3 = \partial M_3$ and the correct lagrangian to consider on $X_3$ turns out to be 
\begin{equation}\label{eq:Gamma3}
\Gamma_3(U,A) = \langle  ( U \dd U^{-1})^3 \rangle_{\rm CS} - \dd \langle (U \dd U^{-1} ) A  \rangle_{\rm CS} \, .
\end{equation}
All together, the Dirichlet boundary conditions become
\begin{empheq}[box=\fbox]{align}\label{DirichletEven3d}
D_k(G)^{(\text{3d})}:~\quad S_{\rm bdry} =& 
 - \frac{i}{2\pi} \int_{\partial M_{3}} \left\langle A - \cA^{(U^{-1})} , B_{1} -\dd_A\lambda_{0}\right\rangle_{\BF}  -\frac{i}{2\pi}\int_{\partial M_{3}} \langle \lambda_{0}, F  \rangle_{\BF} \cr 
&
- \frac{i}{2\pi} \int_{\partial M_{3}}   \frac{k}{2} \left\langle A^{(U)}, \cA \right\rangle_{\rm CS} 
- \frac{i k}{2(2\pi)} \int_{X_3} \Gamma_3(U,A) \, .
\end{empheq}
with $\mathcal{F} = 0$.
An analogous procedure is carried for the action defining the Neumann boundary condition $N(G,H)$. In this case, the characterization of topological boundary conditions depends on the structure of the CS inner product restricted to the components of the split $\mathfrak{g} = \mathfrak{h} \oplus \mathfrak{m}$. The corresponding boundary condition exists only if $\langle \mathfrak{h}, \mathfrak{h} \rangle_{\rm CS} = 0$ \footnote{In the absence of BF terms, $\mathfrak{h}$ we must also impose $\mathfrak{h}$ to be Lagrangian, i.e.~$\text{dim}\,\mathfrak{h} = \text{dim} \mathfrak{g}/2$.}. We will then work with the following boundary action
\begin{equation}
N_k(G,H)^{(\text{3d})}:~\quad S_{\rm bdry} = - \frac{i}{2\pi} \int_{\partial M_3} \left\{ \langle A_{\mathfrak{m}} , B_{1} \rangle_{\rm BF} + \frac{k}{2} \langle A_{\mathfrak{m}} , A_{\mathfrak{h}} \rangle_{\rm CS} \right\} \, .
\end{equation}
This boundary action includes an improvement term which was was also obtained e.g.~in \cite{Arvanitakis:2024vhz} without BF terms. The variational problem for this action gives the same boundary conditions as in \eqref{eq:BC_Neumann}, and it is evidently $H$-gauge invariant \footnote{The gauge variation proportional to the WZW term vanishes due to $\langle \mathfrak{h}, \mathfrak{h} \rangle_{\rm CS} =0$ even if it's integrated over a manifold with boundary and large $H$-gauge transformations might be present.} 
Once again, the reductive structure \eqref{eq:reductive} guarantees the whole system to be invariant under gauge transformations which take values in $H$ on $\partial M_3$. 

Equivalently, we can again derive $N_k(G,H)^{(3d)}$ from an $H$-gauging applied to the Dirichlet boundary condition $D_k(G)^{(3d)}$ analogous to the discussion in \ref{sec:gaugings}. The requirement for $\langle \mathfrak{h}, \mathfrak{h} \rangle_{\rm CS} =0$ is necessary and equivalent to gaugeability of the subgroup $H$. 

We can restore full gauge symmetry by introducing St\"uckelbergs as follows:
\begin{empheq}[box=\fbox]{align}\label{NeumannEven}
N_k(G,H)^{(\text{3d})}:~\quad S_{\rm bdry} =& 
- \frac{i}{2\pi} \int_{\partial M_{3}} \langle  A_{\mathfrak{m}}, B_{d-1} - \dd_A \lambda_0 \rangle_{\BF} - {i\over 2 \pi} \int \langle \lambda_0, F \rangle\cr 
&+ \frac{k}{2(2\pi)} \int_{\partial M_3} \left\langle A^{(U)}_{\fm}, A^{(U)}\right\rangle_{\rm CS}  - \frac{ik}{2(2\pi)} \int_{X_3} \Gamma_3(U,A) \, .
\end{empheq}

\paragraph{Five-Dimensional Bulk.} 
\label{sec:5dBCS}

These boundary conditions can be generalised to higher dimensions. In $d=5$, the CS-functional is defined with a tri-linear adjoint-invariant product. The total boundary variation of the BF+ CS system reads 
\begin{align}
\delta S_{\rm}|_{\partial M_5} &= \frac{i}{(2\pi)} \int_{M_5} \left\{ \langle \delta A, B_3 \rangle_{\rm BF} + \frac{k}{3(2\pi)} \int_{\partial M_5} \left\langle \delta A, A, \left( F - \frac{1}{4} A^2 \right) \right\rangle_{\rm CS} \right\} \,.
\end{align}
Going through similar steps as in 3d, in particular requiring that the combination of bulk and boundary terms are gauge invariant, the Dirichlet boundary condition takes the form 
\footnote{We drop terms proportional to $\mathcal{F}$ as the boundary configuration must be flat.}
\begin{empheq}[box=\fbox]{align}\label{Dirichletodd5d}
 &{D_k(G)^{(\text{5d})}} \cr 
& = 
 - \frac{i}{2\pi} \int_{\partial M_{5}} \left\langle A - \cA^{(U^{-1})} , B_{3} -\dd_A\lambda_{2}\right\rangle_{\BF}  -\frac{i}{2\pi}\int_{\partial M_{5}} \langle \lambda_{2}, F  \rangle_{\BF} \cr 
&- \frac{i k}{6(2\pi)^2} \int_{\partial M_5} \left\langle A^{(U)} , \cA , \dd A^{(U)} + \dd \cA + \frac{1}{2} A^{(U)} \wedge A^{(U)} + \frac{1}{2} \cA \wedge \cA + \frac{1}{4} [A^{(U)},\cA] \right\rangle_{\CS}
\cr
&- \frac{i k}{6(2\pi)^2} \int_{X_5} \Gamma_5(U,A) \, , \quad \text{with} \quad \mathcal{F} = 0
\end{empheq}
Note that the second line comes from the transgression terms (\ref{TransgressionTerms}). Notice that generically quiche partition functions are not gauge invariant under $G$:
\begin{equation}
\mathcal{Z}_D[g^{-1} \cA g + g^{-1} \dd g] \neq \mathcal{Z}_D[\cA] \, .
\end{equation}
Thus the gauging of this boundary condition to a full Neumann $N_k(G,G)$ is obstructed. However, subgroups $H<G$ such that $\langle \fh, \fh , \fh\rangle = 0$ can be gauged as in section \ref{sec:gaugings} \footnote{This is due to the fact that we did not  include an auxiliary CS-functional of the background $\cA$ as one should to define in the transgression (\ref{Tterm})}. 
The result of such gauging in $5d$ defines the partial Neumann condition as follows:~
\begin{empheq}[]{align}
N_k(G,H)^{(\text{5d})}:~&\cr 
 S_{\rm bdry} =& - \frac{i}{2\pi} \int_{\partial M_{5}} \Bigg\{ \left\langle A_{\mathfrak{m}} , B_{3} \right\rangle_{\rm BF}  \Bigg\} \nonumber\\
&- \frac{i k}{6(2\pi)^2} \int_{\partial M_5} \left\langle A , A_{\mathfrak{h}} , \dd A + \dd A_{\mathfrak{h}} + \frac{1}{2} A^2 + \frac{1}{2} A_{\mathfrak{h}}^2 + \frac{1}{4} [A,A_{\mathfrak{h}}] \right\rangle \, .
\end{empheq}
The fully dressed analog with St\"uckelberg terms is 
\begin{empheq}[box=\fbox]{align}
\label{Neumann5d}
N_k(G,H)^{(\text{5d})} &=
 - \frac{i}{2\pi} \int_{\partial M_{5}} \langle  A_{\mathfrak{m}}, B_{3} - \dd_A \lambda_2 \rangle_{\BF} - {i\over 2 \pi} \int \langle \lambda_2, F \rangle\cr  
&- \frac{i k}{6(2\pi)^2} \int_{\partial M_5} 
\left\langle A^{(U)} , A^{(U)}_{\fh}  , \dd A^{(U)} + \dd A^{(U)}_\fh  \right.\cr 
 &\left.\qquad \qquad \qquad\  
+ \frac{1}{2} A^{(U)} \wedge A^{(U)} + \frac{1}{2} A^{(U)}_\fh \wedge A^{(U)}_\fh + \frac{1}{4} [A^{(U)},A^{(U)}_\fh ] 
\right\rangle 
\cr
&- \frac{i k}{6(2\pi)^2} \int_{X_5} \Gamma_5(U,A) 
\end{empheq}

\subsection{{Example:~SymTFT Compactification for Compact Groups}}
\label{sec:CompEx}

Let us evaluate one of the SymTFTs with the choice of BCs given above. As we will discuss the spacetime symmetries in detail in subsequent sections, it is worthwhile considering an application to internal symmetries (to which this analysis is applicable as well). 
Let's consider the following SymTFT sandwich:
\be
\begin{tikzpicture}
\begin{scope}[shift={(0,0)}]
\draw [cyan,  fill=cyan, opacity = 0.8] 
(0,0) -- (0,3) --(4,3) -- (4,0) -- (0,0) ; 
\draw [white](0,0) -- (0,3) --(4,3) -- (4,0) -- (0,0) ;  ; 
\draw [very thick] (0,0) -- (0,3)  ;
\draw [very thick] (4,3) -- (4,0) ;
\node at (2,1.5) {$\BF(G)$} ;
\node[above] at (0,3) {$\Bsym = D(G)$}; 
\node[above] at (4,3) {$\Bphys = N^*(G,H) $}; 
\end{scope}
\end{tikzpicture}
\ee
The physical boundary condition is chosen here, to be a gapless modified Neumann boundary introduced in section \ref{sec:GaplessBC}. 
We will consider the case with even bulk dimension $d+1 = 2n$, so that we have the BF-terms only. The total system has action that is the combination of the Dirichlet on the left and partial modified Neumann on the right. Note that all the St\"uckelberg fields drop out, due to the Dirichlet BC and the flatness $\cF=0$. The reduced action is then 
\be
S_{\text{total}} = S_{\BF} + S_{\Bsym} + S_{\Bphys}\,, 
\ee
where 
\be\label{DNGHstar}
\ba
S_{\Bsym = D(G)} =& - \frac{i}{2\pi} \int_{\partial M_{2n}} \left\langle A - \cA^{(U_L^{-1})} , B_L \right\rangle_{\BF} \cr 
S_{\Bphys= N^*(G,H)}=&   - \frac{i}{2\pi} \int_{\partial M_{d+1}} \left\langle U_R\, A^{(U^R)}_\mathfrak{m}\, U_R^{-1} , B_R  \right\rangle_{\BF}  +{f^2\over 2} \left \langle B_R , * B_R \right \rangle_{\kappa}  \,,
\ea
\ee
where $\langle \; , \; \rangle_{\kappa}$ is the quadratic pairing with the Killing form $\kappa_{ab}={\rm Tr}(T_aT_b)$ and generators $T_a,T_b\in \mathfrak{g}$. Note that we use the modified Neumann BC (\eqref{eq:modNeu})
with the Hodge star as we are dealing in this example with internal symmetries. In the next section we will generalize to spacetime symmetries. 
Finally, $B_{L/R}$ are the values of the $B$-field on the boundaries, and likewise $U_{L/R}$ the gauge transformations,  on the left and right boundaries, respectively. 

The gauge fields $A$ are flat and thereby solve already the bulk equations of motion. We are left the the equations for the $B_{L/R}$ which read 
\be
A=\cA^{(U_L)} \,,\qquad A^{(U^R)}_{\mathfrak{m}} = * B_R \,.
\ee
In particular we then find 
\be
\cA^{(V)}_\mathfrak{m}  = * B_R\,,
\ee
where the residual gauge transformation is the combination
\be
V= U_{L}^{-1} U_R \,.
\ee
Recall that the gauge symmetries on the various boundaries are 
\be
\ba
U_L & \rightarrow g^{-1} U_L \,,\qquad g \in G  \cr 
U_R & \rightarrow g^{-1} U_R\,, \qquad g\in G \cr 
U_R & \rightarrow h^{-1} U_R h \,,\qquad h\in H \,,
\ea
\ee
so that the residual one precisely the one expected for a field valued in $G/H$
\be
V \rightarrow V h \,,\qquad h\in H \,.
\ee
We then get the $d$-dimensional action 
\be
S_{\SSB} = {1\over 2} \int_{M_d} \left\langle \cA^{(V)}_{\mathfrak{m}}, * \cA^{(V)}_{\mathfrak{m}} \right \rangle_{\kappa} 
\ee
This is precisely the action  of the $G/H$ Goldstone boson, coupled to a $G$ background field $\cA$.

\section{SymTFT for the Conformal Symmetry}
\label{sec:conformal-anomaly}

We now apply the formalism to Euclidean conformal symmetry in $d$ dimensions, described by the symmetry group $SO(d+1,1)$.\footnote{The Lorentzian version of the conformal group is $SO(d,2)$.} Conventions and definitions for conformal symmetry groups and their associated Lie algebras are collected in appendix \ref{sec:App_algebras}.\footnote{Similar computations appear also in the supersymmetrized context of conformal supergravity in \cite{Fradkin:1985am}.}

{The key differences to the case of internal symmetries, in particular compact symmetry groups, is that now we will have non-compact groups, and more importantly, the components of the gauge field have an interpretation as e.g.~the vielbein on the boundary. I.e. the gauge fields have now a geometric spacetime interpretation. }

\subsection{SymTFT for Conformal Symmetry in $d$ Dimensions}

\paragraph{BF-theory.}
In the absence of conformal and gravitational anomalies, the SymTFT for conformal symmetry is given by the $(d+1)$-dimensional BF-theory based on the gauge group $SO(d+1,1)$, defined on a manifold $M_{d+1}$. 
This gauge group is precisely the conformal group of a $d$-dimensional Euclidean CFT living on a boundary of $M_{d+1}$. The gauge connection can be decomposed into the generators of conformal algebra as
\begin{equation}
  \label{eq:A-expansion}
A = \frac{1}{2} w^{ab} L_{ab} + e^a P_a + f^a K_a + b D~.
\end{equation}
These generators (for the Wick rotated Lorentz algebra) include rotations $L_{ab}$, translations $P_a$, special conformal transformations $K_a$ and dilatation $D$. The components $(w^{ab}, e^a, f^a, b)$ are loosely referred to as the Lorentz gauge field, the vielbein, the special conformal gauge field and the dilatation gauge field, respectively. In particular, the Lorentz gauge field $w^{ab}$ is related to the more familiar spin connection $\omega^{ab}$ by \cite{Fradkin:1985am}
\ie
\omega_\mu^{ab}=w^{ab}_\mu+b^{[a} e^{b]}_\mu~.
\fe
To construct the BF Lagrangian, we introduce a $(d-1)$-form field $B$ valued in the Lie algebra $\mathfrak{so}(d+1,1)^*$, which can be decomposed as 
\ie
\label{eq:B-decomposition}
B=\frac{1}{2} j^{ab}L_{ab}^*+t^a P_a^*+s^a K_a^*+\phi D^*~.
\fe
Choosing the canonical pairing between the algebra and dual algebra of $\mathfrak{so}(d+1,1)$ we  define a BF-theory with Lagrangian 
\ie\label{eq:BF_Lagrangian}
\mathcal{L}_{\text{BF}}=\ &i\langle B,  F\rangle_{\BF}
\\
=\ &\frac{i}{2}j_{ab}( \dd \omega^{ab}+\omega^{a}{}_c\wedge \omega^{cb}-2e^{[a}\wedge f^{b]})+2it_a (\dd e^a+\omega^{a}{}_{b}\wedge e^b+b\wedge e^a)
\\
&+2is_a (\dd f^a+\omega^{a}{}_b\wedge f^b-b\wedge f^a)-i\phi (\dd b-2e_a\wedge f^a)~.
\fe
The equation of motion of the $B$ field, $F=0$, in components reads:
\ie\label{eq:EOM_A}
0 &= \dd w^{ab} + w^{a}_{\,\,\, c} \wedge w^{cb} -
e^{[a} \wedge f^{b]}~, \\
0 &= \dd e^a + w^a_{\,\,\, b} \wedge e^b + 
b \wedge e^a~, \\
0 &= \dd f^a + w^a_{\,\,\, b} \wedge f^b - b \wedge f^a ~,  \\
0 &= \dd b - 2 e_a \wedge f^a~.
\fe
On the other hand, the equation of motion of the $A$ field, $\dd_A B=0$, implies
\ie\label{eq:EOM_B}
&0=\dd j^{ab}+w^{[a}{}_c\wedge j^{cb]}-2e^{[a}\wedge t^{b]}-2f^{[a}\wedge s^{b]}~,
\\ 
&0=\dd t^a+w^a{}_b\wedge t^b+f^b\wedge j_{b}{}^a+f^a\wedge\phi -b\wedge t^a~,
\\ 
&0=\dd s^a+w^a{}_b\wedge s^b+e^b\wedge  j_b{}^a-e^a\wedge \phi+b\wedge s^a~,
\\
&0=\dd \phi-2e^a\wedge t_a +2f^a\wedge s_a~.
\fe

\paragraph{Linking in the $SO(d+1,1)$-BF.} 
Unitary irreducible representations (irreps) of the conformal group are labeled by the primary state $|\Delta,R \rangle$ with $R$ an irrep of $SO(d-1)$. For simplicity, we will take $R$ to be the spin $\ell$ symmetric representation of $SO(d-1)$. The full representation is generated from the primary state as   
\ie
\mathcal{M}_{(\Delta,\ell)} &= \text{span}\left\{  P_{\mu_1} ... P_{\mu_n} | \Delta , \ell \rangle \, | \, n \geq 0   \right\},
\\
D | \Delta, \ell \rangle &= i \Delta |\Delta, \ell \rangle ,
\\
J^2 | \Delta, \ell \rangle &= \ell(\ell+d-2) | \Delta , \ell \rangle \,.
\fe
We are interested in the irreps that are relevant for CFT i.e.~those satisfy $D=D^\dagger$, $P_a^\dagger =K_a$ and $L_{ab}^\dagger=-L_{ab}$.
Using these irreps, we can define Wilson lines as
\begin{equation}
\mathcal{W}_{(\Delta,\ell)}[\gamma] = \text{Tr}_{\mathcal{M}_{(\Delta,\ell)}} \mathrm{P}\, e^{\oint_\gamma A}\,.
\end{equation}
The topological operators $\mathcal{U}^{[g=e^X]}[\Sigma_{d-2}]$ link non-trivially with these Wilson lines.  
A relevant set of classes is the one where we take the representative to be $X = \tau D$ with $\tau \in \mathbb{R}_{\geq 0}$. The corresponding operator which links with a Wilson line $\mathcal{W}_{(\Delta,\ell)}$ correctly measures its scaling dimension, since the linking factor is
\begin{align}
\text{Tr}_{\mathcal{M}_{(\Delta,0)}}\left\{ e^{i \tau D} \right\} &= \sum_{n=0}^\infty e^{-\tau (\Delta+n)}  \binom{d+n-1}{d-1}  =  \frac{e^{- \Delta \tau}}{(1-e^{-\tau})^d}\, .
\end{align}
The sum over $n$ is the sum over descendant in the multiplet, and the binomial factors counts their rotational $O(d)$-degeneracy.

\subsection{Symmetry Generators from the Dirichlet BC}
\label{app:Charges}

We now construct explicit expression of the symmetry charges on the Dirichlet background $\cA$ corresponding to flat space:
\begin{equation}\label{eq:flatSpaceConn}
\cA = \delta^a_\mu \dd x^\mu \otimes P_a \, .
\end{equation}
We carry this analysis out with the BF-term only. Moreover, we take $B_{d-1} \in \Omega^{d-1}(M_d, \mathfrak{g})$ rather than in $g^*$, defining the BF-action with the $\mathfrak{so}(d+1,1)$ Killing form in \eqref{eq:kappatr}. As we will see, two of the BF equations of motion are redundancies in the components of $A$, which allow us to express the Lorentz gauge field $w^{ab}_\mu$ and special conformal gauge field $f^a_\mu$ in terms of the vielbein $e^{a}_\mu$ and dilatation gauge field $b_\mu$.

First, we use the second equation of \eqref{eq:EOM_A} to solve for $w^{ab}_\mu$. Physically, this equation is the torsion free condition. In component form, it is written as
\ie
w^{ab}_{[\mu} e_{\nu]b}=- (\partial_{[\mu} e^a_{\nu]}+b_{[\mu}e_{\nu]}^a)\equiv-\hat\partial_{[\mu}e^a_{\nu]}~.
\fe
Multiplying both sides of the equation by $e_{a\rho}$, we obtain
\ie
w_{\mu \rho\nu} -w_{\nu \rho\mu} =-e_{a\rho} \hat\partial_{[\mu}e^a_{\nu]}\,,
\fe
where $w_{\mu \rho\nu}=w^{ab}_{\mu} e_{a\rho}e_{b\nu}=-w_{\mu\nu\rho}$. We now consider the following combination
\ie
2w_{\mu\rho\nu}=(w_{\mu \rho\nu} -w_{\nu \rho\mu})+(w_{\rho\mu\nu} -w_{\nu \mu\rho}) -(w_{\mu\nu\rho} -w_{ \rho\nu\mu})=-e_{c\rho} \hat\partial_{[\mu}e^c_{\nu]}-e_{c\mu} \hat\partial_{[\rho}e^c_{\nu]}+e_{c\nu} \hat\partial_{[\mu}e^c_{\rho]}~.
\fe
Multiplying both sides of the equation by $e^{\rho a}e^{\nu b}$ gives
\ie
w^{ab}_{\mu}=-\frac{1}{2}e^{a\rho}e^{b\nu}\left[e_{c\rho} \hat\partial_{[\mu}e^c_{\nu]}+e_{c\mu} \hat\partial_{[\rho}e^c_{\nu]}-e_{c\nu} \hat\partial_{[\mu}e^c_{\rho]}\right]=\omega^{ab}_\mu
-b^{[a}e^{b]}_\mu~, 
\fe
where $\omega^{ab}$ is the spin connection
\ie
\omega^{ab}_\mu=-\frac{1}{2}\left[e^{b\nu} \partial_{[\mu}e^a_{\nu]}-e^{a\rho} \partial_{[\mu}e^b_{\rho]}+e^{a\rho}e^{b\nu}e_{c\mu} \partial_{[\rho}e^c_{\nu]}\right]~.
\fe
Next, we use the first equation of \eqref{eq:EOM_A} to solve for $f^{a}_\mu$. Physically, this equation is a generalization of the flatness condition of the spin connection. In component form, it is written as
\ie
e^{a}_{[\mu} f^{b}_{\nu]}-e^{b}_{[\mu}  f^{a}_{\nu]} = \partial_{[\mu} w^{ab}_{\nu]} + w^{a}{}_{c[\mu} w^{cb}_{\nu]} \equiv \mathcal{R}^{ab}_{\mu\nu}~.
\fe
Multiplying both sides by $e^\nu_b$, we obtain
\ie\label{eq:e_equation_1}
 e^{a}_{\mu}e^\nu_bf^{b}_{\nu}+(d-2)f^{a}_{\mu} = \mathcal{R}^{ab}_{\mu\nu}e^\nu_b~.
\fe
Further, multiplying both sides by $e^\mu_a$, one gets
\ie\label{eq:e_equation_2}
f^{a}_{\mu}e^\mu_a = \frac{1}{2(d-1)}\mathcal{R}^{ab}_{\mu\nu}e^\mu_a e^\nu_b~.
\fe
Substituting it back to \eqref{eq:e_equation_1}, we obtain
\ie
f^{a}_{\mu} = \frac{1}{(d-2)}\left[\mathcal{R}^{ab}_{\mu\nu}e^\nu_b- \frac{1}{2(d-1)}e^{a}_{\mu}\mathcal{R}^{bc}_{\mu\nu}e^\mu_b e^\nu_c\right]~.
\fe
The flat spacetime background \eqref{eq:flatSpaceConn} which solves the equation of motion \eqref{eq:EOM_A} correspond to the choice
\ie
e^a_\mu=\delta^a_\mu~,\quad b_\mu=w^{ab}_\mu=f^a_\mu=0~.
\fe
in the equations above. In this background, the gauge symmetry associated with $g$ in \eqref{eq:gauge_trans} is frozen on the boundary, so the remaining gauge symmetry is \footnote{Recall that as $B_{d-1}$ takes values in $\mathfrak{g}$ this is just the ordinary adjoyint action.}
\ie
B_{d-1}\rightarrow B_{d-1}+\dd \sigma_{d-2} +A\wedge \sigma_{d-2} -(-1)^d \sigma_{d-2} \wedge A ~.
\fe
Let us decompose the gauge parameter $\lambda$ as follows
\ie
\sigma_{d-2} = \frac{1}{2} \sigma^{ab}L_{ab}+\alpha^a P_a+\beta^a K_a+\gamma D \, ,
\fe
omitting the form degree for all the components. In this flat spacetime background the components of $B$ \eqref{eq:B-decomposition} transforms  as
\ie\label{eq:gauge_symmetry_boundary}
&j^{ab}\rightarrow j^{ab}+\dd\sigma^{ab}-2e^{[a}\wedge \beta^{b]}~,
\\
&s^a\rightarrow s^a+\dd\alpha^a-e^a\wedge\gamma+e^b\wedge\sigma_b{}^a~,
\\
&t^a\rightarrow t^a+\dd\beta^a~,
\\
&\phi\rightarrow \phi+\dd\gamma-2e^a\wedge\beta_a~.
\fe

On the boundary, the flatness condition \eqref{eq:EOM_B} of $B$ simplifies to
\ie\label{eq:EOM_B_boundary}
&0=\dd j^{ab}-2e^{[a}\wedge t^{b]}~,
\\ 
&0=\dd t^a~,
\\ 
&0=\dd s^a+e^b\wedge  j_b{}^a-e^a\wedge \phi~,
\\
&0=\dd \phi-2e^a\wedge t_a~.
\fe
In this background $e^a = \delta_\mu^a \dd x^\mu \equiv \dd x^a$ so these equations can be recasted as various closeness conditions
\ie\label{eq:closeness}
&0=\dd (j^{ab}-2x^{[a} t^{b]})~,
\\ 
&0=\dd t^a~,
\\ 
&0=\dd (s^a+x^b  j_b{}^a-x^a \phi-x^b x_{b} t^{a}+2  x^{a}x^b t_{b})~,
\\
&0=\dd (\phi-2x^a t_a)~.
\fe
To recover the original first and last equation, we need to use the second equation $\dd t^a=0$. Finally, to recover the original third equation, we need to use $\dd j^{ab}=2e^{[a}\wedge t^{b]}$ and $\dd \phi=2e^a\wedge t_a$:
\ie
&\dd (s^a+x^b  j_b{}^a-x^a \phi-x^b x_{b} t^{a}+2  x^{a}x^b t_{b})
\\
& = \dd s^a+e^b\wedge  j_b{}^a-e^a\wedge \phi+x^b  (\dd j_b{}^a-2 e_{b} \wedge t^{a}+2  e^{a}\wedge t_{b})-x^a (\dd\phi-2  e^b\wedge t_{b})
\\
&=\dd s^a+e^b\wedge  j_b{}^a-e^a\wedge \phi~.
\fe
Because of the closedness condition \eqref{eq:closeness}, we can build the following topological charges that generate the conformal symmetry on the boundary 
\ie\label{eq:charges}
\mathcal{P}^a&=\oint t^a~,
\\
\mathcal{J}^{ab}&=\frac{1}{2}\oint \left(j^{ab}-2x^{[a}t^{b]}\right)~,
\\
\mathcal{K}^a&=\oint (s^a+x^b  j_b{}^a-x^a \phi+2  x^{a}x^b t_{b}-x^b x_{b} t^{a})~,
\\
\mathcal{D}&=\frac{1}{2}\oint \left(2x^at_a-\phi\right)~.
\fe
Under the gauge transformation \eqref{eq:gauge_symmetry_boundary}, these charges are invariant
\ie
\mathcal{P}^a&\rightarrow \mathcal{P}^a+\oint \dd\beta^a~,
\\
\mathcal{J}^{ab}&\rightarrow \mathcal{J}^{ab}+\frac{1}{2}\oint\dd\left(\sigma^{ab}-2x^{[a}\wedge \beta^{b]}\right)~,
\\
\mathcal{K}^a&\rightarrow \mathcal{K}^a+\oint \dd(\alpha^a-x^a\gamma-x^b x_{b} \beta^{a}+2  x^{a}x^b \beta_{b}+x^b\sigma_{b}{}^a)~,
\\
\mathcal{D}&\rightarrow \mathcal{D}+\oint \dd\left( x^a\beta_a-\frac{1}{2}\gamma\right)~.
\fe

We now discuss the physical meaning of these charges. Locally, let us pick the gauge
\ie
j^{ab}=s^a=\phi=0\,.
\fe
In this gauge, the component $t^a$ should be identified with the stress tensor $T_{\mu\nu}$ via the relation
\ie
t^a=\star T^a,\quad T^a\equiv e^{a\mu}T_{\mu\nu}\dd x^\nu~.
\fe
Then, the first equation of \eqref{eq:EOM_B_boundary} implies that the stress tensor is symmetric
\ie
e^{[a}\wedge t^{b]}e_{a\mu}e_{b\nu}=e_{[\mu} T_{\nu]\rho}\wedge\star \dd x^\rho=T_{[\nu\mu]}\Omega=0,
\fe
where $\Omega$ is the volume form on the boundary, using the fact that $\star \dd x^\rho$ is a $(d-1)$-form so $e_\mu \wedge \star \dd x^\rho=\delta_{\mu}^\rho \Omega$.
The second equation of \eqref{eq:EOM_B_boundary} reduces to the stress tensor conservation
\ie
\dd t^a e_{a\mu}=\dd (\star T_{\mu\nu}\dd x^\nu)=\partial^{\nu} T_{\mu\nu}\Omega=0~.
\fe
The third equation of \eqref{eq:EOM_B_boundary} implies that the stress tensor is traceless
\ie
e^a\wedge t_a=e^{\mu} T_{\mu\nu}\wedge\star \dd x^\nu=T^\mu{}_{\mu}\Omega=0~.
\fe
These are precisely the conditions a stress tensor in conformal field theories in flat space background should satisfy. The charges in \eqref{eq:charges}, expressed in terms of the stress tensor $T_{\mu\nu}$, take the familiar form
\ie
\mathcal{P}_\mu&=\oint \star\, T_{\mu\nu} dx^\nu~,
\\
\mathcal{J}_{\mu\nu}&=\oint \star\, x_{[\nu} T_{\mu]\rho} dx^\rho~,
\\
\mathcal{K}_\mu&=\oint \star (2  x_{\mu}x^\nu T_{\nu\rho} -x^2 T_{\mu\rho}) dx^\rho~,
\\
\mathcal{D}&=\oint \star\, x^\mu T_{\mu\nu} dx^\nu~,
\fe
where $\mathcal{P}_\mu,\mathcal{J}_{\mu\nu}, \mathcal{K}_\mu,\mathcal{D}$ are the translation generator (momentum), the rotation generator (angular momentum), special conformal generator and the dilation generator, respectively.

\subsection{SSB from SymTFT:~Conformal to Poincar\'e SSB, $d$ odd}\label{sec:Sandwichodd}

We now turn to realizing various symmetry breaking setups using the SymTFT for the conformal group. The first example will be the odd $d$ (boundary) dimensions where the SymTFT is simply the BF-theory.  
Our goal is to derive the effective description of the SSB that breaks conformal to Poincaré.

\paragraph{Conventions.}
Our conventions for the commutators of the $d$-dimensional conformal algebra $\mathfrak{g} \equiv \mathfrak{so}(d+1,1)$ is summarized in \ref{sec:App_algebras}. When defining partial Neumann BC with this algebra, we need to choose a split $\mathfrak{g} = \mathfrak{h} \oplus \mathfrak{m}$.  We will then choose as subalgebra the Lorentz group $\mathfrak{h} = \mathfrak{so}(d)$. 
Notice that the choice $\mathfrak{h} = \mathfrak{iso}(d)$ would not give rise to a reductive coset, i.e.~does not satisfy  \eqref{eq:reductive}. { This is not an issue, as Goldstone modes for broken translations can be fixed to specific configurations recovering the breaking pattern $\mathfrak{so}(d+1,1) \rightarrow \mathfrak{iso}(d)$ \cite{Low:2001bw, Delacretaz:2014oxa}. }

The choice of generators of the complement $\mathfrak{m}$ is not unique and it depends on the coordinates used on the $SO(d+1,1)$ group. Ultimately, this choice is immaterial as the action for the Goldstone bosons will be invariant under change of coordinates in the target. A convenient choice is 
\begin{equation}\label{eq:m_split}
\mathfrak{m} = \text{span}_{\mathbb{R}} \left\{ T_a^+ \coloneqq \frac{P_a+K_a}{2}, \, T_a^- \coloneqq \frac{P_a-K_a}{2} \, , D  \right\} 
\end{equation}
on which algebra commutators read
\begin{equation}
[D, T_a^{\pm}] = T_a^{\mp} , \quad [T_a^+ , T_b^-] =  \eta_{ab} D \, , \quad [T^{\pm}_a, T^{\pm}_b] = \pm L_{ab} \,.
\end{equation}

\paragraph{Analysis for $d$ odd.}
We now consider the following SymTFT sandwich configuration:~
\begin{itemize}
\item $\Bsym$ is fixed to be the Dirichlet bondary condition for the $SO(d+1, 1)$ conformal group. 
\item {$\Bphys$ is the partial modified Neumann BC of section \ref{sec:GaplessBC}, where the $SO(d)$ subgroup has Neumann. This is a gapless BC.}
\end{itemize}
We decompose the Lie algebra as follows
\be
\mathfrak{g} = \mathfrak{so}(d+1, 1)  = \mathfrak{h} \oplus \mathfrak{m}\,,\ \text{where} \,,\ \mathfrak{h}= \mathfrak{so}(d)\,.
\ee
This is depicted as follow 
\be
\begin{tikzpicture}
\begin{scope}[shift={(0,0)}]
\draw [cyan,  fill=cyan, opacity = 0.8] 
(0,0) -- (0,3) --(4,3) -- (4,0) -- (0,0) ; 
\draw [white](0,0) -- (0,3) --(4,3) -- (4,0) -- (0,0) ;  ; 
\draw [very thick] (0,0) -- (0,3)  ;
\draw [very thick] (4,3) -- (4,0) ;
\node at (2,1.5) {$\BF(SO(d+1,1))$} ;
\node[above] at (-0.5,3) {$\Bsym = D(SO(d+1,1))$}; 
\node[above] at (5.5,3) {$\Bphys = N^*(SO(d+1,1), SO(d)) $}; 
\end{scope}
\end{tikzpicture}
\ee
We expect this to break the dilatation generator, but be symmetric under the $H=SO(d)$ Lorentz subgroup. 
The sandwich is again the sum of the bulk and two boundary terms. As the gauge fields solve the bulk equations of motions 
we can focus on the boundary terms. Again, the St\"uckelberg fields drop out due to the Dirichlet condition and flatness, and the reduced action is
\be \label{eq:DandNGH}
\ba
S_{\Bsym = D(G)} =& - \frac{i}{2\pi} \int_{\partial M_{d+1=2n+2}} \left\langle A - \cA^{(U_L^{-1})} , B_L \right\rangle_{\BF} \cr 
S_{\Bphys= N^*(G,H)}=&   - \frac{i}{2\pi} \int_{\partial M_{d+1=2n+2}} \left\langle U_R\, A^{(U_R)}_\fm\, U_R^{-1} , B_R  \right\rangle_{\BF}  +{f^2\over 2}   \left\langle B_R , \Hod (A^{(U_R)}, B_R)  \right\rangle_{\BF} \,.
\ea
\ee
Again, the physical boundary is given by a {\bf gapless partial modified Neumann BC}, introduced in section \ref{sec:GaplessBC} which is constrained by the following requirements:
\begin{itemize}
\item quadratic in $B_R$ 
\item invariant under $H= SO(d)$. 
\end{itemize}
The additional term in the physical boundary condition proportional to  $f^2$ requires further discussion:~This is similar to the term added in (\ref{DNGHstar}), for internal global continuous symmetries, however for spacetime symmetries, instead of introducing a metric through an explicit Hodge star, we can build one from boundary values of the $P^a$ components of the gauge field, that is decomposed as \eqref{eq:A-expansion}.
Notice that the $P$ component defines a linear map $T_p \Sigma_d \rightarrow \mathfrak{p} \cong \mathbb{R}^d$, which is local coordinates is just a matrix $e^a_{\,\,\, \mu}$.
The resulting operator is denoted by $\Hod(A,  -)$ above and mimicks the properties of the Hodge star without introducing said explicit metric dependence. Let us consider a given 
$\mathfrak{g}$-valued one-form $\omega_p \in \Omega^p(\Sigma_d, \mathfrak{g})$, the Hodge dual operation is defined by
\begin{equation}
\begin{aligned}\label{Hodnew}
\text{Hod}(A,\omega_p) 
&\equiv \frac{1}{(d-p)!} T_i \big(\omega^i_{b_1 \ldots b_p}  
\eta^{b_1 a_1} \cdots \eta^{b_p a_p} \big) 
\varepsilon_{a_1 \ldots a_d} \, 
e^{a_{p+1}} \wedge \cdots \wedge e^{a_d} 
\in \Omega^{d-1}(\Sigma_d , \mathfrak{g}) \,, \\
&= \frac{1}{(d-p)!} T_i \Big(
\omega_{\mu_1\ldots \mu_p}^i 
e^{\mu_1}_{\,\, b_1} \eta^{b_1 a_1} e^{\nu_1}_{\,\, a_1} 
\cdots 
e^{\mu_p}_{\,\, b_p} \eta^{b_p a_p} e^{\nu_p}_{\,\, a_p} 
\Big) \times \\
&\qquad \qquad \qquad \times \text{det}(e^{a_1}_{\,\, \mu_1}) \, 
\epsilon_{\nu_1 \ldots \nu_d} \, \dd x^{\nu_{p+1}} \wedge \cdots \wedge \dd x^{\nu_d}\,,
\end{aligned}
\end{equation}
where the $T^i$ generically indicates the generators of $\mathfrak g$, that is $\{P_a,K_a, L_{ab},D\}$. The same Hod operation applies to $p$-forms valued $\mathfrak g^{*}$. See appendix \ref{app:Hodge} for more details about the Hod operation.

We will show that the interval compactification leads to the Goldstone boson for the SSB. The computation is in fact very similar to the one for the example in section \ref{sec:CompEx}, i.e.~
\be
\ba
A &= \cA^{(U_L^{-1})} \cr 
U_R A^{(U_R)}_\fm U_R^{-1} &=- f^2 \Hod (A^{(U_R)}, B_R)  \,,
\ea
\ee
where the second equation can be solved for $B_R$ 
\be
B_R = -{\epsilon \over f^2} \Hod (\cA^{(V)}, U_R \,  \cA^{(V)}_{\mathfrak{m}} \, U_R^{-1}) \,,
\ee
where $\epsilon= (-1)^{p (d-p)}$ for $p$ the degree of $B_R$ was used, which is the sign appearing in (\ref{HODHOD}) and $V= U_L^{-1} U_R$. 
Integrating out the $B_{L/R}$ results in 
\be\label{SandwichResultodd}
S_{\rm Sandwhich} = -{\epsilon \over 2 f^2}\int_{\partial M_{d+1}}
\left\langle \Hod \left(\cA^{(V)}, \cA^{(V)}_\fm\right), \cA^{(V)}_\fm \right\rangle_{\kappa} \,.
\ee
This holds true in any spacetime dimension, and is the complete answer for $(d+1)$ even. Using the results in \ref{sec:App_algebras}, one can expand this action in components for the conformal group. For simplicity, we take $V \in SO(d+1,1)/SO(d)$ to have only $D$-components and choose the $\cA$ background to correspond to flat space. In this case, the Maurer-Cartan form simply reads
\begin{equation}
\mathcal{A}^{(V)} = e^\sigma \delta_{\mu}^a \dd x^\mu - \dd \sigma D
\end{equation}
in local coordinates. Then the sandwich action reduces to the known leading-order in derivative expansion of the dilaton action
\begin{equation}
S_{\rm Sandwich} = \frac{f^2}{2} \int_{M_d}  \dd^d x\sqrt{g}\,  e^{-(d-2) \sigma} g^{\mu\nu} \partial_\mu \sigma \partial_\nu \sigma \, , \quad g_{\mu\nu} = \eta_{ab} e^a_{\,\,\, \mu} e^b_{\,\,\, \nu}\,.
\end{equation}
This is the same action as previously obtained using the coset construction in \cite{Monin:2016jmo}.

\subsection{SSB from SymTFT:~Conformal Symmetry in $d$ even}

In $d=2n$ dimensions, the bulk is odd-dimensional and has not only BF-term but also a CS-term for the symmetry $SO(d+1,1)$:~
\be
\begin{tikzpicture}
\begin{scope}[shift={(0,0)}]
\draw [cyan,  fill=cyan, opacity = 0.8] 
(0,0) -- (0,3) --(4,3) -- (4,0) -- (0,0) ; 
\draw [white](0,0) -- (0,3) --(4,3) -- (4,0) -- (0,0) ;  ; 
\draw [very thick] (0,0) -- (0,3)  ;
\draw [very thick] (4,3) -- (4,0) ;
\node at (2,1.5) {$\BF (SO(d+1,1)) + \CS$} ;
\node[above] at (-0.5,3) {$\Bsym = D(SO(d+1,1))$}; 
\node[above] at (5.5,3) {$\Bphys = N^*(SO(d+1,1), SO(d)) $}; 
\end{scope}
\end{tikzpicture}
\ee
One has to be careful about the choice of CS-term.

\subsubsection{Chern-Simons Terms for $\mathfrak{so}(d+1,1)$}
To define Chern-Simons terms for a group $G$ in $d=2n+1$ one needs an adjoint-invariant product on $\mathfrak{g}$ with $n+1$ entries. In general, there can be more than one of such products, and these are classified by the degree-$(n+1)$ casimirs of $\mathfrak{g}$. Casimirs are elements of the center of the universal enveloping algebra of $\mathfrak{g}$, and via the Harish-Chandra isomorphism this is related to $\mathcal{Z}(\mathcal{U}[\mathfrak{g}]) \cong S(\mathfrak{h})^W$, the algebra of symmetric polynomials on the Cartan algebra which is invariant under the Weyl group $W[\mathfrak{g}]$. In the specific case we are interested in,
\begin{equation}
W[\mathfrak{so}(d+1,1)] = W[\mathfrak{so}(2n+2)] = \mathbb{Z}_2^{n} \rtimes S_{n+1}
\end{equation}
independently on signature, which act on the Cartan algebra of $(n+1)$ elements by even sign flips and permutations with the composition rule $( \epsilon_1 , \sigma_1 ) (\epsilon_2 , \sigma_2) = ( \epsilon_1 \sigma(\epsilon_2) , \sigma_1 \sigma_2 )$.
Since $\mathfrak{so}(2n+2)$ is semisimple, it has exactly $n+1$ independent Casimirs, all higher ones can be built from them. For $n>1$ the algebra is simple, so there is a unique quadratic casimir. To build a invariant product with $(n+1)$ entries, we could use an independent degree-$(n+1)$ Casimir as well as powers of lower-degree ones, thus is necessary to know all the lower-degree independent ones. For the cases we are interested, independent Casimirs are related to the following generating elements of $S(\mathfrak{h})^W$:
\begin{align}
n=1:~&\quad   C_2 \sim \sum_{i=1}^2 h_i^2, \quad C_2' \sim \prod_{i=1}^2 h_i \nonumber\\
n=2:~&\quad  C_2 \sim \sum_{i=1}^3 h_i^2 , \quad C_3 \sim \prod_{i=1}^3 h_i , \quad C_4 \sim \sum_{i=1}^4 h_i^4 \nonumber\\
n=3:~&\quad C_2 \sim \sum_{i=1}^4 h_i^2, \quad C_4 \sim \sum_{i=1}^4 h_i^4, \quad C_4' \sim \prod_{i=1}^4 h_i, \quad C_6 = \sum_{i=1}^4 h_i^6 
\end{align}
In general, for $\mathfrak{so}(2n+2)$ one has the following set of independent casimirs
\begin{equation}
C_2, C_4, C_6 , ... , C_{2n} , C'_{n+1} \, .
\end{equation}
As lower-degree independent casimirs can be used to build higher degree ones, the number of possible Chern Simons terms grows with dimension. For $n=1\, (d+1=3)$ there two and for $n=2\, (d+1=5)$ there is one, while for $n=3\, (d+1=7)$ there are two, which matches with the counting of type-$a$ anomaly and gravitational anomaly in $d= 2+4k,\, k \in \mathbb{Z}_{\geq0}$. 

The two Chern-Simons terms present in $d=2$, are referred to as Tr and Tr$*$ in \cite{Witten:2007kt}, see also the (\ref{eq:epsilontr}) and (\ref{eq:kappatr}). The latter is known to capture the conformal anomaly of $d=2$ CFTS, while the former encodes the gravitational anomalies present for $c_L \not= c_R$. If both are added in the SymTFT, then the symmetry is anomalous and there is only the Dirichlet BC. Instead if we add only the CS with the Tr$*$, then we also have a partial Neumann BC for the $SO(2)$ Lorentz group.  These two CS-terms have different quantization conditions. 
 Note that in (super)conformal $d=4$ theories, once can derive conformal anomlies from 5d CS-theory, using an inflow or BRST approach as well \cite{Imbimbo:2023sph, Imbimbo:2025ffw}.
 We thus expect to reproduce these anomalies by CS-dressing our SymTFTs, as we will explicitely verify in $d=2,4$. 

More generally, for any dimension CS-functionals built from different multilinear products might have different quantization conditions. In the case of $\mathfrak{so}(d+1,1)$, the  CS-functionals built out of trace-like Casimirs, $C_2, C_4, \cdots $, do not vanish when restricted onto  the maximally compact $\mathfrak{so}(d)$ subalgebra. Group elements, generated from this subalgebra might admit large-gauge transformations, and therefore the corresponding coupling must be quantized. Instead, CS-functionals built from the $C_{n+1}'$ Casimir do vanish on $\mathfrak{so}(d)$ and are non-vanishing only when one of the non-compact algebra elements are involved. Thus, their coupling does not need any quantization condition in general.

\subsubsection{3d SymTFT for 2d SSB from Conformal to 
Poincar\'e}
\label{sec:3dSandwichCS}

Let us first consider this explicitly for $d=2$. 
The left, symmetry boundary is chosen to be Dirichlet, which gives rise to the conformal symmetry, including the WZ term $\Gamma_3$:
\begin{equation}
D_k(G)^{(\text{3d})}:~\quad S = - \frac{i}{2\pi} \int_{\partial M_{3}} \Bigg\{ \left\langle A - \cA^{(U_L^{-1})} , B_L \right\rangle_{\rm BF} + \frac{k}{2} \left\langle A^{(U_L)}, \cA \right\rangle_{\epsilon} \Bigg\} + \frac{i}{2\pi} \Gamma_3(U_L,A) \, .
\end{equation}
where $\langle \; , \; \rangle_{\CS}=\langle \; , \; \rangle_{\epsilon}$ defined in \eqref{eq:epsilontr}.
The right boundary is the physical BC and we chose it to be the partial Neumann where we have flat-gauged the subgroup $SO(2)$ and added again the singleton mode 
\be\ba \label{eq:Nstar3d}
N_k^*(G,H)^{(\text{3d})}:~\quad S_{\Bphys} =& - \frac{i}{2\pi} \int_{\partial M_{3}}  \left\langle U_R\, A^{(U_R)}_\fm U_R^{-1} , B_{R} \right\rangle_{\rm BF} \cr 
&+ \left\langle A^{(U_R)}_\fm, A^{(U_R)} \right\rangle_{\rm \epsilon} + \frac{i}{2\pi}   \Gamma_3(U_R,A) \cr 
&+ {f^2\over 2}   \left\langle B_R , \Hod (A^{(U_R)}, B_R)\right\rangle_{\kappa} \,.
\ea\ee

Solving again as before for $A$ and $B_R$ and reinserting this we obtain the following contributions to the effective action of the SymTFT compactification 
\be
\ba
&\frac{i}{2\pi}\int_{\partial M_3} {k\over 2} \left\langle  \cA^{(V)}_\fm, \cA^{(V)} \right\rangle_{\epsilon} + \frac{f^2}{2} \left\langle \Hod \left(\cA^{(V)}, \cA^{(V)}_\fm\right), \cA^{(V)}_\fm \right\rangle_{\BF} \cr 
&+ \frac{i}{2\pi} \left\{ \Gamma_3 \left(U_R, \cA^{(U_L^{-1})} \right) -\Gamma_{3} \left(U_L, \cA^{(U_L^{-1})} \right) \right\}  \,.
\ea\ee

We now furthermore would like the two $\Gamma_3$ action contributions from the two boundaries to combine. 
From the properties of the $\Gamma_3$ action \eqref{eq:Gamma_properties} it is evident that this is the case:
\begin{equation}
\Gamma_3\left(U_L, \cA^{(U_L^{-1})} \right) - \Gamma_3\left(U_R , \cA^{(U_L^{-1})} \right) = - \Gamma_3\left(U_L^{-1} U_R, \cA \right) 
= -\Gamma_3 (V, \cA) \,.
\end{equation}
The BF part of this action has already been computed in the previous section, while the remaining contribution will match the conformal anomaly:
\begin{align} \label{eq:sandcomp}
S_{\rm anomaly} [V,\cA]&= \frac{k}{2} \int_{\partial M_3} \left\langle  \cA^{(V)}_\fm , \cA^{(V)} \right\rangle_{\epsilon} + \Gamma_3(V, \cA) \,.
\end{align}
The anomaly is detected by performing gauge transformations of $\cA$, which is treated as a background for the anomalous symmetry. As $V$ is path integrated over, we can also simulatenously redefine $V \mapsto g^{-1} V$. Then one obtains 
\begin{equation}
\Delta^{(g)} S_{\rm anomaly}[V, \cA] = \Delta^{(g)} \int_{X_3} \Gamma_3(V, \cA) = \Delta^{(g)} \int_{X_3} \text{CS}_3(\cA, \cF ) \, ,
\end{equation}
where in the last equation one uses the relation to the Chern-Simons transgression form \ref{sec:CS_conventions}. As expected, the anomaly is a functional of the $\cA$ background only. If we focus on the scale anomaly $g= e^{-\tau D}$ Then one finds
\begin{align}\label{eq:2daAnomaly}
\Delta^{(g)}S_{\rm anomaly} = - \frac{i k}{2(2\pi)} \int_{\partial M_3} \langle \dd \tau D, \cA \rangle_{\epsilon} =  \frac{i k}{2(2\pi)} \int_{\partial M_3} \tau \bar{E}_2 \,,
\end{align}
where $E_2 = \epsilon_{ab} \bar{R}^{ab} / 2 = \epsilon_{ab} \dd \bar{\omega}^{ab} / 2$ reproduces the type-A conformal anomaly, i.e.~anomaly proportional to the Euler density \cite{Zamolodchikov:1986gt, Schwimmer:2010za}. 

\subsubsection{3d SymTFT with full Gravitational Anomaly}

So far we have ignored the possibility of including into the SymTFT the contribution corresponding to the gravitational anomaly whose coefficient is\footnote{Up until now we have worked with the assumption $k_L=k_R$.}
\begin{equation}
    k'=k_L-k_R \neq 0. 
\end{equation}
This is because we have considered only one type of Chern-Simons functional in the SymTFT action. However, as discussed at the start of this section, $G=SO(3,1)$ admits the possibility of another invariant quadratic bilinear form which is the quadratic Casimir i.e.~the standard Killing form given by the traces $
\kappa_{ab}=\text{Tr}(T_a T_b)$. The SymTFT action with both of these terms added is 
\be\label{SymTFTBFCS3dfull}
\ba
S_{\SymTFT} &= S_{\BF} + S_{{\CS}_{\epsilon}} + S_{{\CS}_{\kappa}} \cr 
&= {i\over 2 \pi} \int_{M_{d+1}} \langle B_{d-1} , F \rangle_{BF} +  \frac{i k}{2(2\pi)} \left\langle A, F - \frac{1}{3} A^2 \right\rangle_{ \epsilon} +   \frac{i k'}{2(2\pi)} \left\langle A, F - \frac{1}{3} A^2 \right\rangle_{\kappa}\,,
\ea
\ee
where the pairing $\langle \;\rangle_{\epsilon}$ is defined in \eqref{eq:epsilontr} and correspond to the ${\rm Tr}^*$ in \cite{Witten:2007kt}. The pairing $\langle \; \rangle_{\kappa} $ is defined in \eqref{eq:kappatr} and corresponds to the standard ${\rm Tr}$ in \cite{Witten:2007kt}. The standard Killing form will reduce to the Chern-Simons for $\omega \in \frak{so}(3)$ as pointed out in \cite{Witten:2007kt}. For instance if we consider a transformation $g=e^{\alpha_{ab}L^{ab}}$ the gauge variation of the $S_{\rm CS}^{\kappa}$ action in the presence of a boundary reads,
\begin{equation}
    \Delta^{(g)}S_{\rm CS}^{\kappa}= - \frac{i k'}{2(2\pi)} \int_{\partial M_3} d\alpha^{ab} \, \frac{\omega^{cd}}{2} \langle L_{ab}, L_{cd}\rangle =- \frac{i k'}{2(2\pi)} \int_{\partial M_3} d\alpha^{a}_{\;b} \, \omega^{b}_{\;a} = - \frac{i k'}{2(2\pi)} \int_{\partial M_3} {\rm Tr} (d\alpha  \, \omega)
\end{equation}
that exactly corresponds to the gravitational anomaly. 

Finally, adding the $S_{\CS}^{\kappa}$ term to the SymTFT implies that the set of gapped boundary condition is modified. In particular we are not allowed to take any Neumann boundary condition that preserves any subgroup $H \in SO(3,1)$. If we would like to repeat the analysis to get the SSB action of Goldstone bosons we can only work with the full breaking and hence $D(G)$ on one boundary and $D^*(G)$ on the other one.

\subsubsection{Consistent Weyl Anomaly and WZ Condition}

Let us now analyze for $d=2$ whether the Chern--Simons functional $S_{{\rm CS}_{\epsilon}}$ provides a consistent anomaly, i.e.~satisfies the Wess--Zumino (WZ) consistency condition. In SymTFT language this means that the functional \eqref{eq:sandcomp}, obtained by compactifying the interval in the sandwich construction, obeys
\begin{equation}\label{eq:WZcond}
S_{\rm anomaly}[g^{-1}V, \cA^{(g)}] - S_{\rm anomaly}[V, \cA] = - S_{\rm anomaly}[g, \cA]\,,
\end{equation}
where the form of $S_{\rm anomaly}$ is given in \eqref{eq:sandcomp} and its anomalous variation in \eqref{eq:2daAnomaly}.  
This anomaly functional indeed satisfies the WZ condition for dilatations.  

It is important to stress the distinction between dilatation and Weyl transformations also in the SymTFT setup, recalling that the background $\cA$ obeys the flatness condition $\cF=0$ (see appendix~\ref{sec:App_algebras}). In particular, consider the $SO(d+1,1)$ flat connection for AdS geometry:
\be
\ba
  \cA &= \tfrac{1}{2} \bar{e}^a P_a + \tfrac{1}{2} \omega^{ab}(\bar{e}) L_{ab} - \tfrac{1}{2} \bar{e}^a K_a\cr 
 \bar{R}^{ab} &= - \bar{e}^a \wedge \bar{e}^b = d\omega^{ab} + \omega^{ac} \wedge \omega_{c}^{\; b}\, .
\ea
\ee
A dilatation $g=e^{\tau D}$ acts on $\cA$ as
\begin{equation}
\cA \mapsto (\cA)^{(g)} = \tfrac{1}{2} \bar{e}^a \big(e^{ \tau}  P_a - e^{-\tau} K_a \big) + \tfrac{1}{2} \omega^{ab}(\bar{e}) L_{ab} + \dd \tau D \, ,
\end{equation}
leaving both the spin connection and curvature $\bar{R}^{ab}$ unchanged, since $[L_{ab},D]=0$.  

In contrast, Weyl transformations in $d=2$ act also on the metric, spin connection, and curvature:
\begin{equation} \label{eq:2dweyl}
    ds^2= e^{2 \tau } \overline{ds}^2, \qquad 
    \omega^{ab}(e^\tau \bar{e}) = \bar{\omega}^{ab} - (\partial^{[a} \tau ) e^{b]}, \qquad 
    R= e^{-2 \tau} (\bar{R}-2 \overline{\square} \tau)\, .
\end{equation}
The bulk gauge symmetry does not capture this transformation. We can, however, implement it on $\cA$ as
\begin{equation}\label{eq:ProperWeyl}
    \begin{aligned}
    V &\mapsto e^{- \tau D}V, \\
    \cA &\mapsto \cA(e^\tau \bar e^a)= \tfrac{1}{2} \bar{e}^a e^{ \tau} (  P_a - K_a ) + \tfrac{1}{2} \omega^{ab}( e^{ \tau} \bar{e}) L_{ab} \, .
    \end{aligned}
\end{equation}

For dilatations, the non-trivial contribution in \eqref{eq:sandcomp} arises from $\Gamma_3$, while the other term is invariant.  
For Weyl transformations \eqref{eq:2dweyl}, however, $\Gamma_3(V,\cA)$ fails the WZ condition \eqref{eq:WZcond} due to the shift in $\omega$:
\begin{equation}
    \Gamma_3(e^{- \tau D}V, \cA(e^\tau \bar e^a))= \Gamma_3(V, \cA) - \Gamma_3(e^{\tau D}, \cA) - \tfrac{1}{2} \langle  \dd \tau D,  (\partial^{[a} \tau ) e^{b]} L_{ab}  \rangle_{\epsilon}\,.
\end{equation}
The last term can be rewritten as
\begin{equation}
    \langle  \dd \tau D,  (\partial^{[a} \tau ) e^{b]} L_{ab} \rangle 
    = \dd \tau \wedge  \epsilon_{ab} (\partial^{[a} \tau ) e^{b]} 
    = \dd \tau \wedge \text{Hod}(\cA^{(V)}, \dd\tau)\,,
\end{equation}
where we used the definition of the Hodge dual in $d=2$ \eqref{HodnewnoT}. Therefore,
\begin{equation}
     \Gamma_3(e^{- \tau D}V, \cA(e^\tau \bar e^a))- \Gamma_3(V, \cA) 
     = \Gamma_3(e^{\tau D}, \cA) + \frac{i k}{2(2 \pi)} \int_{M_2} \dd\tau \wedge \text{Hod}(\cA^{(V)}, \dd\tau)\, .
\end{equation}

The other term in \eqref{eq:sandcomp},
\begin{equation}
    I[V,\cA]=\frac{i k}{2 (2\pi)}\int_{M_2}\left\langle \cA^{(V)}_\fm , \cA^{(V)} \right\rangle_{\epsilon}\, ,
\end{equation}
also transforms under \eqref{eq:ProperWeyl}. Restricting to dilatation components in $V$ (i.e.~$V=e^{\sigma D}$), we find
\begin{equation}\label{eq:weyltransfA0}
   (\cA)^{(e^{-\tau D}V)}(e^\tau \bar e^a) = \alpha \bar{e}^a (e^{ \sigma}  P_a + e^{-\sigma+2\tau} K_a ) + \tfrac{1}{2} \omega^{ab}(e^{-\tau}\bar{e}) L_{ab} + \dd \sigma  D ,
\end{equation}
with
\begin{equation}
    \omega^{ab}(e^\tau \bar{e}) = \bar{\omega}^{ab} - (\partial^{[a} \tau ) e^{b]}\, ,
\end{equation}
so that
\begin{equation}
 I[e^{- \tau D}V, \cA(e^\tau \bar e^a)]-I[V,\cA]= -\frac{i k}{2(2\pi)} \int_{M_2} \dd\sigma \wedge \text{Hod}(\cA^{(V)},\dd\tau)\,.
\end{equation}

Combining the transformations of $\Gamma_3$ and $I$, one sees that $S_{\rm anomaly}$ does not satisfy the WZ condition for Weyl transformations, which would require
\begin{equation}\label{eq:WZcondweyl}
   \Delta_{\rm Weyl} S_{\rm dilaton}=  S_{\rm dilaton}[e^{-\tau D}V, \cA(e^{\tau} \bar{e}^a)] - S_{\rm dilaton}[V, \cA] = -S_{\rm dilaton}[e^{\tau D}, \cA]   \,.
\end{equation}

Following the strategy of \cite{Antinucci:2024bcm}, a consistent boundary anomaly can be restored by adding
\begin{equation}   
S_{\rm bt}=\frac{ik}{4(2\pi)}\int_{M_2} \left\langle \cA^{(V)}_\fm, \text{Hod}\!\left(\cA^{(V)},\cA^{(V)}_\fm \right)\right\rangle_{\kappa}\, ,
\end{equation}
which for $\langle D,D\rangle_{\kappa}=-1$ evaluates to
\begin{equation}
    S_{\rm bt}=-\frac{ik}{4(2\pi)}\int_{M_2} \dd\sigma \wedge \text{Hod}(\cA^{(V)},\dd\sigma)\, .
\end{equation}
Its Weyl variation is
\begin{equation}
\begin{aligned}
    S_{\rm bt}[e^{- \tau D}V, \cA(e^\tau \bar e^a)]&=S_{\rm bt}[V,\cA]\\
    &-\frac{ik}{4(2\pi)}\int_{M_2}  2\, \dd\sigma \wedge \text{Hod}(\cA^{(V)},\dd\tau) +\frac{ik}{4(2\pi)} \int_{M_2} \dd\tau \wedge \text{Hod}(\cA^{(V)}, \dd\tau) \, .
\end{aligned}
\end{equation}

Putting everything together, the combined action
\begin{equation}
    S_{\rm dilaton}= I+\Gamma_3+S_{\rm bt}
\end{equation}
satisfies the WZ consistency condition for Weyl transformations \eqref{eq:WZcondweyl}, and thus reproduces the dilaton action in $d=2$.  
This boundary term can also be derived prior to sandwich compactification by using the modified Neumann boundary condition $N^{*}_k(G)^{(\text{3d})}$ in \eqref{eq:Nstar3d}. The WZ condition fixes the coefficient to $f^2=\tfrac{i k}{4 \pi}$. This feature is special to $d=2$, where the dilaton kinetic term participates in anomaly matching. In higher even dimensions the condition instead constrains coefficients of possible new terms in $N^{*}_k(G)^{(d>2)}$, producing higher-derivative corrections to the dilaton action.

\subsubsection{5d SymTFT for 4d Conformal-SSB}

We now consider the case of $d=4$, 
where we determined the partial Neumann BC in (\ref{Neumann5d}). We break again the conformal symmetry group $G$ to the Lorentz group $H=SO(d)$, using the SymTFT sandwich $\langle D_k(G)^{(5d)}| N_k(G,H)^{(5d)}\rangle$. Most of the calculations follow from the three-dimensional analog case, with the boundary conditions in 5d obtained in section \ref{sec:SymTFTST}. The end result for the anomaly-matching part of the closed sandwich is
\begin{align}
S_{\rm anomaly}[V,A] =& \frac{ik}{6(\pi)^2} \int_{M_4} \Big\langle \cA^{(V)} , \cA^{(V)}_{\fh} , \dd \cA^{(V)} + \dd \cA^{(V)}_\fh \\
&+ \frac{1}{2}  \cA^{(V)} \wedge  \cA^{(V)} + \frac{1}{2}  \cA^{(V)}_\fh \wedge  \cA^{(V)}_\fh + \frac{1}{4} \left[ \cA^{(V)} , \cA^{(V)}_\fh \right]  \Big\rangle \nonumber\\
&+ \frac{ik}{6(2\pi)^2} \int_{X_5} \Gamma_5(V,\mathcal{A}) \, .
\end{align}
The anomaly is completely captured by the $\Gamma_5$ action as follows
\begin{equation}
\Delta^{(g)} S_{\rm anomaly}[V, \cA] = \Delta^{(g)} \int_{X_5} \Gamma_5(V, \cA) = \Delta^{(g)} \int_{X_5} \text{CS}_5(\cA, \cF ) \, .
\end{equation}
From this one can show -- e.g.~see the analysis in \cite{Imbimbo:2023sph,Imbimbo:2025ffw} -- that evaluated on the boundary this gives rise to the Euler density $E_4$, thus reproducing the $a$-anomaly. Notice that in our case, the anomaly will be evaluated on the particular geometry defined by the background $\cA$ specified in appendix \ref{sec:App_algebras}.

\section{Spacetime Symmetry Action:~How to Move a Point}
\label{sec:moving-points}

We have identified the generators of the boundary conformal group on the Dirichlet boundary condition of $A$
in terms of components of the $B$ field in section \ref{app:Charges}. The boundary conformal group includes,
in particular, translations, 
so that the
$B$ holonomies in the bulk should generate translations of operator
insertions in the boundary theory. This is perhaps surprising: how can
an operator in a topological theory move an operator insertion?

The simplest way of understanding the action of spacetime symmetries is as follows: consider for instance a correlator of the form
\[
\langle \cU_X(\Sigma) \cO(x) \ldots\rangle\,,
\]
where $\Sigma$ links with $x$, $\cU_X(\Sigma)$ is the operator in the SymTFT implementing translations $x\to x+X$ (other conformal transformations can be studied similarly), $\cO(x)$ lives at the endpoint of a line in the BF SymTFT we propose, and the dots stand for other possible insertions outside $\Sigma$. If we contract $\Sigma$ to $x$, the action on $\cO(x)$ can be read from the bulk action, as described in previous sections:
\[
  \cU_X(\Sigma) \cO(x) = e^{X^\mu\partial_\mu} \cO(x) = \sum_{n=0}^\infty \frac{1}{n!}(X^\mu\partial_\mu)^n \cO(x)\, ,
\]
an infinite sum of operators inserted at $x$. This infinite sum can of course be interpreted as an insertion of $\cO$ at $x+X$.

In this section we want to understand the spacetime action from a boundary perspective instead: if we first push the symmetry generators to the Dirichlet boundary, how do the resulting topological operators realize spacetime symmetries? We answer this question using two approaches: 
first from a Hamiltonian point of view, and then using a path integral formulation.

\subsection{Hamiltonian Approach}

Consider the bulk operator that implements boundary
translations in \eqref{eq:charges}:
\[
  \cP_a(\Sigma_{d-1}) = \int_{\Sigma} t_{a}\, .
\]
We want to analyze the effect of
pushing this operator to the 
 Dirichlet boundary, and in particular
 identify how the boundary conditions are modified by the insertion of this operator.

In order to do this, we will first work in a Hamiltonian framework, and
consider the bulk to have the form $\cM_d\times \bR$ near the
boundary, where $\cM_d$ is the boundary and we treat $\bR$ as
time. Furthermore, we take $\Sigma$ to be placed at a fixed time in
the $\bR$ direction (that is, $\Sigma\subset \cM_d$). 
From this point of view, the Dirichlet boundary
condition is a specific state in the Hilbert space of the SymTFT on
$\cM_d$, and we are trying to understand the action of the operator
$\cP_a(\Sigma_{d-1})$ on this state. We can determine this, already in
the classical theory, by computing the Poisson bracket:
\be 
  \label{eq:A-P-PB}
  \{e^b, \cP_a(\Sigma_{d-1})\} = \delta^{b}{}_a\, \delta^{(1)}(\Sigma_{d-1})\,,
\ee
where $\delta^{(1)}(\Sigma_{d-1})$ denotes the distributional 1-form that is Poincaré dual to $\Sigma_{d-1}$ in $\cM_{d}$.
This relation follows immediately from the fact that $e^a\in A$ and $t^a\in B$ are
canonically conjugate variables. 
The effect of a finite
translation $\exp(X\cP_a(\Sigma_{d-1}))$ is therefore to shift
$e^b\to e^b + X\delta^{b}{}_a\, \delta^{(1)}(\Sigma_{d-1})$.

\begin{figure}
  \centering
  \begin{subfigure}{0.4\textwidth}
    \centering
    \includegraphics[height=4cm]{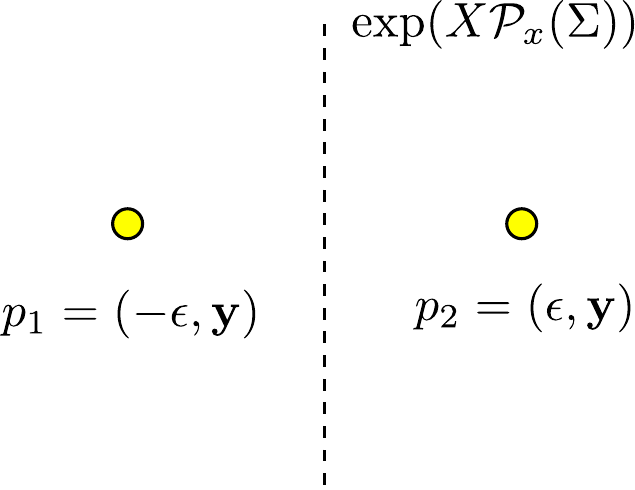}
    \caption{Operator insertion at $x=0$.}
    \label{fig:displacementLocally}
  \end{subfigure}
  \hspace{0.1\textwidth}
  \begin{subfigure}{0.4\textwidth}
    \centering
    \includegraphics[height=4cm]{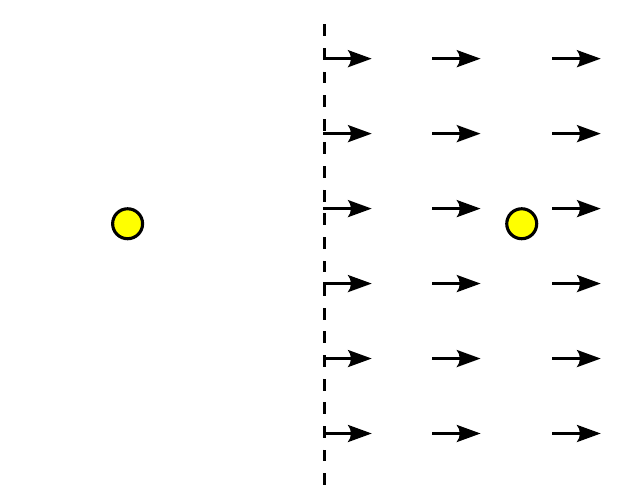}
    \caption{Diffeomorphism representation.}
    \label{fig:displacementAsADiff}
  \end{subfigure}
  \caption{Local picture of an operator implementing translations in
    the horizontal ($x$ axis) direction inserted between two operators
    separated in the $x$ direction. As discussed in the text, the
    effect of this operator on the background vielbein can be undone
    by a diffeomorphism generated by a step-function vector field
    $\xi^\mu=X\delta^\mu_x\theta(x)$. }
\end{figure}

Let us analyse a concrete example  to develop some
intuition. We take $\cM_d=\bR^d$ with a flat metric and no other
backgrounds. In the notation of~\eqref{eq:A-expansion}, this corresponds to
taking $e^a = \delta^a_\mu dx^\mu$ with all other components of $A$
vanishing, so that $A=\delta^a_\mu \dd x^\mu\otimes P_a$. Denote the coordinates of $\bR^d$ by
$(x,y^1,\ldots,y^{d-1})$. We take $\Sigma_{d-1} = \{x = 0\}$ and choose to
generate translations in the $x$ direction. The situation is then
effectively one-dimensional; we sketch it in
figure~\ref{fig:displacementLocally}. This kind of configuration is
what we will see if we zoom into the neighbourhood of a displacement
operator.

Consider two marked points $p_1=(-\epsilon,\mathbf{y})$ and
$p_2=(\epsilon,\mathbf{y})$ with $\epsilon>0$, and $\mathbf{y}$
fixed. Given that we are starting from the flat metric, before
introducing the defect, the distance between the two points is
$2\epsilon$. Introducing the defect $\exp(X\cP_x(\Sigma_{d-1}))$ modifies
the $x$ vielbein as \mbox{$e^x\to e^x + X\delta(x)\dd x$} and keeps the other
vielbeine invariant. (We are being cavalier with smoothness here:~all of our statements about $\delta$ and $\theta$ functions should be regularised, so that the relevant vector fields are smooth. We elaborate on this point below.) 
After this modification, the distance between $p_1$ and $p_2$ becomes $2\epsilon+X$. This
action is indeed consistent with moving $p_2$ to
$(\epsilon+X, \mathbf{y})$, or $p_1$ to $(-\epsilon-X,
\mathbf{y})$. 

We can make this displacement action more concrete in the
following way. The translation generator can be expressed as
\[
  \label{eq:Px-Stokes}
  \cP_x(\Sigma_{d-1}) =\int_{\bR^d} \delta(x)\dd x\wedge t_x= \int_{\bR^d} \dd(\theta(x))\wedge t_x= -\int_{\bR^d} \theta(x)\delta^a{}_x\dd t_a\, .
\]
where 
$\theta(x)$ is the Heaviside
step function. Computing the Poisson bracket as above we get
\[
  \{e^a, \cP_x(\Sigma_{d-1})\} = \dd(\theta(x)\delta^a{}_x )\, .
\]
The result is just as in~\eqref{eq:A-P-PB}, but this
formula has a nice interpretation. Define the adjoint-valued 0-form
$\lambda = \theta(x)\delta^a{}_x\otimes P_a$, and note that
$\dd\lambda = D_A\lambda$, with $D_A\lambda$ the covariant derivative of
$\lambda$ with respect to the background  $A=\delta^a_\mu \dd x^\mu\otimes P_a$. We can then
write:
\[
  \{A, \cP_x(\Sigma_{d-1})\} = D_A\lambda\, .
\]
That is, $\cP_x(\Sigma_{d-1})$ generates an infinitesimal gauge
transformation of $A$ with gauge parameter $\lambda$. This is not a
surprise:~given the index structure, we could have chosen to
write~\eqref{eq:Px-Stokes} in terms of $D_At_a$ instead of $\dd t_a$. When the SymTFT is a BF-theory without any CS term, $D_A t_a$ is a component of $A$'s  equation of motion $D_A B=0$, which generates the gauge transformations of $A$ in the BF-theory
\cite{DiracLectures,Henneaux:1992ig}. In the presence of CS terms, the $A$'s equation of motion is modified to $D_A B + (\text{some power of } F)=0$, which generates the gauge transformation in the BF + CS system. In this case, because of $B$'s equation of motion $F=0$, we can again replace $\dd t_a$ by $A$'s equation of motion in \eqref{eq:Px-Stokes} and interpret $P_x(\Sigma_{d-1})$ as generating a gauge transformation.

This interpretation of the action of $P_x$ allows us to re-interpret
the situation in a more geometric way, using the relation between diffeomorphisms and gauge transformations in BF theory described below~\eqref{eq:inf-diffs-as-gauge}:~recall that the algebra of
diffeomorphisms is given by vector fields $\xi$, which generate
diffeomorphisms via the Lie derivative $\cL_\xi$. Acting on the
connection $A$, we have
\[
 \label{eq:diff-on-A}
  \cL_\xi A = \iota_\xi(F) + D_A(\iota_\xi A)
\]
with $F=dA+A\wedge A$ and $\iota_\xi$ the interior product of forms
with $v$. Since $F=0$ is the other constraint in the SymTFT, the first
term on the right hand side can be ignored. What this equation is then
telling us is that diffeomorphisms generated by a vector field $\xi$ act
on a flat connection $A$ as gauge transformations with gauge parameter $\iota_\xi A$. Given
our choice of Dirichlet boundary condition, we can choose
$\xi^\mu=\theta(x) \delta_x^\mu$ so that we have $\iota_\xi A =
\theta(x) \delta^{a}{}_x\otimes P_a=\lambda$. This means that we can undo the effect of $\cP_x$ on $A$
by a diffeomorphism, generated by a vector field in the $x$ direction,
with magnitude $\theta(x)$, as in
figure~\ref{fig:displacementAsADiff}. This diffeomorphism will, by
construction, undo the effect of $\cP_x$ on $A$, leaving us with our
original $\bR^d$ with the standard flat metric, but it acts
non-trivially on the points of the manifold:~$p_1$ will stay where it
was, but $p_2$ will be shifted precisely to $(\epsilon+X,\mathbf{y})$.

We now turn to briefly discuss rotations with generators given in \eqref{eq:charges}:
\[
  \cJ_{ab}(\Sigma_{d-1}) = \frac{1}{2}\oint (j_{ab} - 2 x_{[a}t_{b]})\, .
\]
We want to show that this operator
implements rotations on the boundary. To this end, we proceed as above and 
compute the relevant Poisson bracket:
\[
  \{e^c, \cJ_{ab}\} &= (x_b\delta^{c}{}_a- x_a\delta^{c}{}_b) \delta^{(1)}(\Sigma_{d-1}) \, ,
\\
\{\omega^{cd}, \cJ_{ab}\} &=  \frac{1}{2}(\delta^{c}{}_{a}\delta^d{}_b-\delta^{c}{}_{b}\delta^d{}_a) \delta^{(1)}(\Sigma_{d-1})\,.
\]
Combining these two equations  gives
\[
   \{A, \cJ_{ab}\} &= (x_b\delta^{c}{}_a- x_a\delta^{c}{}_b) \delta^{(1)}(\Sigma_{d-1})\otimes P_c+\frac{1}{2}(\delta^{c}{}_{a}\delta^d{}_b-\delta^{c}{}_{b}\delta^d{}_a) \delta^{(1)}(\Sigma_{d-1})\otimes L_{ab}\,.
\]
The right hand side can be organized into a gauge transformation $D_A\lambda$ with the background $A=\delta^a_\mu \dd x^\mu\otimes P_a$ and the adjoint-valued field
\ie
\lambda= (x_b\delta^{c}{}_a- x_a\delta^{c}{}_b) \theta(D_d)\otimes P_c+\frac{1}{2}(\delta^{c}{}_{a}\delta^d{}_b-\delta^{c}{}_{b}\delta^d{}_a) \theta(D_d)\otimes L_{cd}
\fe
where $\partial D_d = \Sigma_{d-1}$, and $\theta(D_d)$ denotes a generalised Heaviside
function that equals to 1 inside $D$ and vanishes outside, so that
$\dd\theta(D_d)=\delta(\Sigma_{d-1})$. We can trade the first half of $\lambda$ by a diffeomorphism associated with the vector field $\xi^\mu=(x_b\delta^{\mu}_a- x_a\delta^{\mu}_b) \theta(D_d)$ such that $\iota_\xi A=(x_b\delta^{c}{}_a- x_a\delta^{c}{}_b) \theta(D_d)\otimes P_c$. This vector field is precisely what generates rotations on the $(a,b)$ plane within the domain $D_d$. The second half of $\lambda$ is still treated as a gauge transformation, and implements the expected action of rotation on
the internal indices, again acting only inside $D_d$. In summary, $\mathcal{J}_{ab}$ effectively implements the rotation associated with $\xi^\mu$ up to a gauge transformation that rotate the internal indices.

Finally, let us come back to the issue of smoothness:~what is shown
infinitesimally in~\eqref{eq:inf-diffs-as-gauge} and~\eqref{eq:diff-on-A},
and is explored
more in detail in appendix~\ref{sec:App_finite_transform}, is that we
can represent diffeomorphisms in terms of gauge transformations. The
arguments that we have just given show, in turn, that the relevant
gauge transformations can be constructed in terms of the $B$ holonomy
operators in the bulk, so we should be able to generate the
diffeomorphisms of the boundary theory (up to a subtlety described in
appendix~\ref{sec:App_finite_transform}) by acting on the boundary
with $B$ operators. As we have seen, the vector field resulting from a
single finite symmetry generator localised on a submanifold is
singular, so if we want to represent more familiar smooth vector
fields we need to consider suitable superpositions of bulk operators,
by a (physically, at least) straightforward generalisation of the
previous discussion:~divide the boundary into small simplices, such
that the vector field is approximately constant inside each simplex,
and only changes as we cross from one simplex to the next. Then
introduce symmetry generators on the faces of the simplices that
implement the changes in the vector field that occur when moving
across neighbouring simplices. In the limit of vanishing volume for
the simplices, we end up with a smooth network of symmetry generators.

There seems to be nothing from the bulk point of view, though, that
forces us to choose such smooth configurations of symmetry generators,
and it is interesting to explore in more detail what happens for
localised, finite symmetry generators. We explore this topic in the next section.

\subsection{Path Integral Approach}
\label{sec:path-integral}

\begin{figure}
$$
\begin{tikzpicture}
\scalebox{1}{
\draw [cyan, fill= cyan, opacity =0.8]
(0,0) -- (0,4) -- (2,5) -- (7,5) -- (7,1) -- (5,0)--(0,0);
\draw [white, thick, fill=white,opacity=1]
(0,0) -- (0,4) -- (2,5) -- (2,1) -- (0,0);
\begin{scope}[shift={(5,0)}]
\draw [white, thick, fill=white,opacity=1]
(0,0) -- (0,4) -- (2,5) -- (2,1) -- (0,0);
\end{scope}
\begin{scope}[shift={(0,0)}]
\draw [cyan, thick, fill=cyan, opacity= 0.2]
(0,0) -- (0, 4) -- (2, 5) -- (2,1) -- (0,0);
\draw [black, thick]
(0,0) -- (0, 4) -- (2, 5) -- (2,1) -- (0,0);
\node at (1.2,3.7) {$\Bsym_{G}$};
\end{scope}
 \node at (3.5, 3.7) {$\SymTFT_{d+1}(G)$};
\begin{scope}[shift={(5,0)}]
\draw [black, thick, fill= blue,opacity=0.1] 
(0,0) -- (0, 4) -- (2, 5) -- (2,1) -- (0,0);
\draw [black, thick]
(0,0) -- (0, 4) -- (2, 5) -- (2,1) -- (0,0);
\node at (1,3.7) {$\Bphys_d$};
 \draw [black,fill=black] (1,1.3) ellipse (0.07 and 0.07);
  \draw [black,fill=black] (1,2.5) ellipse (0.07 and 0.07);
  \draw [thick] (1,1.3) -- (-4,1.3);
    \draw [ultra thick, red ] (-4,2.5) ellipse (0.75 and 0.8);
    \draw [thick] (1,2.5) -- (-4,2.5);
     \draw [black,fill=black] (-4,1.3) ellipse (0.07 and 0.07);
  \draw [black,fill=black] (-4,2.5) ellipse (0.07 and 0.07);
\node[left, red] at (-4.95, 2.5) {$\cU_{e^{x^aP_a}}[\Sigma_{d-1}]$};
\node[above] at (-1.5,1.3) {$W_1$};
\node[above] at (-1.5,2.5) {$W_2$};
\draw [red, thick, ->] (-4, 1.8) -- (-4+ 0.2, 1.8+ 0.4) ;
\draw [red, thick, ->] (-4-0.2, 1.8) -- (-4, 1.8+ 0.4) ;
\draw [red, thick, ->] (-4-0.4, 1.8) -- (-4-0.2, 1.8+ 0.4) ;
\end{scope}
\node at (3.3, 2.5) {};
\draw[dashed] (0,0) -- (5,0);
\draw[dashed] (0,4) -- (5,4);
\draw[dashed] (2,5) -- (7,5);
\draw[dashed] (2,1) -- (7,1);  
}
\end{tikzpicture}
$$
\caption{Action of the symmetry generators $\cU_{e^X} (S_{d-1})$ on the charges, which are the end-points of bulk operators $W_i$. \label{fig:blanki}}
\end{figure}
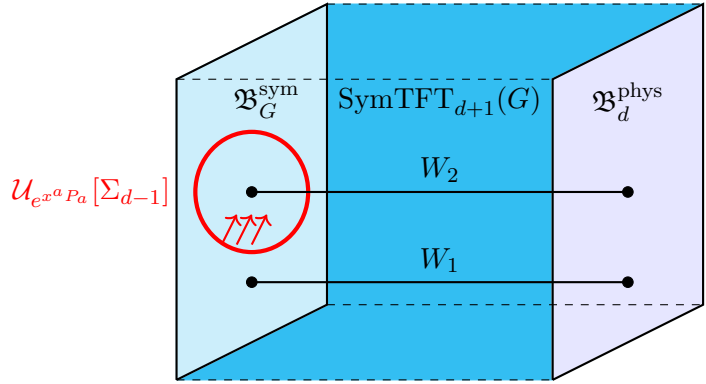

Let us first discuss the path integral perspective on symmetry operators.
As an example, consider a SymTFT quiche
configuration with Dirichlet boundary 
\begin{equation}
A|_{\partial M_{d+1}}=\cA = \delta_{\mu}^a \dd x^\mu \otimes P_a \equiv h^{-1} \dd h \,, \quad h =  e^{ \delta_\mu^a x^\mu P_a} \, ,
\end{equation}
which describes a flat metric on $\partial M_{d+1}$.
To study the action of the symmetry generators, we insert a symmetry generator $\mathcal{U}_{e^X}[\Sigma_{d-1}]$ along the  Dirichlet boundary condition. It is inserted such that it separates the endpoints of two  Wilson lines that stretch from the physical to the symmetry boundary, i.e.~charged  local operators, see figure \ref{fig:blanki}.

The insertion of the symmetry operator associated with element $g = e^X$ modifies the boundary condition to 
\begin{equation}
A|_{\partial M_{d+1}} = h^{-1} \dd h + (h^{-1} X h) \, \delta^{(1)}(\Sigma_{d-1})  \,.
\end{equation}
As a check, we show that this new boundary condition is flat and thus compatible with the bulk equation of motion:
\be
\ba
F|_{\partial M_{d+1}} &= \dd_{h^{-1} \dd h}(h^{-1} X h) \wedge \delta^{(1)}(\Sigma_{d-1}) \nonumber\\
&= h^{-1} (\dd X ) h \wedge \delta^{(1)}(\Sigma_{d-1}) = 0 \,,\ea
\ee
where the term proportional to $\delta^{(1)}(\Sigma_{d-1})\wedge \delta^{(1)}(\Sigma_{d-1})$ vanishes when $\Sigma$ is not self-intersecting.

Let us consider the case of symmetry operators associated to translations $X = X^a P_a$ in the background $\cA = \delta_\mu^a \dd x^\mu \otimes P_a$. The new metric, locally around $\Sigma_{d-1}$ reads
\begin{equation}\label{eq:translationmetric}
    g_{\mu\nu}dx^\mu dx^\nu = \eta_{\mu\nu} \dd x^\mu \dd x^\nu + 2 X_\mu \delta(r-r_0) \dd r  \dd x^\mu +  |X|^2 \delta(r-r_0)^2 \dd r  \dd r  \,.
\end{equation}
{The vielbein underlying this metric will generically become non-invertible in a region near the defect, rendering the metric not everywhere positive-definite. (See appendix~\ref{sec:singmetric} for a careful analysis in terms of a regulated solution.) This is a manifestation of the phenomenon, common in the first-order formulations of gravity, that the field space of these formulations naturally includes configurations with non-invertible vielbeine, which do not have a simple interpretation in terms of Riemannian geometry. Although we expect that we can avoid such configurations by considering superpositions of smooth families of defects, as sketched in the previous section, in our context this smoothing is not a very natural operation. Regardless of our attitude towards such backgrounds, we can adapt the results in the previous section to construct a diffeomorphism that turns these backgrounds back into flat space.} 
In detail, the new boundary condition  after acting with this translation is gauge equivalent to $\mathcal{A}$ and can be written as
\begin{equation}
A|_{\partial M_{d+1}} = A^{(e^\alpha)} = e^{-\alpha} \cA e^{\alpha} + e^{-\alpha} \dd e^{\alpha}, \quad \alpha = X^a P_a \theta(r-r_0) \,.
\end{equation}
The appropriate diffeomorphism corresponding to the gauge
transformation above is generated by a vector field 
\be\label{Diffeoxi}
\xi^\mu = X^\mu\theta(r-r_0)\,.
\ee

Once we include the physical boundary condition  $\mathcal{B}_{\rm phys}$, we obtain local operators $\mathcal{O}$ as the endpoints of topological lines in the bulk, which end on both boundaries. Consider the correlator with the symmetry defect inserted:
\begin{equation}\label{eq:correlator_A}
\langle \mathcal{O}(x_1) \mathcal{U}_{e^X}(\Sigma_{d-1}) \mathcal{O}(x_2) \rangle_{\cA}\,,
\end{equation}
where $\Sigma_{d-1}$ encloses $x_1$. The effect of the topological operator is to change the background, so that
\begin{equation}
\langle \mathcal{O}(x_1) \mathcal{U}_{e^X}(\Sigma_{d-1}) \mathcal{O}(x_2) \rangle_{\cA} = 
\langle \mathcal{O}(x_1) \mathcal{O}(x_2) \rangle_{\cA^{(\alpha)}}\,.
\end{equation}
We can restore the background from $\cA^{(\alpha)}$ to $\cA$ by performing a diffeomorphism with $\xi$ in (\ref{Diffeoxi}), so that one has
\begin{equation}
\langle \mathcal{O}(x_1) \mathcal{O}(x_2) \rangle_{\cA^{(\alpha)}} =  \langle (e^{\mathcal{L}_{\xi}}\mathcal{O}(x_1) (e^{\mathcal{L}_{\xi}}\mathcal{O}(x_2) \rangle_{\cA} \,.
\end{equation}
The vector  $\xi$ implements a constant translation outside $\Sigma_{d-1}$. 
The net effect inside the correlator above is to leave invariant the operator insertion at $x_1$ and move the one at $x_2$ to $x_2 - X$ (even if operators have spin, all Jacobians are trivial away from the surface $\Sigma_{d-1}$). Notice that the fact that $\mathcal{U}_{e^X}(\Sigma_{d-1})$ is topological is evident from the fact that if $\Sigma_{d-1}$ does not enclose any insertion point, correlators are unaffected as they depend on relative distances.

\section{Relation to Gravity }
\label{sec:gravity}

In this final section, we address the obvious question:~what is the relation between the SymTFT for the conformal symmetry and gravity with negative cosmological constant? The latter is also where the connection to standard holography of AdS spacetime becomes relevant.

For internal symmetries, it is by now well-established that the SymTFT is captured in the standard holographic setting in terms of certain topological operators, usually realized in terms of branes (in a topological limit) \cite{Apruzzi:2022rei, GarciaEtxebarria:2022vzq, Antinucci:2022vyk, Bah:2023ymy, Cvetic:2023pgm, Cvetic:2025kdn, Antinucci:2024bcm, Heckman:2025wqd, Heckman:2025lmw, Heckman:2025isn, Heckman:2024zdo, Heckman:2024oot, Bah:2024ucp}.  It is therefore natural to ask how our spacetime SymTFT relates to the standard holographic paradigm. 


\subsection{First-Order Formulation of Gravity}
\label{sec:Palatini}

Since the SymTFT is formulated as a gauge theory, it is more natural to connect it with gravity in the first order (or Palatini) formulation. In the standard second order formulation, the fundamental degree of freedom is the metric $g_{\mu\nu}$, which is not obviously related to a gauge field. By contrast, in the first order formulation, the fundamental degrees of freedom are the vielbein one-form $e^a=e^a_\mu dx^\mu$ and the spin connection one-form $\omega^{ab}=\omega^{ab}_\mu dx^\mu=-\omega^{ba}$, with $a,b=1,...,d+1$. Here, the bulk spacetime dimension is taken to be $d+1$. 
These one-forms are similar to gauge fields. They are subject to an $SO(d+1)$ gauge symmetry
 \ie\label{eq:vielbein_spin_connection}
 e^a&\rightarrow (\Lambda^{-1}e)^a=({\Lambda}^{-1})^{a}{}_b e^b\,,
 \\
 \omega^{a}{}_{b}&\rightarrow (\Lambda^{-1}\omega\Lambda)^{a}{}_b+(\Lambda^{-1} \dd\Lambda)^{a}{}_{b}\,,
 \fe
 where $\Lambda^{a}{}_b\in SO(d+1)$ obeys $ \eta_{ab}\, \Lambda^{a}{}_{c}\Lambda^b{}_d=\eta_{cd}$ with $\eta_{ab}$ the Euclidean flat metric.\footnote{We deliberately introduce the flat metric $\eta_{ab}$ and distinguish the upper and lower indices so that the formulae are also applicable to Lorentzian signature. All the omitted index contraction is between one upper and one lower index, as how the index is contracted in the first line of \eqref{eq:vielbein_spin_connection}.}  Expanding $\Lambda=\exp(\lambda)$ around the identity, we obtain the infinitesimal gauge transformation
 \ie\label{eq:small_SO(d)_gauge}
 \delta_\lambda e^a=-(\lambda \, e)^a\,,
 \qquad
 \delta_\lambda\omega^{ab}=(\dd_\omega\lambda)^{ab}\equiv(\dd\lambda+[\omega,\lambda])^{ab}\,.
 \fe
 From the vielbein $e^a$ and spin connection $\omega^{ab}$, one can construct the gauge-invariant metric $g_{\mu\nu}$ and the gauge-covariant curvature two-form $R^{ab}$ and  torsion two-form $T^a$:
 \ie\label{eq:first-order-to-second-order-relation}
 &g_{\mu\nu}\equiv e^a{}_\mu \eta_{ab} e^b{}_\nu\,,
 \\
 &R^{ab}\equiv \dd\omega^{ab}+(\omega\wedge\omega)^{ab}\,,
 \\
 &T^a\equiv \dd_\omega e^a\equiv \dd e^a+(\omega\wedge e)^a\,.
 \fe
 These gauge-covariant forms transform as $R^{ab}\rightarrow (\Lambda^{-1} R \Lambda)^{ab}$ and $T^a\rightarrow (\Lambda^{-1} T)^a$.

 In the first-order formulation, the Einstein-Hilbert action with a cosmological constant $\Lambda_{\text{c}}$ is given by\footnote{We use $\Lambda_{\text{c}}$ to denote the cosmological constant, which should not be confused with the $SO(d)$ gauge parameter $\Lambda^{a}{}_b$.}
 \ie\label{eq:first-order-EH-action}
 S_{\text{EH}}[e,\omega]=-\frac{1}{16\pi G_N}\int \epsilon_{{a_1}\cdots{a_{d+1}}}\left(\frac{1}{(d-1)!} R^{a_1 a_2}\wedge e^{a_3}\wedge \cdots\wedge e^{a_{d+1}}-\frac{2\Lambda_{\text{c}}}{(d+1)!}e^{a_1}\wedge\cdots\wedge e^{a_{d+1}}\right)\,.
 \fe
 Using the relation \eqref{eq:first-order-to-second-order-relation}, we recover the Einstein-Hilbert action in the familiar form\footnote{We work in the Euclidean signature. In the Lorentzian signature, the Einstein-Hilbert action is $S[g_{\mu\nu}]=\frac{1}{16\pi G}\int d^{d+1}x \sqrt{-g}( R-2\Lambda_{\text{c}})$ and the first-order action differ from the Euclidean one by an overall minus sign.}
 \ie\label{eq:second-order-EH-action}
 S_{\text{EH}}[g_{\mu\nu}]=-\frac{1}{16\pi G_N}\int d^{d+1}x \sqrt{g}( R-2\Lambda_{\text{c}})\,.
 \fe
 Here, we omit the boundary Gibbons-Hawking-York term.
 The equation of motion from \eqref{eq:first-order-EH-action} is 
 \ie
 &\omega^{ab}:\quad T^a=de^a+(\omega\wedge e)^a=0\,,
 \\
 &e^a:\quad \ \, 
 \epsilon_{a{a_1} \cdots{a_{d}}} R^{a_1 a_2}\wedge e^{a_3}\wedge{\cdots}\wedge e^{a_{d}}=\frac{2\Lambda_{\text{c}}}{d(d-1) }\epsilon_{a{a_1}\cdots{a_{d}}}e^{a_1}\wedge{\cdots}\wedge e^{a_{d}}\,.
 \fe
 The first equation is the torsion-free condition, while the second equation is the Einstein equation in vacuum. Note that in the first equation we use the invertiblity of the vielbein.

Gravity is the first-order formulation appears to be a gauge theory. However, there are several important subtleties, which we now emphasize:
\begin{itemize}
    \item For a well-defined geometry, we demand that the metric is non-degenerate, or equivalently that the vielbein is invertible i.e.~$\det(e)=\sqrt{g}\neq0$. This imposes a non-trivial restriction on the space of integration in the first-order formulation, making the theory deviate from the naive gauge theory.
    \item In gravity, diffeomorphisms are also gauge symmetries, so they should be modded out in the path integral. An infinitesimal diffeomorphism acts on the vielbein $e^a$ and spin connection $\omega^{ab}$ as a shift by Lie derivatives
    \ie
    \delta_\xi e^a=\mathcal{L}_\xi e^a&=\iota_\xi T^a +\dd_\omega (\iota_\xi e^a)-(\iota_\xi\omega^a{}_b)e^b\,,
    \\
    \delta_\xi \omega^{ab}=\mathcal{L}_\xi \omega^{ab}&=\iota_\xi R^{ab}+\dd_\omega(\iota_\xi\omega^{ab})\,,
    \fe
    where $\xi^\mu$ is the vector field parameterizing the infinitesimal diffeomorphism. In a standard theory of gravity, torsion vanishes on-shell, so we can ignore the first term in $\mathcal{L}_\xi e^a$. Moreover, the last term in $\mathcal{L}_\xi e^a$ and $\mathcal{L}_\xi \omega^{ab}$ can be undone by an $SO(d)$ gauge transformation \eqref{eq:small_SO(d)_gauge} with $\lambda^{a}{}_{b}=-\iota_\xi \omega^{a}{}_b$, so diffeomorphisms effectively act as
    \ie\label{eq:diffeo_bulk}
    \delta_\xi e^a&=\dd_\omega (\iota_\xi e^a)\,,\qquad
    \delta_\xi \omega^{ab}=\iota_\xi R^{ab}\,.
    \fe
    This expression of infinitesimal diffeomorphisms will be useful in the coming section.
    In addition to these diffeomorphisms, there can also be large diffeomorphisms that are disconnected from the identity. They also need to be modded out in a theory of gravity. These disconnected diffeomorphisms are captured by the mapping class group
    \ie
    \text{MCG}(\Sigma)=\frac{\text{Diff}^+(\Sigma)}{\text{Diff}_0(\Sigma)}~,
    \fe
    where $\text{Diff}^+(\Sigma)$ denotes orientation preserving diffeomorphisms of the manifold $\Sigma$, while $\text{Diff}_0(\Sigma)$ denotes those diffeomorphisms continuously connected to identity.
    \item In general, gravity requires summing over topology,
    which does not seem necessary
    if we treat the first-order formulation of gravity as a gauge theory.
\end{itemize}


\subsection{Gravity versus SymTFT}

After reviewing the first-order formulation of gravity, we are now ready to discuss the connection between gravity with negative cosmological constant and the proposed SymTFT for conformal symmetry. Recall that in even dimensions $d+1=2n$, the SymTFT is simply a BF-theory for the conformal group, whereas in odd dimension $d+1=2n+1$, it includes additional CS-terms.  

\subsubsection{Gauge-theoretic Formulation}
To make the connection more transparent, one can package the vielbein $e^a$ and spin connection $\omega^{ab}$ into an $\mathfrak{so}(d+1,1)$-valued one-form field as
\ie\label{eq:bulk_SO(d,2)_gauge_field}
A=\frac{1}{\ell} e^a M_{a,d+2} +\frac{1}{2}\omega^{ab} M_{ab}\,,
\fe
where $M_{AB}=-M_{BA}$ with $A,B=1,...,d+2$ are the generators of $\mathfrak{so}(d+1,1)$ Lie algebra 
\ie
[M_{A B}, M_{EF}] =  (\eta_{AE} M_{BF} - \eta_{BE} M_{AF} - \eta_{AF} M_{BE} + \eta_{BF} M_{AE})\,,
\fe
with $\eta_{AB}$ the flat metric in $(+,...,+,-)$ signature. We can embed the $SO(d+1)$ gauge parameter $\Lambda=\exp({\lambda})$ into  an $SO(d+1,1)$ matrix as $\mathbf{\Lambda}=\exp(\lambda^{ab}M_{ab})$. With this embedding, the $SO(d+1)$ gauge symmetry acts on $A$ as a standard gauge transformation 
\ie
A\rightarrow  \mathbf{\Lambda}^{-1} A \mathbf{\Lambda} + \mathbf{\Lambda}^{-1} \dd\mathbf{\Lambda}\,.
\fe
It is tempting to enlarge this $SO(d+1)$ gauge symmetry to $SO(d+1,1)$, making the one-form field $A$ a full-fledged $\mathfrak{so}(d+1,1)$ gauge field. However, this is generally not possible, so in general $A$ behaves more like an $\mathfrak{so}(d+1,1)$ gauge field coupled to a Higgs field that Higgses the gauge symmetry down to $SO(d+1)$. Only in some special cases can the full $SO(d+1,1)$ gauge symmetry be realized. To see when this happens, let us spell out the action of the additional would-be $SO(d+1,1)$ gauge transformations associated with $\mathbf{\Lambda}=\exp(v^a M_{a,d+2}/\ell)$ for infinitesimal $v^a$\,:
\ie
\delta_v e^a&=\dd_\omega v^a\,,
so\\
\delta_v \omega^{ab}&=-\frac{1}{\ell^2} (v^a e^b -v^b e^a)\,.
\fe
Comparing this with diffeomorphisms in \eqref{eq:diffeo_bulk}, we see that they coincide when  $v^a=\iota_\xi e^a$ and the curvature is constant, $R^{ab}=-\ell^{-2} e^a\wedge e^b$. In this case, diffeomorphisms make up the missing $SO(d+1,1)$ gauge transformation, provided the on-shell geometry is restricted to constant curvature spaces. As we will show below, this is what happens in the SymTFT for conformal symmetry, as well as in 2d Jackiw-Teitelboim gravity and 3d Einstein-Hilbert gravity. In general, diffeomorphisms differ from the $SO(d+1,1)$ gauge transformations, and $A$ should be interpreted as a Higgsed $\mathfrak{so}(d+1,1)$ gauge field. 

When comparing SymTFT for conformal symmetry with gravity, it is natural to decompose the $\mathfrak{so}(d+1,1)$ gauge field $A$ in the SymTFT as in \eqref{eq:bulk_SO(d,2)_gauge_field}. This differs from the decomposition into $d$-dimensional conformal generators in  as \eqref{eq:A-expansion}. In this decomposition, the field strength of $A$ takes a simple form in terms of the curvature 2-form $R^{ab}$ and torsion 2-form $T^a$ defined in \eqref{eq:first-order-to-second-order-relation}:
\begin{equation}
F = \dd A + A^2 = \frac{1}{\ell}T^a M_{a,d+2} 
+ \frac{1}{2}\!\left(R^{ab} + \frac{1}{\ell^2} e^a \wedge e^b\right) M_{ab}\,.
\end{equation}
In the SymTFT, the gauge field $A$ obeys the flatness constraint $F=0$, which enforces vanishing torsion $T^a=0$ and  constant negative curvature $R^{ab}=-\ell^{-2} e^a\wedge e^b$. This restricts the geometry to be locally AdS space with AdS radius $\ell$. In dimensions $d+1\geq 4$, this condition is stronger than the vacuum Einstein equation, which admits a much broader set of solutions, including gravitational waves. 

In what follows, we elaborate on the comparison between SymTFT and gravity with increasing dimension.

\subsubsection{1d Gravity} 
We start with the lowest possible dimension with $d+1=1$, i.e.~1d bulk SymTFTs.  Although this is a somewhat degenerate case, it does fit into the progression of dimensions, and we will briefly discuss it first. 
The naive specialisation to $d=0$ of the general conformal group $SO(d+1,1)$ is $SO(1,1)$, which is abelian. So the 
putative SymTFT would be an abelian BF-theory, with the $B$ field formally a $(-1)$-form. (Note that there is no conformal anomaly in 0d, so we don't add additional terms to the 1d SymTFT.) To make sense of this BF-theory, we formally integrate by parts to write $\int (dB)_0 \, A_1$ instead, where $(dB)_0$ is a constant, and we have omitted the $\langle...\rangle_{\mathrm{BF}}$ inner product for notational simplicity. This coupling is analogous to the Romans mass. According to \eqref{eq:bulk_SO(d,2)_gauge_field}, the gauge field $A_1$ is identified with the ein-bein $e$ (the one-legged vielbein, where we omit the index). 
Assuming $e>0$ everywhere, this action can be written as $\int(dB)_0 \sqrt{g}$ where $g=e^2$, which is indeed the (not very interesting) action of 1d gravity with cosmological constant $(dB)_0$.

\subsubsection{2d Jackiw-Teitelboim Gravity}

We now move up to $d+1=2$ dimensions. The Einstein-Hilbert action $S_{\text{EH}}$ in \eqref{eq:second-order-EH-action} with $\Lambda_\text{c}=0$ is proportional to the Euler characteristic, so the theory is purely topological with no dependence on the geometry. Furthermore, because the Einstein tensor vanishes identically in 2d, the vacuum Einstein equation only has $g_{\mu\nu}=0$ as its solution when $\Lambda_\text{c}\neq0$.

To obtain a more interesting 2d theory of gravity, we consider Jackiw-Teitelboim (JT) gravity, which is a 2d dilaton gravity theory, with the action
\ie
S=-\frac{1}{16\pi G_N}\int d^2x\,\phi\sqrt{g}(R+2)~,
\fe
where $\phi$ is the dilaton whose equation of motion constrains the space to have negative constant  curvature. In the first-order formulation, JT gravity action can be reorganized into a BF-action based on $SL(2,\mathbb{R})\simeq SO(2,1)$ using the combination of $A$ in \eqref{eq:bulk_SO(d,2)_gauge_field} and $B=\phi J_{ab}+\phi^a P_a$ with $\phi$ the dilaton and $\phi^a$ the Lagrange multiplier for the torsion free condition
\cite{Fukuyama:1985gg,Isler:1989hq,Chamseddine:1989yz,Jackiw:1992bw} (see also \cite{Saad:2019lba,Iliesiu:2019xuh} for recent applications). This BF-action is precisely the action for the proposed SymTFT for conformal symmetry in $d=1$!

However, we want to emphasize that this equivalence is only true at the classical level. As discussed in section \ref{sec:Palatini} that there are various subtleties in the first-order formulation of gravity, which make them deviate from ordinary gauge theory. First of all, because of the invertibility condition of the vielbein, 
the space of integration is restricted from the space of flat $SL(2,\mathbb{R})$ connections to the Teichm\"uller space space, which is a disconnect component inside the space of flat $SL(2,\mathbb{R})$ connections \cite{Verlinde:1989ua}. 
Second, in a theory of gravity, we need to mod out diffeomorphisms. As explained section \ref{sec:Palatini}, diffeomorphism connected to the identity i.e.~those in $\text{Diff}_0$ are already included in the $SL(2,\mathbb{R})$ gauge symmetry of $A$, so we only need to mod out the mapping class group, which then restrict the space of integration to the moduli space of Riemann surfaces. Lastly, we need to perform a sum over topology.

Another difference between SymTFT and the usual treatment of JT gravity is in the type of boundary conditions that we study:~instead of the Schwarzian, 
which is relevant for the holographic duality to SYK, we consider here gapped boundary conditions, that allow us to generate spacetime symmetries on the boundary (the conformal symmetry to be precise). See \cite{Goel:2020yxl} for other possible boundary conditions in JT graivty.

\subsubsection{3d Gravity and Virasoro TQFT}

In $d+1=3$ dimensions, gravity defined by the Einstein-Hilbert action is again topological without any propagating degrees of freedom. With a negative cosmological constant, its first-order action can be reorganized into the CS action based on $SO(3,1)\simeq SL(2, \mathbb{C})$ in Euclidean signature and $SO(2,2)\simeq SL(2,\mathbb{R})\times SL(2,\mathbb{R})$ in Lorentzian signature using the combination in \eqref{eq:bulk_SO(d,2)_gauge_field} \cite{Witten:1988hc}. It however does not mean 3d gravity is identical to these CS-theories because crucially not every gauge field configurations in the CS-theory correspond to a physical geometry with an invertible vielbein. 
This problem is severe in Euclidean signature \cite{Witten:2007kt} (see \cite{Collier:2025lux} for some recent progress). However, in Lorentzian signature, it was resolved  by restricting the phase space of $SL(2,\mathbb{R})$ CS-theory (a chiral half of $SO(2,2)$ CS-theory) from flat $SL(2,\mathbb{R})$ connections to Teichm\"uller space \cite{Verlinde:1989ua,Collier:2023fwi}. 
Surprisingly, this restriction yields a consistent theory upon quantization, named Virasoro TQFT in \cite{Collier:2023fwi}. The Hilbert space of the Virasoro TQFT on a Riemann surface is spanned by the Virasoro conformal blocks. Virasoro TQFT itself is still not yet a theory of gravity. To promote it to the full-fledged 3d gravity, we need a chiral and anti-chiral copies of Virasoro TQFTs (see \cite{Hartman:2025cyj}) for a dual formulation of $\text{Virasoro}\times \overline{\text{Virasoro}}$ TQFTs in terms of conformal Turaev-Viro  theory) and further incorporate the gauge constraints from the mapping class group. This allows one to compute the partition functions on a fixed topology. Lastly, we need to sum over topology.

It is tempting to identify the $\text{Virasoro}\times \overline{\text{Virasoro}}$ TQFT 
as the SymTFT for conformal symmetry. However, it is incorrect. The $\text{Virasoro}\times \overline{\text{Virasoro}}$ TQFT should be interpreted as the SymTFT for the continuous non-invertible Virasoro-preserving topological lines \cite{Drukker:2010jp} in the corresponding Liouville CFT \cite{Bao:2024ixc}. In comparison, the proposed SymTFT for conformal symmetry captures a completely different set of lines that generate the conformal symmetry. These lines, despite topological, generally do not commute with stress tensors and therefore breaks the Virasoro symmetry. 

The difference between $\text{Virasoro}\times \overline{\text{Virasoro}}$ TQFT and the SymTFT for conformal symmetry also shows up at the classical level. $\text{Virasoro}\times \overline{\text{Virasoro}}$ TQFT is classically equivalent to $SL(2,\mathbb{R})\times SL(2,\mathbb{R})$ CS-theory while the SymTFT for conformal symmetry is a (BF+CS)-theory based on the $SL(2,\mathbb{R})\times SL(2,\mathbb{R})$ conformal group.

It is instructive to draw an analogy with $SU(N)_1$ Wess-Zumino-Witten model. The theory has an extended $\mathfrak{su}(N)_1$ chiral algebra. It has $N$ topological lines that preserves this extended chiral algebra and their SymTFT is $SU(N)_1\times SU(N)_{-1}$ CS-theory. In contrast, the theory has a wealth of topological lines that preserve only the Virasoro symmetries. They include for example the $G=(SU(N)_L\times SU(N)_R)/\mathbb{Z}_N$ global symmetry. The SymTFT of this global symmetry is a BF+CS theory based on $G$ with CS-level 1 capturing the anomaly.

In summary, the SymTFT for conformal symmetry is distinct from the $\text{Virasoro}\times \overline{\text{Virasoro}}$ TQFT. The former captures conformal symmetry on the boundary, while the latter captures the Virasoro-preserving topological lines in Liouville CFTs. Furthermore, the former is a BF theory with topological defects from both the holonomies of $B$ and of $A$, while the latter is classically equivalent to a CS theory with only topological defects from $A$.

\subsubsection{Topological Limit of 4d Gravity}

In general, gravity and the SymTFT for conformal symmetry are distinct. We 
have argued thus far, that the $(d+1)$-dimensional SymTFT for the
conformal symmetries of a $d$-dimensional CFT is a
BF-theory (plus CS-couplings for odd bulk dimensions) for the group $SO(d+1,1)$. It is natural to ask how one could motivate this result holographically {in cases where gravity is not the same as the SymTFT}.  In the case of internal
symmetries of theories with a holographic dual, a number of works
\cite{Witten:1998wy, Apruzzi:2021phx, Apruzzi:2022rei,
  GarciaEtxebarria:2022vzq, Antinucci:2022vyk} have argued that the
SymTFT arises from studying the dynamics of the bulk fields at
infinity. We will now argue that the same is true for
spacetime transformations in $d=3$, or equivalently 4d gravity in the bulk:~the gravitational dynamics at infinity are described effectively by a $G_N\to 0$ limit, and the bulk gravity theory reduces to the SymTFT in this limit.
The restriction to
$d=3$, which we do not believe is due to any fundamental principle, is
because the formulation of gravity that we use in our argument seems
to be currently only known for $d+1 = 4$.

\paragraph{Gravity in AdS.}
We start by fixing some notation and conventions. In the Poincaré
patch, the metric of Euclidean AdS$_{d+1}$ can be written as
\be
  ds^2 = g_{\mu\nu}^{\text{AdS}} dx^\mu dx^\nu \df \frac{\ell^2}{z^2}\bigl(dz^2 + \delta_{ij}dx^idx^j\bigr)\,,
\ee
with $\delta_{ij}$ the flat metric on $\bR^d$. Here we are interested in the case $d=3$. The boundary is at
$z=0$. The cosmological constant in AdS$_{4}$ is 
$\Lambda=-3/\ell^2$. Gravitational dynamics are described by the
Einstein-Hilbert action
\[
  \label{eq:Einstein-Hilbert}
  S_{\text{EH}}[g_{\mu\nu}] = -\int_{\mathrm{AdS}_4} dx^4 \frac{\sqrt{g}}{16 \pi G_N} \left(R - 2\Lambda\right)\, .
\]
Here $G_N$ is the gravitational constant, which we have written inside
the integral for reasons that will become clear momentarily.

In the Fefferman-Graham gauge
\cite{FeffermanGraham,Fefferman:2007rka}, any asymptotically AdS
metric can be written as
\[
  \label{eq:Fefferman-Graham}
  ds^2 = g_{\mu\nu}dx^\mu dx^\nu =
  \frac{\ell^2}{z^2}(dz^2 + \tilde g_{ij}(x,z)dx^idx^j)
\]
with
$\tilde g(x,z) = \tilde g_{(0)}(x) + z^2 \tilde g_{(2)}(x) + \ldots$,
where we omit higher powers of $z$. This kind of rewriting is familiar
from studies of holographic renormalisation
\cite{Henningson:1998gx,Henningson:1998ey,deHaro:2000vlm,Bianchi:2001kw}
(see \cite{Skenderis:2002wp} for a review). In
fact, our analysis in this section is in some sense the most trivial
aspect of this program:~we will show that the effective dynamics for
the non-singular combination $(z^2/\ell^2) g_{\mu\nu}$, which we
identify with the metric degrees of freedom appearing in the SymTFT
construction, becomes gapped as we approach the boundary (i.e.~there
are no local dynamics associated with this field on the asymptotic
boundary). Nevertheless, the precise way in which bulk gravity
approaches a gapped system is interesting, and we argue that in the IR
it is described by the BF-theory above.
It would be very interesting to bring the holographic renormalisation
analysis closer to the BF/SymTFT language, and in particular to our
analysis of the conformal anomaly in section \ref{sec:conformal-anomaly}, but we will not attempt to do so in this paper. 

We are interested in understanding the dynamics governing the Weyl
rescaled metric $\sg_{\mu\nu}$
\[
  \sg_{\mu\nu} \df e^{-2\varphi} g_{\mu\nu}
\]
with $g_{\mu\nu}$ the original metric, and
$e^{\varphi} \df \ell/z$. The Weyl rescaled field
$\sg_{\mu\nu}$ is no longer divergent near the boundary, at the cost
of introducing an explicit $r$-dependence in the Einstein-Hilbert
action~\eqref{eq:Einstein-Hilbert} when expressed in terms of
$\sg_{\mu\nu}$. We have \cite{Wald:1984rg}:
\[
  \sqrt{g} R = e^{2\varphi}\sqrt{\sg}(\sR - 6e^{-\varphi}\Delta e^{\varphi})\, .
\]
In our notation, $\sqrt{g}$ and $R$ on the left hand side are computed in terms of
$g_{\mu\nu}$, and $\sqrt{\sg}$, $\sR$ and $\Delta$ on the right hand side are
computed in terms of $\sg_{\mu\nu}$. Here $\Delta$ is the Laplace-Beltrami operator
\[
    \Delta f \df \frac{1}{\sqrt{\sg}}\partial_\mu \left(\sqrt{\sg}\sg^{\mu\nu}\partial_\nu f\right)\, .
\]
Since $e^\varphi$ depends only on $z$, and in the Fefferman-Graham gauge the metric~\eqref{eq:Fefferman-Graham} is block diagonal in the $z$ components we have 
\[
  e^{-\varphi}\Delta e^{\varphi} = \frac{1}{\sqrt{\sg}}e^{-\varphi}
  \partial_z\left(\sqrt{\sg}\, \partial_z e^{\varphi}\right) = \frac{2}{z^2} - \tr(\tilde g_0^{-1} \tilde g_2) + \cO(z)\, ,
\]
where the omitted terms vanish as $z\to 0$. The action describing the dynamics of $\sg_{\mu\nu}$ is therefore
\[
  \label{eq:EH-conformal-g}
  S_{\varphi}[\sg_{\mu\nu}] \df
  S_{\text{EH}}[e^{2\varphi}\sg_{\mu\nu}] = -\int_{\mathrm{AdS}_4}
  dx^4\frac{\sqrt{\sg} \,\ell^2}{16\pi G_N z^2} \left[\sR-2\Lambda -
    6e^{-\varphi}\Delta e^{\varphi} \right]
\]

Let us momentarily ignore the last, divergent term inside the brackets
(we will come back to it soon). Ignoring this term, the resulting
action describes Einstein gravity for the rescaled metric $\sg$ with
an effective $G_N' \df G_N z^2/\ell^2$ that vanishes near the
boundary. So a reasonable guess, given that the symmetries of the
system naturally act (and arise) on asymptotic infinity \cite{Brown:1986nw,Coussaert:1995zp},
is that the SymTFT for spacetime symmetries is what remains of
Einstein gravity as we take $G_N'\to 0$.

This situation is analogous to what happens with continuous internal
symmetries:~the SymTFT can be understood as the zero (or infinite,
depending on the duality frame) coupling limit of Maxwell or
Yang-Mills theory \cite{Apruzzi:2024htg,Bonetti:2024cjk}.  

\paragraph{Relation to MacDowell-Mansouri Formulation.}
To understand this limit in the case of
gravity, we switch to a classically equivalent alternative formulation
of four-dimensional Palatini gravity:~the BF
reformulation 
\cite{Smolin:2003qu,Freidel:2005ak} of MacDowell-Mansouri gravity
\cite{MacDowell:1977jt}. This formulation is based on a gauge group
$G$ that depends on the cosmological constant, and the signature\footnote{In
this argument we will not be careful about the global form of $G$.}.
For instance, with $\Lambda<0$ and Euclidean signature,
$G=SO(4,1)$. Other possibilities are summarised in
table~\ref{table:BF-G}.

\begin{table}
  \centering
  \begin{tabular}{|c|cc|}\hline
    & Lorentzian & Euclidean\\
    \hline
    $\Lambda > 0$ & $SO(4,1)$ & $SO(5)$\\
    $\Lambda < 0$ & $SO(3,2)$ & $SO(4,1)$\\
    \hline
  \end{tabular}
  \caption{Choice of gauge group $G$ in the BF formulation of four
    dimensional gravity, depending on the signature and sign of the
    cosmological constant.}
  \label{table:BF-G}
\end{table}

Once we have chosen the gauge group $G$ adequately, we construct a
connection $A$ in $G$, with field strength $F$, and then write the
MacDowell-Mansouri action \cite{MacDowell:1977jt} 
\be
  \label{eq:MM-gravity}
  S_{\text{MM}} = \frac{3}{64\pi^2\Lambda G_N'} \int F_{IJ}\wedge F_{KL}
  \epsilon^{IJKL5}\, .
\ee
We are interested in the BF reformulation of this theory introduced
in \cite{Smolin:2003qu,Freidel:2005ak}
\[
  \label{eq:BF-gravity}
  S_{BF} = \frac{i}{2\pi}\int_{X^4} B^{IJ}\wedge F_{IJ} - \frac{1}{2}\int_{X^4} B_{IJ} \wedge
  B_{LM} \epsilon^{IJKLM} v_M\, ,
\]
where $B$ is a 2-form valued in the adjoint of $G$, $v$ is in the
vector representation of $G$, and $F$ is as
in~\eqref{eq:MM-gravity}. Indices in this expression are raised with
the natural bilinear form on $G$. If we take
\be 
v=(0,0,0,0,8\pi\Lambda G_N'/3)\, ,
\ee 
the resulting theory has the same local
dynamics as Einstein gravity (in the Palatini formalism) with coupling
constant $G_N'$ and cosmological constant $\Lambda$
\cite{Smolin:2003qu,Freidel:2005ak,Freidel:2006hv}.\footnote{It is very tempting to try to explain a non-zero value of $v$ by some sort of dynamic
  mechanism. See for example \cite{Wilczek:1998ea} for an early
  proposal in this direction.} 

For non-zero values of $v$, the $G$ gauge group is reduced to a
subgroup, but when $v=0$, which is the relevant value for our
asymptotic analysis, the full gauge symmetry $G$ is present. In this
latter case, it is not difficult to show (see \cite{Freidel:2012np}
for example) that the theory has no local dynamics, once we quotient
by gauge transformations.\footnote{See also \cite{Smolin:1992wj,Escalante:2009nxa,Escalante:2015oua} for previous work on the zero coupling limit of gravity. One important difference between these works and ours is that for us the zero coupling limit arises as an effective description near the boundary of AdS space.} As soon as $v\neq 0$, on the other hand, the
gauge group is reduced, and one has physical excitations in the
spectrum, corresponding to the physical polarisations of the
graviton.\footnote{This discontinuity in the gauge group makes the
  situation challenging to analyse in detail in the continuum
  \cite{Cattaneo:1997eh,Rovelli:2005qb,Cattaneo:2023cnt}, although it
  seems reasonable to hope that a lattice formulation, where gauge
  fixing is not necessary, would behave in a better way. Perhaps it
  might also be worth pointing out that this issue is not specific to
  gravity:~if one formulates Maxwell theory as a $B^2$ deformation of
  abelian BF-theory, as in \cite{Witten:1995gf} for example, there
  is an enhancement of the gauge group when the electric coupling $e$
  vanishes, which similarly has the effect of making the photon pure
  gauge, while the photon is physical for $e\neq 0$.} More in detail,
the four dimensional vierbein $e^a$ is identified with $\ell A^{a5}$,
and the spin connection $\omega^{ab}$ is given by $A^{ab}$.

In this way, the BF formulation of MacDowell-Mansouri gravity allows
us to derive the SymTFT for spacetime transformations for the cases
covered by our derivation ($d=3$ CFTs with a holographic dual):~it is
obtained by taking the $G_N'\to 0$ limit of gravity, which in
\eqref{eq:BF-gravity} corresponds to taking $v\to 0$.

There is a loose end we need to tie up. The final, divergent term
$6e^{-\varphi}\Delta e^{\varphi}$ in~\eqref{eq:EH-conformal-g} might
seem to invalidate everything we have said so far:~if included, it
effectively introduces an effective cosmological constant
$\Lambda' \df \Lambda + 12/z^2 + \ldots$, where the omitted terms do
not diverge as $z\to 0$. These omitted terms do not affect the
argument above, which in the $z\to 0$, $G_N'\to 0$ limit will still
lead to the same BF-theory even if the cosmological constant is
$x$-dependent, but the $12/z^2$ piece would result in a cosmological
constant which diverges close to the boundary (with the wrong sign, in
fact!). If we include this term, $v$ no longer vanishes as we go to
the boundary, but rather becomes a non-zero constant. The reason that
we encounter this divergent term is because we have not yet
regularised the Einstein-Hilbert action:~the divergent $\Lambda'$ near
the boundary also arises if we evaluate the
action~\eqref{eq:EH-conformal-g} with $\sg$ the (rescaled) AdS$_4$
metric. 
{This is
  a well studied phenomenon, which can be solved by adding boundary
  counterterms to the Einstein-Hilbert action.\footnote{The full
    analysis also involves the Gibbons-Hawking-York term
    \cite{York:1972sj,Gibbons:1976ue}, which we have ignored, and
    which suffers from similar divergences.} We refer the reader to
\cite{Henningson:1998gx,Balasubramanian:1999re,Emparan:1999pm,Kraus:1999di,deBoer:1999tgo,deHaro:2000vlm} for further details.}

\paragraph{Relation to Palatini Formulation.}
{Rather than going into the details of this procedure, let
  us point out an interesting alternative approach, which relies on
  some beautiful properties of the BF formulation
  in~\eqref{eq:BF-gravity}. We claimed above that this action, with
  $v$ chosen adequately, leads to the same local dynamics as the
  Palatini formulation of Einstein gravity {as summarized in section \ref{sec:Palatini}}. This is true, but
  \eqref{eq:BF-gravity} differs from the Palatini action by a term
  proportional to the Euler density
  \cite{MacDowell:1977jt,Smolin:2003qu,Freidel:2005ak},
  which implements the subtraction of the divergent $12/z^2$ for us
  \cite{Anastasiou:2020zwc}. In fact, there is a beautiful geometric
  way of understanding this subtraction of the divergence, using the
  Cartan geometry picture of MacDowell-Mansouri explained in
  \cite{Wise:2006sm,Wise:2009fu}. From this point of view, the
  curvature $F$ in~\eqref{eq:MM-gravity} encodes the deviation from
  the model AdS$_4$ geometry, so the MacDowell-Mansouri action
  necessarily vanishes in the AdS$_4$ vacuum, and is more generally
  automatically finite in asymptotically AdS spaces.}

\paragraph{Relation to Plebański Formulation.}
Finally, let us very briefly comment on an alternative way of
discussing gravity as a BF-theory, introduced by Plebański
\cite{Plebanski:1977zz}. (For more details on this, and alternative
formulations of gravity more generally, see
\cite{Celada:2016jdt,Krasnov_2020}.) Plebański also presents Palatini
gravity in a BF form, but the details are rather different to what
we have discussed so far, and the potential connection to the SymTFT
is much less clear to us:~the gauge group is the local Lorentz group
and the ``deformation'' from pure BF-theory is a Lagrange
multiplier, which does not seem to disappear in the $G_N'\to 0$ limit
discussed above.

\section{Conclusions and Outlook}
\label{sec:Yipee}

In this paper, we have started the exploration of the SymTFT for continuous spacetime symmetries in $d$ dimensions. Our proposal is that the SymTFT is given in terms of the BF-theory for the spacetime symmetry, e.g.~the conformal group, and in the case of odd bulk dimensions, it may include additional CS-couplings. We tested this proposal in various ways. Firstly, we constructed the topological defects of the SymTFT and showed that they give rise to the symmetry generators as expected. Secondly, we also checked that they reproduce the correct spacetime symmetry action on local operators. 
We also initiated the study of spontaneous symmetry breaking in this context, exploring the SymTFT configurations that break the conformal symmetry to a subgroup. 
In particular, we have to consider gapless BCs, which we call modified Neumann, that include the leading non-topological terms.  
This reproduced  the dilaton effective theories when breaking from the conformal to Poincar\'e group.

The SymTFT can in certain instances be understood as a topological limit of gravity. In $d=1$,  JT gravity in 2d is classically the same as the BF-theory for the conformal group $SL(2,\R)$. In higher dimensions, one has to consider various limits of gravity to recover the SymTFT. 

Future applications of this framework are numerous, and we list a few:~

\begin{enumerate}
\item {\bf Relation between SymTFT and Gravity in $d>3$.} We discussed the relation of the SymTFT to gravity. In particular, in the case of 4d gravity, we showed that the SymTFT, i.e.~BF-theory, can be thought of as a topological limit of gravity. This relied on the existence of a formulation of gravity in 4d in terms of MacDowell-Mansouri action. In higher dimensions, such a BF-formulation of gravity is not known. It would be very interesting therefore to be able to establish a similar relation between gravity and the SymTFT. 

Another obvious relation is to holography, which is closely connected, but again the precise relation in general needs to be further sharpened. For internal global symmetries, a holographic picture exists and connects directly with the SymTFT picture \cite{Apruzzi:2022rei, GarciaEtxebarria:2022vzq, Antinucci:2022vyk, Cvetic:2025kdn, Antinucci:2024bcm}. We have made some steps towards connecting the SymTFT to gravity in AdS spactime, in particular in low dimensions. It would be important to develop this relation in higher dimensions as well.

\item  {\bf Other Spacetime Symmetries.} 
We focused our application to the conformal symmetry and conformal symmetry breaking. 
Of course, there are endlessly many other interesting applications:~
an interesting avenue is to explore more exotic phases of spacetime symmetry breaking \cite{Nicolis:2015sra}. 
Another obvious application is to other spacetime symmetries, such as supersymmetry. 
Clearly, it is desirable to also find a formulation that combines internal and spacetime symmetries into one complete framework, allowing also non-trivial inter-dependences of these.

\item {\bf Applications of Non-Abelian BF-theory.} We have remarked before that our analysis can be carried out equally for compact internal symmetries -- in particular the analysis in section \ref{sec:SymTFTST} is equally applicable to internal symmetries, as exemplified in section \ref{sec:CompEx}. 
A systematic characterization of boundary conditions and SymTFT sandwich compactifications in order to describe symmetry breaking etc have thus far not been analyzed in the literature. Some examples can be found in \cite{Antinucci:2024bcm}. 
Our analysis should be a good starting point for further explorations. 

\item {\bf Mathematical Formulation of Continuous Symmetries.} For finite symmetries, the SymTFT and the braided  category of its topological defects is very well understood both for fusion categories and fusion 2-categories. It would be very interesting to develop a mathematical framework to incorporate continuous symmetries, internal and spacetime into this framework. Some mathematical results studying continuous symmetries and their mathematical properties have appeared in \cite{Marin-Salvador:2025stc}.

\end{enumerate}

\subsection*{Acknowledgments}

We thank Andrea Antinucci, Leron Borsten, Davide Cassani, Yichul Choi, Chris Couzens, Scott Collier, Lorenz Eberhardt, Andrea Grigoletto, Jonathan Heckman, Gary Horowitz, Adam Kmec, Albert Law, Javier Magán, Lionel Mason, Mark Mezei, Paul McFadden, Tomás Ortín, Lorenzo di Pietro, Pavel Putrov, Rajath Radhakrishnan, Brandon Rayhaun, Simon Ross, Romain Ruzziconi, Tin Sulejmanpasic, Zimo Sun and Alison Warman for discussions. 
FA, HTL, and SSN thank the KITP for hospitality, while some of this work was being carried out. 
The work of FA is partially supported by the University of Padua under the 2023 STARS Grants@Unipd programme (GENSYMSTR – Generalized Symmetries from Strings and Branes) and in part by the Italian MUR Departments of Excellence grant 2023-2027 ``Quantum Frontiers”. N.D. acknowledges the receipt of the joint grant from the Abdus Salam International Centre for Theoretical Physics
(ICTP), Trieste, Italy and INFN, sede centrale Frascati, Italy. 
The work of IGE is supported by the ``Global Categorical Symmetries'' Simons Collaboration grant (award number 888990) and the STFC ST/X000591/1 consolidated grant. 
HTL is supported by the U.S. Department of Energy,
Office of Science, Office of High Energy Physics of U.S. Department of Energy under grant Contract Number DE-SC0012567 (High Energy Theory research) and by the Packard Foundation award for
Quantum Black Holes from Quantum Computation and Holography. 
The work of SSN is supported by the UKRI Frontier Research Grant, underwriting the ERC Advanced Grant ``Generalized Symmetries in Quantum Field Theory and Quantum Gravity''.
This research was supported in part by grant NSF PHY-2309135 to the Kavli Institute for Theoretical Physics (KITP). 
The MIT preprint number of this paper is MIT-CTP/5921.

\appendix

\section{Finite Diffeomorphisms vs Finite Gauge Transformations}
\label{sec:App_finite_transform} 

Following up on the infinitesimal discussion in
section~\ref{sec:NA-BF-review}, in this appendix we want to match
gauge transformations and diffeomorphisms beyond leading order. We
focus on the action of gauge transformations on $A$, since this is the
case most relevant to section~\ref{sec:moving-points}. A finite
diffeomorphism generated by a vector field $\xi$ acts on $A$ as
\[
  A\to e^{t\cL_\xi} A
\]
with $\cL_\xi$ the Lie derivative associated to $\xi$, and $t\in\mathbb{R}$ a
constant that measures how far along the flow we move.\footnote{\label{fn:flows-vs-diffs}
  It is
  important to note at this point that our arguments in this section
  do not apply to the whole group of diffeomorphisms, only to those
  generated by exponential flows. Not all diffeomorphisms can be
  generated in this way, not even all those in a small neighbourhood
  of the identity, see for example
  \cite{Grabowski1984,Grabowski1988}.} Given that
$t\cL_\xi = \cL_{t\xi}$, in the following we will redefine $\xi$ so
that $t=1$. Our goal in this section is to find a gauge representation
of this diffeomorphism. That is, we want to find some $g$ such that for a flat connection $A$,
\[\label{eq:GaugeDiffMatch}
  e^{\cL_\xi}A = g^{-1} (A + d) g\, .
\]

Let us first consider the case that $A$ takes values in a commuting
algebra. This case is relevant, for example, if we are considering
diffeomorphisms acting on flat $\bR^d$, where the only components we
turn on are the vielbein, which (recall~\eqref{eq:A-expansion}) take
values along the $P_a$ translation components, which commute among
themselves. We take the ansatz
\[
  g=e^{\iota_\xi \alpha},\quad\text{ with }\
  \alpha = \sum_{k=0}^\infty \alpha_k\, ,
\]
where $\alpha_k$ is of degree $k$ in $\xi$. Furthermore, given that we
are in the commuting $A$ case, we expect that $\alpha$ will belong to
the same commuting subalgebra, so that
$g^{-1}(A+d)g=A + d\iota_\xi\alpha$. We claim that the following
choice for $\iota_\xi\alpha$ represents the finite diffeomorphism
$\exp(\cL_\xi)$:
\[
  \iota_\xi\alpha = \int_0^1 ds \, e^{s\cL_\xi} \iota_\xi A\, ,
\]
or more explicitly
\[
  \label{eq:alpha_k transforms}
  \alpha_k = \frac{1}{(k+1)!}\cL_\xi^k A\, .
\]
To see this, use Cartan's magic formula
$\cL_\xi=\iota_\xi d+d\iota_\xi$ and the equation of motion $dA=0$:
\[
  d\iota_\xi \alpha = \int_0^1 ds\, e^{s\cL_\xi}d\iota_\xi A = \int_0^1 ds\, e^{s\cL_\xi}\cL_{\xi} A = (e^{\cL_\xi} - 1)A\, .
\]

Unfortunately we don't know of a similar closed form for the general
non-abelian case, but it seems possible (if somewhat labour intensive)
to find solutions for $g$ order by order in $\xi$. The result is
relatively concise up to $\mathcal{O}(\xi^4)$, so we record here for the benefit of the reader:
\[
  g = \exp\left(\sum_{k=0}^3\iota_\xi\alpha_k + \frac{1}{6}[\iota_\xi\alpha_0,\iota_\xi\alpha_1] + \frac{1}{4}[\iota_\xi\alpha_0,\iota_\xi\alpha_2]+\cO(\xi^5)\right)
\]
with $\alpha_k$ as in~\eqref{eq:alpha_k transforms}. We have also verified that a solution exists to fifth order in $\xi$, but the resulting expression is more involved so we omit it.

Note that what we have shown in this appendix is that the action of every diffeomorphism on $A$ (subject to the subtlety in footnote~\ref{fn:flows-vs-diffs}) can be rewritten as a gauge transformation. But the opposite does not hold:~not every gauge transformation can be obtained from a diffeomorphism. Consider, as a simple example, a starting configuration with (abelian) $A=0$, and gauge transform it to $A=d\alpha\neq 0$. The action of diffeomorphisms is linear in $A$, so we cannot reproduce the effect of the gauge transformation from a diffeomorphism. In the main text we will encounter subtleties related to this fact when we try to give a ``classical" interpretation, in terms of diffeomorphisms, of some of the symmetry operators arising from the bulk.

\section{Conventions for Chern-Simons Terms}
\label{sec:CS_conventions}

We collect here our conventions and useful formulas concerning Chern-Simons (CS) actions. For more details, see the textbooks \cite{Nakahara:2003nw, Bertlmann:1996xk}.

\paragraph{Chern-Simons Actions.}
~\\
In $d+1=2n+1$ dimensions, the CS actions can be defined starting from a totally symmetric, non-degenerate, adjoint-invariant $(n+1)$-linear form
\begin{align}\label{eq:CS_innerproduct}
\langle \, ... \, \rangle :~\quad \text{Sym}^{n+1}(\mathfrak{g}) \rightarrow \mathbb{C} ,\quad \langle g^{-1} X_1 g, ... \,, g^{-1} X_{n+1} g \rangle = \langle X_1, ...\,, X_{n+1} \rangle, \quad X_i \in \mathfrak{g} \, .
\end{align} 
Adjoint invariance can also be written for infinitesimal transformations $g \sim 1 + \epsilon^a T_a$ as
\begin{equation}
\sum_{i=1}^{n+1} \langle X_1 ,\cdots  [T^a, X_i] , \cdots X_{n+1} \rangle  = 0  \, .
\end{equation}
An appropriate graded version of the equation above holds when entries are replaced by $\mathfrak{g}$-valued differential forms. 
For compact groups, a natural candidate for such a multilinear form is given by the symmetric trace 
\ie
\langle X_1, \cdots\, ,X_{n+1} \rangle=\frac{1}{(n+1)!}\sum_{\sigma\in S_{n+1}}\text{Tr}[X_{\sigma(1)} \cdots X_{\sigma(n+1)}]\,,
\fe
which however can be trivial in cases.
In general, there is one such multilinear form for each $(n+1)$-Casimir of $\mathfrak{g}$.
With this multilinear form, we can define the symmetric polynomial $P_{n+1}(F) \coloneqq \langle F^{n+1}\rangle=\langle F,...\,, F \rangle$ \footnote{We sometimes simplify the formula by writing powers in the multilinear form, these powers should be distributed appropriately into different entries of the multilinear form.} and the CS-functional is defined as $P_{n+1}(F)=\dd\text{CS}_{2n+1}(A)$, which can be solved by the following integral
\ie
\text{CS}_{2n+1}(A) &= (n+1)\int_{0}^1 \dd t \, \langle A ,\left[ t F + t(t-1) A^2 \right]^{n} \rangle \, .
\fe
Our convention for the CS action is as
\begin{align}
S_{\rm CS}^{(d+1=2n+1)} &= \frac{i k}{(2\pi)^n(n+1)!} \int_{M_{2n+1}} \text{CS}_{2n+1}(A)\,.
\end{align}
It is normalized such that for compact groups with the $(n+1)$-linear form given by the symmetric trace, the CS level $k$ is integral quantized.
Explicitly, in $d+1=3,5$, one has 
\begin{align}
S_{\rm CS}^{(3)} &= \frac{i k}{2(2\pi)} \int_{M_3} \left[\left\langle A, F\right\rangle - \frac{1}{3} \left\langle A, A^2 \right\rangle\right]\,, \nonumber\\
S_{\rm CS}^{(5)} &= \frac{i k}{6(2\pi)^2} \int_{M_5} \left[\left\langle A, F,F\right\rangle - \frac{1}{2} \left\langle A, A^2, F\right\rangle + \frac{1}{10} \left\langle A, A^2, A^2\right\rangle\right]\,.
\end{align}
The general variation of the CS-functional is
\begin{equation}
\delta \text{CS}_{2n+1} = (n+1) \langle \delta A, F^n \rangle + \dd \left\{  (n+1) n \int_0^1 \dd t \left\langle \delta A ,t A, [t F + t(t-1) A^2]^{n-1} \right\rangle \right\} \, .
\end{equation}
From which follows in $d+1 =3,5$:
\begin{align}
\delta S_{\rm CS}^{(3)} &= \frac{i k}{(2\pi)} \int_{M_3} \left\langle \delta A, F \right\rangle + \frac{ik}{2(2\pi)} \int_{\partial M_2} \left\langle \delta A ,A\right\rangle \nonumber\,,\\
\delta S_{\rm CS}^{(5)} &= \frac{i k}{2(2\pi)^2} \int_{M_5} \langle \delta A, F,F \rangle + \frac{i k}{3(2\pi)^2} \int_{\partial M_5} \left\langle \delta A, A, F - \frac{1}{4} A^2  \right\rangle  \,.
\end{align}
Finally, the general finite gauge transformation of the CS-functional reads 
\begin{equation}
\begin{aligned}
\Delta^{(g)} \text{CS}_{2n+1} &=\text{CS}_{2n+1}(A^g)-\text{CS}_{2n+1}(A)= \text{WZW}_{2n+1}(g) + \dd \alpha_{2n}\,,
\\
\text{WZW}_{2n+1}(g)&=(-1)^n \frac{n!(n+1)!}{(2n+1)!} \left\langle (\dd g g^{-1})^{2n+1}\right\rangle\,,
\end{aligned}
\end{equation}
where $\text{WZW}_{2n+1}(g)$ is the Wess-Zumino-Witten (WZW) functional whose integral on a close manifold is integer multiple of $(2\pi)^{n+1}(n+1)!$ for compact groups with the $(n+1)$-linear form given by the symmetric trace, and $\alpha_{2n}$ is a $2n$-form built out of $A,F,\dd g g^{-1}$. Rather than provide a general expression for it, we specify the gauge variation of the CS-functional for $d+1=3,5$:
\begin{align}
\Delta^{(g)} S_{\rm CS}^{(3)} &= -\frac{i k}{2(2\pi)}  \int_{M_3} \frac{1}{3}\left\langle (\dd g g^{-1})^{3} \right\rangle - \frac{ik}{2(2\pi)} \int_{\partial M_2} \left\langle \dd g g^{-1}, A \right\rangle\,, \nonumber\\
\Delta^{(g)} S_{\rm CS}^{(5)} &= \frac{ik}{6(2\pi)^2} \int_{M_5} \frac{1}{10}\left\langle (\dd g g^{-1})^5\right\rangle \\
&- \frac{ik}{6(2\pi)^2} \int_{\partial M_5} \left\langle \dd g g^{-1} , A, \left[ F - \frac{1}{2} A^2 - \frac{1}{2}(\dd g g^{-1})^2 - \frac{1}{4} (\dd g g^{-1} A + A \dd g g^{-1}) \right] \right\rangle\,.\nonumber
\end{align}

\paragraph{Transgression forms.}
~\\
When studying the BF+CS system with boundary conditions that explicitly break    the $G$ gauge symmetry, one introduces a set of St\"uckelberg fields $U :~\partial M_{d+1} \rightarrow G$ to restore it. The (topological) action of these fields, however, must also absorb boundary gauge variations coming from the bulk CS-functional. Such actions can be constructed starting from the transgression form ${T}_{2n+1}$ of the Chern-Simons functional $\CS_{2n+1}$. See \cite{Nakahara:2003nw, Bertlmann:1996xk} for  standard references and \cite{Mora:2006ka} for an application in building gauge-invariant actions. The transgression $T_{2n+1}$ is defined in term of two connections $A_0, A_1$ as follows
\begin{align}
T_{2n+1}(A_1,A_0) &= (n+1) \int_0^1 \dd t \left\langle A_1 - A_0, F_t^n \right\rangle  \,,\nonumber\\
F_t &= t F_1 + (1-t) F_0 -t (1-t)(A_1-A_0)^2 .
\end{align}
From this definition, it is evident that $T_{2n+1}$ is an exactly gauge-invariant functional under the gauge transformation $A_1 \mapsto A_1^{(g)}, \, A_0 \mapsto A_0^{(g)}$. The transgression functional is related to the Chern-Simons functional as follows:
\begin{equation}\label{Tterm}
T_{2n+1}(A_1, A_0) = \text{CS}_{2n+1}(A_1 ) - \text{CS}_{2n+1}(A_0) - \dd Q_{2n}(A_1,A_0) \, ,
\end{equation}
 where the $2n$-form $Q_{2n}$ can be computed systematically. For example:
\begin{align}\label{TransgressionTerms}
Q_{2} &= - \langle A_0, A_1 \rangle \nonumber\\
Q_4 &= - \left\langle A_0, A_1, \left[ F_0 + F_1 - \frac{1}{2} A_0^2 - \frac{1}{2} A_1^2 + \frac{1}{4} ( A_0 A_1 + A_1 A_0 ) \right] \right\rangle
\end{align}

\begin{figure}
$$
\begin{tikzpicture}[scale=1.4]
  \def\R{2.2}       
  \def\tilt{75}     
  \pgfmathsetmacro{\ry}{\R*cos(\tilt)} 

\begin{scope}[rotate=90,yscale=1.4,xscale=0.8]
  \path[fill=blue!35, opacity= 0.8]
    (-\R,0)
      arc[start angle=180, end angle=0, x radius=\R, y radius=\ry] 
      arc[start angle=0, end angle=-180, radius=\R];               
  \path[fill=red!45,opacity=0.8]
    (-\R,0)
      arc[start angle=180, end angle=360, x radius=\R, y radius=\ry] 
      arc[start angle=0, end angle=180, radius=\R];                  
  \draw[line width=0.6pt] (0,0) circle (\R);
  \draw[line width=0.6pt]
    (-\R,0) arc[start angle=180, end angle=360, x radius=\R, y radius=\ry];
  \draw[dashed, line width=0.6pt]
    (-\R,0) arc[start angle=180, end angle=0, x radius=\R, y radius=\ry];
\end{scope}
\node at (-2.3,-2) {$M_{2n+1}$};
\node at (2.3,-2) {$\overline{M}_{2n+1}$};
\node at (0,2.5) {$\partial M_{2n+1} = -\partial\overline{M}_{2n+1}$};
\node at (-2.2,0) {$\text{CS}_{2n+1}(A_1)$};
\node at (2.2,0) {$\text{CS}_{2n+1}(A_0)$};
\node at (0,0) {$Q_{2n}(A_1,A_0)$};
\end{tikzpicture}
$$
\caption{The transgression functional describes Chern-Simons functional for two $G$-connections $A_0, A_1$ interacting on a topological interface via the $Q_{2n}$ form.
\label{fig:TransgressionAction}}
\end{figure}
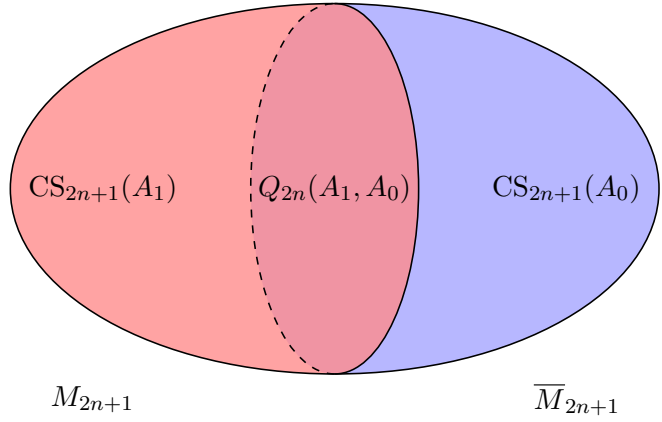

The most natural way to build a gauge-invariant action is to take $A_1, A_0$ as $G$-connections on different manifolds $M_{2n+1}, \overline{M}_{2n+1}$ with a common boundary $\partial M_{2n+1}=-\partial \overline M_{2n+1}$:
\begin{equation}
S_{T} = \int_{M_{2n+1}} \text{CS}_{2n+1}(A_1) + \int_{\overline{M}_{2n+1}} \text{CS}_{2n+1}(A_0) - \int_{\partial M_{2n+1}=-\partial \overline M_{2n+1}} Q_{2n}(A_1,A_0) \, .
\end{equation}
This action describes two $G$-connections with CS action interacting at a topological interface defined by $Q_{2n}$, see figure \ref{fig:TransgressionAction}. Notice that if $\overline{M}_{2n+1} \equiv - M_{2n+1}$ then $S_T$ reduces to $T_{2n+1}$ integrated over $M_{2n+1}$, and it is exactly gauge invariant. Otherwise, for gauge transformations acting on the entire system and smoothly gluing at the interface, one finds
\begin{equation}
\Delta^{(g)} S_T = \int_{M_{2n+1} \sqcup \overline{M}_{2n+1}} \text{WZW}_{2n+1}[g] \, .
\end{equation}
Depending on whether the WZW term for $G$ is trivial or not, gauge invariance is still retained by properly quantizing the CS level.  

\paragraph{From interfaces to gapped boundaries.}
~\\
Consider the setup in figure \ref{fig:TransgressionAction} where on the left $A_1 = A$ is a dynamical $G$ connection including a BF term, while on the right $A_0 =\cA$ is a classical background. Moreover, we include an extra term on the topological surface as follows:
\begin{equation}
S_{\rm Dir} = \frac{i}{2\pi} \int_{\partial M_{2n+1}} \langle A- \cA, B_{2n-1} \rangle\, .
\end{equation}
The Dirichlet boundary condition for the BF+CS system is obtained by neglecting the classical functional $\text{CS}_{2n+1}(\cA)$. The terms coming from the $Q_{2n}$ forms are generically allowed improvement terms which vanish on-shell. The resulting system obtained this way has an anomaly
\begin{equation}
\mathcal{Z}_{\rm Dir}[\mathcal{A}^{(g)}] \neq \mathcal{Z}_{\rm Dir}[\mathcal{A}]\,.
\end{equation}
If however we restrict the Dirichlet fixed configuration on a subgroup $\mathcal{A} \in \Omega^1(\partial M_{2n+1}, \fh)$ such that the multilinear product defining the CS-functional vanish when restricted on $\fh$, then we have invariance under $H$-gauge transformations
\begin{equation}
\mathcal{Z}_{\rm Dir}[\mathcal{A}^{(h)}] = \mathcal{Z}_{\rm Dir}[\mathcal{A}] \, .
\end{equation}
The form $Q_{2n}$ represent the correct improvement action which restores $H$-gauge transformations off-shell. This improvement is necessary to define partial Neumann by $H$-gauging. 

\paragraph{St\"uckelberg actions.}
~\\
In this setup, St\"uckelberg fields can be introduced by choosing $A_0 = U \dd U^{-1}$ and $A_1 = A$ (the actual bulk connection). We then define 
\begin{equation}\label{eq:GammaAction_gendim}
\Gamma_{2n+1}(U,A) \equiv \text{CS}_{2n+1}(U \dd U^{-1}) + \dd Q_{2n}(A, U \dd U^{-1} ) \, .
\end{equation}
Thus, $\Gamma_{2n+1}$ is the correct functional of $U,A$ which cancels the boundary gauge variation of the bulk CS functional, where gauge transformations act as $U \mapsto g^{-1} U$. In our work, $U$ is a St\"uckelberg field associated to the boundary $\partial M_{2n+1}$. Generically, this boundary is made of many disconnected pieces $\partial_i M_{2n+1}$ with $i=1,..., b$, and one introduces a different St\"uckelberg $U_i$ for each one. For each, it is then natural to build an action from \eqref{eq:GammaAction_gendim} by integrating over an auxiliary manifold $X^i_{2n+1}$ with $\partial X^i_{2n+1} = \partial_i M_{2n+1}$. The total action in this case reads
\begin{align}
S_{\rm tot} &= S^{(2n+1)}_{\rm CS} - \sum_{i=1}^b\frac{i k}{(2\pi)^n(n+1)!} \int_{X^i_{2n+1}} \Gamma_{2n+1}(U_i,A) \,.
\end{align}
which is gauge invariant upon appropriate quantization of the Chern-Simons level $k$ as $M_{2n+1} \sqcup (-X_{2n+1}^1) \sqcup ... \sqcup (-X_{2n+1}^b)$ is compact by construction. For the specific cases $n=1,2$ we focus on, one gets explicitely
\be\ba
\Gamma_{3}(U,A) =\,&  - \frac{1}{3}  \left\langle (U\dd U^{-1} )^3 \right\rangle - \dd \left\langle U \dd U^{-1} , A \right \rangle\cr 
\Gamma_5(U,A) =\,& \frac{1}{10}   \left\langle (U\dd U^{-1} )^5 \right\rangle
\\
-\,& \dd \left\langle U \dd U^{-1}, A, \left[ F - \frac{1}{2} A^2 - \frac{1}{2} (U\dd U^{-1} )^2 + \frac{1}{4}  (U \dd U^{-1} A+   A U \dd U^{-1} )    \right]\right\rangle \, .
\ea\ee
The action $\Gamma_{2n+1}$ can be equivalently represented as follows
\begin{equation}
\Gamma_{2n+1}(U,A) =  \text{CS}_{2n+1}(A) -\text{CS}_{2n+1}(A^{(U)}) \, .
\end{equation}
From this expression, it is evident that under group inversion and group multiplication are realized as
\begin{align}\label{eq:Gamma_properties}
\Gamma_{2n+1}(U^{-1}, A) &= - \Gamma_{2n+1}(U, A^{(U^{-1})} ) \nonumber\, , \nonumber\\
\Gamma_{2n+1}(U_1 U_2, A ) &= \Gamma(U_1, A) + \Gamma(U_2, A^{(U_1)} ) \, .
\end{align}

\section{Conventions for the Conformal Algebra}
\label{sec:App_algebras}

\subsection{Generators and Algebra}

The $\mathfrak{so}(d+1,1)$
conformal algebra of $\mathbb{R}^d$ (and similarly for the Lorentzian version) 
is:\footnote{The conformal generator $D$ in \cite{Fradkin:1985am} is opposite to the one we choose. Our notation follows \cite{osborn2019lectures} up to $L_{ab} \rightarrow - L_{ab}$.}
\begin{align}
[D, P_a] &= P_a \nonumber\\
[D, K_a] &= - K_a \nonumber\\
[P_a,P_b] &= 0 \nonumber\\
[K_a,K_b] &= 0 \nonumber\\
[K_a, P_b] &= 2 (\eta_{a b} D - L_{a b} )\nonumber\\
[L_{ab}, D] &= 0 \nonumber\\
[L_{a b}, K_e] &= - (\eta_{a e} K_b - \eta_{b e} K_a )\nonumber\\
[L_{a b}, P_e] &=  -(\eta_{a e} P_b - \eta_{b e} P_a )\nonumber\\
[L_{a b}, L_{ef}] &= - (\eta_{ae} L_{bf} - \eta_{be} L_{af} - \eta_{af} L_{be} + \eta_{bf} L_{ae})\,,
\end{align}
where $a,b,e,f = 1 , ... , d$ and $L_{ab}=-L_{ba}$. Here, $\eta_{ab}$ denotes the flat metric in Euclidean signature (or Lorentzian signature). 
The algebra is consistent with the hermiticity condition $D^\dagger = D,\, P_a^\dagger = K_a,\, L_{ab}^\dagger = - L_{ab}$. 
The conformal generators are related to the usual $\mathfrak{so}(d+1,1)$ generators $M_{AB}$ with $A= 1,...,d,d+1,d+2$ as follows
\begin{equation}
M_{ab} = - L_{ab}, \quad M_{a,d+1} = - \frac{P_a+K_a}{2} , \quad M_{a,d+2} = \frac{P_a - K_a}{2}  , \quad M_{d+1,d+2} = - D \, .
\end{equation}
In their commutator appears the flat metric $\eta_{AB}$ with signature $(d+1,1)$. The quadratic casimir is 
\begin{equation}
C_2 = \frac{1}{2} \tr M^2 = -\frac{1}{2} M_{AB} M^{AB} = D(D-d) + \frac{1}{2} \tr L^2 - \eta^{ab} P_a K_b \, .
\end{equation}
The non-degenerate Killing form that follows from the quadratic casimir is (up to normalization):
\begin{equation} \label{eq:kappatr}
\langle X,Y \rangle_{\kappa} = -2 \tr(XY), \quad \langle M_{AB}, M_{CD} \rangle_{\kappa} = \eta_{AC} \eta_{BD} - \eta_{AD} \eta_{BC} \,.
\end{equation}
In terms of conformal generators, this is
\begin{align}
\langle D, D \rangle &= -1 & \langle D, P_\mu \rangle &= 0 &  \langle D, K_\mu \rangle &= 0 \nonumber\\
\langle P_\mu, P_\nu \rangle &= 0 & \langle K_\mu, K_\nu \rangle &=0 & \langle K_\mu, P_\nu \rangle &= 2 \eta_{\mu\nu} \nonumber\\
\langle L_{\mu\nu} , P_\rho \rangle &= 0 & \langle L_{\mu\nu}, K_\rho \rangle &= 0 & \langle L_{\mu\nu} ,L_{\rho \sigma} \rangle &=  \eta_{\mu\rho} \eta_{\nu\sigma} - \eta_{\mu\sigma} \eta_{\nu\rho}\,.
\label{eq:Killing_form_conf}
\end{align}
,%
For $d=2$ there is another adjoint-invariant symmetric bilinear product given by
\begin{equation} \label{eq:epsilontr}
\langle M_{AB}, M_{CD} \rangle_{\epsilon} = \epsilon_{ABCD} \, .
\end{equation}
We can write this in term of conformal generators in the basis \eqref{eq:m_split}, where $T^+_a =-M_{a,d+1},\, T^- = M_{a,d+2}$:
\begin{align}
\langle D, D \rangle &= 0 & \langle D, T_a^{\pm} \rangle &= 0 &  \langle T_a^+, T_b^- \rangle &= \epsilon_{ab} \nonumber\\
\langle L_{ab} , T_c^{\pm} \rangle &= 0 & \langle L_{ab}, D \rangle &=  \epsilon_{ab} & \langle L_{ab} ,L_{cd} \rangle &= 0\,.
\end{align}
In $d=4$, is a tri-linear adjoint-invariant product on the algebra defined as
\begin{equation} \label{eq:epsilontr}
\langle M_{AB}, M_{CD}, M_{EF} \rangle_{\epsilon} = \epsilon_{ABCDEF} \, .
\end{equation}
In terms of conformal generators, the only non-zero entries of this product are
\begin{align}
\langle D, L_{ab}, L_{ab} \rangle_{\epsilon} = - \epsilon_{abcd} , \quad \langle L_{ab} , T_c^+ , T_d^- \rangle = -\epsilon_{abcd} \, .
\end{align}

\subsection{Maurer-Cartan form for $SO(d+1,1)$}

We can compute the Maurer-Cartan form for the conformal group using the split $\mathfrak{so}(d+1,1) = \mathfrak{so}(d) \oplus \text{span}\left\{ T_a^+, T_b^-, D\right\}$ introduced in \eqref{eq:m_split}. $SO(d)$-index contraction is left implicit so that, for example, $x T^- \equiv x^a T_a^-$, $x L y = x^a L_{ab} y^b$, $\omega L = \omega^{ab} L_{ab}$. Furthermore, it is convenient to introduce a two-component notation as follows
\begin{equation}
\Pi = \begin{pmatrix} \pi^+ \\ \pi^ - \end{pmatrix} , \quad T = \begin{pmatrix}
 T^+ \\ T^-   
\end{pmatrix}, \quad \Pi^\dagger T \equiv \pi_- T^- + \pi_+ T^+ \, .
\end{equation}
The conformal algebra commutators can be compactly written as
\begin{align}
[D, X^\dagger T] &=  X^\dagger \sigma_1 T ,  & \nonumber\\  [ X^\dagger T , T^\dagger Y ] &=
X^\dagger (i \sigma_2 D) Y  - X^\dagger (\sigma_3 L) Y  & \nonumber\\
[\omega L, X^\dagger T] &= 2 \omega^{ab} X^\dagger_b  T_a, & \omega^{ab} = - \omega^{ba} \nonumber\\
[\omega L, \omega L] &= 4 (\omega L \omega) & (\omega L \omega) \equiv \omega^{ab} L_{bc} \omega^{ca} 
\end{align}
where $X_a,Y_b$ are auxiliary two-component vectors, $\omega^{ab}$ an auxiliary antisymmetric tensor and $\sigma_i$ are Pauli matrices acting on two-component vectors. We choose the following parametrization for the generic element of $SO(d+1,1)/SO(d)$:
\begin{equation}\label{eq:G/H_groupElement}
U = e^{- \Pi_1^\dagger T} e^{-\Pi_2^\dagger \sigma_3 T} e^{-\sigma D} ,\quad \Pi_i = \begin{pmatrix}
\pi_i \\ \pi_i
\end{pmatrix} \, .
\end{equation}
Notice that $\pi_1, \pi_2$ are Goldstone modes for broken translation and special conformal transformation, respectively.
The corresponding Maurer-Cartan form can be computed using the algebra structure and the formula $e^{x} y e^{-x} = e^{\text{ad}_x} y$ for $x,y \in \mathfrak{g}$ where $\text{ad}_x = [ x, \,\, ]$. The result is
\begin{align}\label{eq:MCformConf}
U^{-1} \dd U &= - \left[ \dd \sigma + (\dd \Pi_1^\dagger \sigma_1 \Pi_2 ) \right] D + (\Pi_2^\dagger \dd \Pi_1)^{[ab]} L_{ab} \nonumber\\
& - \left[ \dd \Pi_{1,b}^{\dagger} + \dd \Pi_{2,b}^{\dagger} \sigma_3 - \frac{1}{2} ( \Pi_2^\dagger \sigma_1 \dd \Pi_1) \Pi_{2,b}^{\dagger} (i \sigma_2) + \frac{1}{2} (\Pi_2^\dagger \dd \Pi_1)^{[ab]} \Pi_{2,a}^{\dagger} \sigma_3 \right] M_\sigma T_b \, ,
\end{align}
where we defined the $SO(1,1)$ matrix
\begin{align}
M_\sigma = \begin{pmatrix} \cosh \sigma & \sinh \sigma \\ \sinh \sigma & \cosh \sigma \, .
\end{pmatrix} \, .
\end{align}
Components of the Maurer-Cartan form \eqref{eq:MCformConf} along broken generators $\{ T_a^\pm, D \}$ are identified with Goldstone Boson covariant derivatives, while components along unbroken generators are identified with an $SO(d)$-connection:
\begin{equation}
U^{-1} \dd U = - (\cD \sigma) D - (\cD \Pi^a)^\dagger T_a - H^{ab} L_{ab}
\end{equation}
From this expression we can recover the corresponding one for the coset $D \ltimes ISO(d)/SO(d)$ by setting $\Pi_2 = 0$. One gets the expected simple result
\begin{equation}
U^{-1} \dd U |_{\Pi_2 = 0} = - e^{\sigma} \dd \pi_1^a P_a - \dd \sigma D \, .
\end{equation}

\paragraph{Addition of backgrounds.}
~\\
The Maurer-Cartan form can be coupled to a background for the global symmetry $U \mapsto g^{-1} U$ as follows
\begin{equation}
U^{-1} \dd U \mapsto   U^{-1} \dd_{\cA} U \equiv U^{-1} (\dd + \cA ) U \, .
\end{equation}
An arbitrary background can be written as follows
\begin{equation}
\cA = \bar{E}^\dagger T + \frac{1}{2} (\bar{\omega} L) + \bar{b} D \, .
\end{equation}
Parameters of the background are constrained by the flatness condition, which in two-component notation reads
\begin{align}
\cF = 0 =& ( \dd \bar{E}^\dagger +  \bar{E}^{\dagger} \bar{\omega}  + \bar{b} \bar{E}^\dagger \sigma_1 ) T \nonumber\\
&+ \left( \dd \bar{b} + \frac{1}{2} \bar{E}^\dagger (i \sigma_2) \bar{E} \right) D \nonumber\\
&+ \frac{1}{2} \left( \dd \bar{\omega}^{ab} + \bar{\omega}^{a}_{\,\, c}\, \bar{\omega}^{cb} - \bar{E}^{\dagger, a} \sigma_3 \bar{E}^b \right) L_{ab} \, .
\end{align}
One can already identify three possible solution to this equation corresponding to known geometries:
\begin{align}
\mathbb{R}^d:~&\quad \bar{E}^\dagger = ( \bar{e}^a, \bar{e}^a ), \quad \dd \bar{e}^a = \bar{b} = \bar{\omega} = 0  \nonumber\\
\text{$EdS_d \equiv S^d$}:~&\quad \bar{E}^\dagger= (\bar{e}^a, 0 ), \quad   \bar{b} = 0, \quad \dd \bar{e}^a + \bar{\omega}^a_{\,\,\, b} \wedge \bar{e}^b = 0, \quad \bar{R}^{ab} = \bar{e}^a \wedge \bar{e}^b \nonumber\\ 
\text{$EAdS_d \equiv \mathbb{H}_d$}:~&\quad \bar{E}^\dagger= (0, \bar{e}^a ), \quad   \bar{b} = 0, \quad \dd \bar{e}^a + \bar{\omega}^a_{\,\,\, b} \wedge \bar{e}^b = 0, \quad \bar{R}^{ab} = -\bar{e}^a \wedge \bar{e}^b
\end{align}
For any of these backgrounds, the corresponding Maurer-Cartan form can be computed similarly as in the previous section.

\section{Properties of Hodge Duals without Introducing a Metric }
\label{app:Hodge}

Consider the $d$-dimensional boundary $\Sigma_d$ of $M_{d+1}$ with some background metric $g = \eta_{ab} e^a \otimes e^b$ where $e^a \equiv e^{a}_{\,\,\, \mu} \dd x^\mu$ are the vielbein in some local coordinate basis. The hodge dual of a $\mathfrak{g}$-valued one-form $\omega \in \Omega^1(\Sigma_d, \mathfrak{g})$ can be written as
\begin{align}\label{eq:hodgestar}
\ast \omega &= \frac{1}{(d-1)!} T_i (\omega_\mu^i e^{\mu}_{\,\,\, a_1}) \varepsilon^{a_1}_{\,\,\, a_2 ... a_{d}} e^{a_2} \wedge ... \wedge e^{a_{d}} \,,
\end{align}
where both $\mu, a_i = 1, ..., d$. The vielbein at each point $p\in \Sigma_d$ is a map $e_p :~T_p \Sigma_d \rightarrow \mathbb{R}^d$ which realizes the non-canonical isomorphism $T_p \Sigma_d \cong \mathbb{R}^d$ at each point $p$. This latter requirement implies the existence of an inverse map, 
in components, $e^{a}_{\,\,\, \mu}$. Both $T_p \Sigma_d$ and $\mathbb{R}_d$ are equipped with inner products $g_{\mu\nu}, \eta_{ab}$ respectively, mapped into each other in the sense that $\eta = e^* g$ and allow for mixed-index objects like $e_{a\mu}, e^{\mu a}$. 
In \eqref{Hodnew} we defined an operation $\Hod$ which is akin to the Hodge star that does not explicitly require a metric, but only the $P$ components of $A$ in the decomposition \eqref{eq:A-expansion}. These components define a linear map $T_p \Sigma_d \rightarrow \mathfrak{p} \cong \mathbb{R}^d$, that in local coordinates is just a matrix $e^a_{\,\,\, \mu}$. In addition we can focus on forms for which these maps are invertible.

An important property to study is what happens if we apply the Hod operation twice. In particular we get,
\begin{equation}
\begin{aligned}
  \text{Hod}(A, \text{Hod}(A,\omega_p) ) &=\\
  &\frac{1}{p!(d-p)!} T_i (\omega^i_{a_1, \ldots a_p}\eta^{b_1 a_1}\cdots \eta^{b_p a_p}) \frac{1}{p!} \epsilon_{a_1,\ldots a_d}  \epsilon^{a_{p+1}\ldots a_{d}}{}_{c_1\ldots c_p}\, e^{a_{c_1}} \wedge ... \wedge e^{c_p}\, .
  \end{aligned}
\end{equation}
We can use the following property of Levi-Civita tensors,
\begin{equation}
  \epsilon_{a_1,\ldots a_d}  \epsilon^{a_{p+1}\ldots a_{d}}{}_{c_1 \ldots c_p}=(-)^{p(d-p)} (d-p)! p! \, \eta_{a_1 [c_1} \ldots   \eta_{a_p |c_p]}
\end{equation}
where the sign factor $\sigma = (-)^{p(d-p)}$ comes from swapping $d-p$ indices $p$ times on the Levi-Civita tensor $\epsilon^{a_{p+1}\ldots a_{d}}{}_{c_1 \ldots c_p}$. Then we get,
\begin{equation}\label{HODHOD}
    \Hod (A, \Hod (A, \omega_p)) = \sigma  \omega_p\, , \qquad \sigma =  (-)^{p(d-p)} \,.
\end{equation}
This operation is completely invariant under gauge transformations $A \rightarrow A^{g}$ in 
\begin{equation}
\mathfrak{so}(d) \ltimes \text{span}\{ K_a \} \cong \mathfrak{iso}(d)\, .
\end{equation}

Finally we can also define the same operator for a general $p$-form $\omega$ that is not a element of $\mathfrak g$, that is
\begin{equation}\label{HodnewnoT}
\begin{aligned}
\text{Hod}(A,\omega_p) 
&\equiv \frac{1}{(d-p)!}  \big(\omega^i_{b_1 \ldots b_p}  
\eta^{b_1 a_1} \cdots \eta^{b_p a_p} \big) 
\varepsilon_{a_1 \ldots a_d} \, 
e^{a_{p+1}} \wedge \cdots \wedge e^{a_d} 
\in \Omega^{d-1}(\Sigma_d , \mathfrak{g}) \,, \\
&= \frac{1}{(d-p)!}  \Big(
\omega_{\mu_1\ldots \mu_p}^i 
e^{\mu_1}_{\,\, b_1} \eta^{b_1 a_1} e^{\nu_1}_{\,\, a_1} 
\cdots 
e^{\mu_p}_{\,\, b_p} \eta^{b_p a_p} e^{\nu_p}_{\,\, a_p} 
\Big)  \text{det}(e^{a_1}_{\,\, \mu_1}) \, 
\epsilon_{\nu_1 \ldots \nu_d} \, 
\dd x^{\nu_{p+1}} \wedge \cdots \wedge \dd x^{\nu_d}.
  \end{aligned}
\end{equation}

 \begin{figure}
 \centering
 \subfloat[$\epsilon = 2$]{\includegraphics[width=0.45\textwidth]{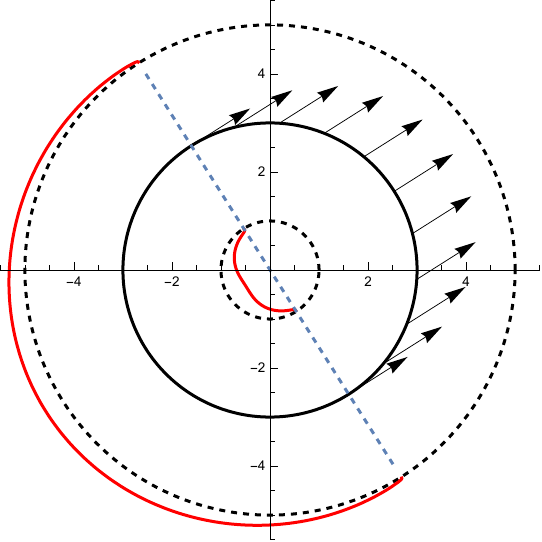}}
 \hfill
 \subfloat[$\epsilon = 0.5$]{\includegraphics[width=0.45\textwidth]{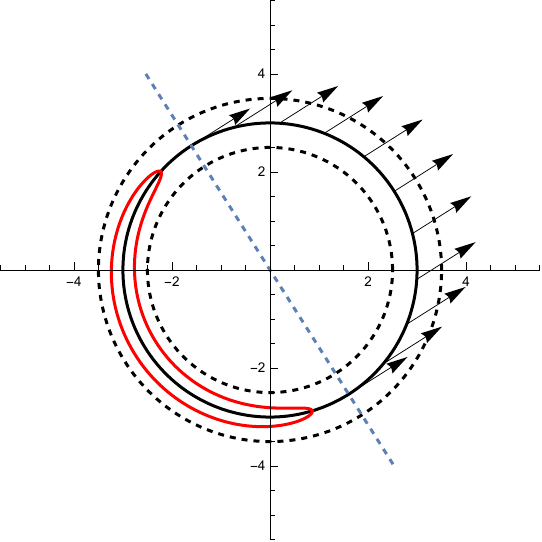}}
 \hfill
 \subfloat[$\epsilon = 0.1$]{\includegraphics[width=0.45\textwidth]{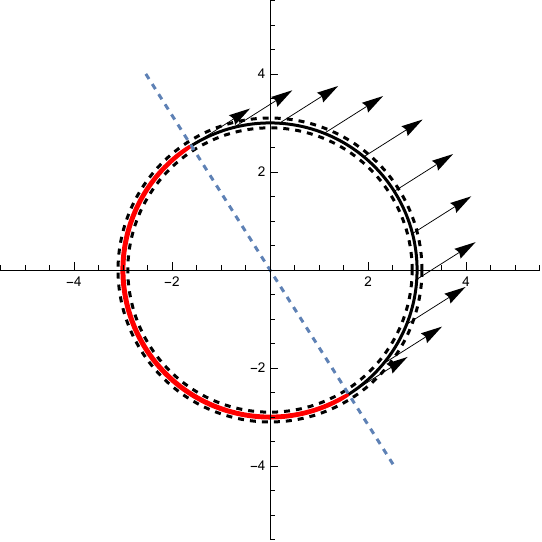}}
 \caption{We take in $d=2$ a topological operator supported on a circle of radius $R=3$ (solid black line) corresponding to the Lie algebra element $X = 1.2 P_1 + 0.76 P_2$ (black arrows). The operator is smeared into a tubular neighbourhood of size $2\epsilon$ (region bounded by dashed black lines). The solid red lines are the regions where the metric becomes degenerate. In (a), these singularities lie outside the tubular neighbourhood, so are not acceptable solutions to \eqref{eq:P_op_horizons}. In this case the metric on the boundary does not have any singularities. If we shrink the neighbourhood down to $\epsilon = 0.5$ in (b), the degenerate regions exists and persist for arbitrary smaller $\epsilon$.  }
\label{fig:P_op_deg}
\end{figure}

\section{Singular Metrics and Topological Operators}
\label{sec:singmetric}

In this appendix, we repeat the calculations of section \ref{sec:moving-points} introducing regularized metrics. 
We can regularize the distributional 1-form $\delta^{(1)}(\Sigma_{d-1})$ by replacing it with
\begin{equation}
\delta^{(1)}(\Sigma_{d-1}) \mapsto \rho(r) \dd r \, ,
\end{equation}
where $\rho(r)$ is a bump function depending on the coordinate $r$ perpendicular to $\Sigma_{d-1}$ and has finite support within a tubular neighbourhood $D_\epsilon(\Sigma_{d-1}) \cong \Sigma_{d-1} \times [-\epsilon,\epsilon]$, which shrinks to $\Sigma_{d-1}$ in the $\epsilon \rightarrow 0^+$ limit.
A prototypical bump function for a spherical $\Sigma_{d-1}$ of radius $r_0$ is
\begin{equation}
\rho(r) = \frac{1}{N(\epsilon)} \begin{cases}
\exp\left( \dfrac{1}{(r-r_0)^2-\epsilon^2} \right) \quad &|r-r_0|<\epsilon \\ 0 \quad &|r-r_0| \geq \epsilon
\end{cases}\,,
\end{equation}
where $N(\epsilon)$ is an appropriate normalization. In section \ref{sec:moving-points}, we showed that topological operators acts on boundary conditions by modifying the induced metrics. For example, in the case of translation operators, the induced metric \eqref{eq:translationmetric} becomes, in the regularization discussed above,
\begin{equation}
    g_{\mu\nu}dx^\mu dx^\nu = \eta_{\mu\nu} \dd x^\mu \dd x^\nu + 2 X_\mu \rho(r) \dd r  \dd x^\mu +  |X|^2 \rho(r)^2 \dd r  \dd r  \, ,
\end{equation}
If we consider this metric around points of $\Sigma_{d-1}$ where the perpendicular component is aligned with the translation vector $X$, Its determinant reads
\begin{equation}
\det g \simeq \left( 1 + X_r \rho \right)^2 + ... ,
\end{equation}
where corrections are related to the components of $X$ laying on $\Sigma_{d-1}$. From this expression follows that wherever $X_r <0$ the metric degenerates. This formally happens at:
\begin{equation}
r_{\text{deg}} \simeq r_0 \pm \sqrt{\epsilon^2 - \frac{1}{\log( |X_r|/N(\epsilon))} } \, .
\label{eq:P_op_horizons}
\end{equation}
These solutions are only valid when $|r_{\rm deg}-r_0| < \epsilon$ or equivalently $|X_r| > N(\epsilon)$. Since $N(\epsilon) \rightarrow 0^+$ as $\epsilon \rightarrow 0$, for any fixed $X_r$, it is always possible to find a tubular neighborhood large enough such that $|r_{\rm deg}-r_0|> \epsilon$ and the metric does not degenerate. The topological operator ``smeared" in this way does not induce any singular metric, see figure \ref{fig:P_op_deg} for an example.

Removing the regulator $\epsilon \rightarrow 0^+$, any nontrivial translation will make the vielbein non-invertible in some regions of the spacetime, and the metric will degenerate there. 
However, as mentioned in section \ref{sec:moving-points}, this geometry is diffeomorphic to ordinary flat space. In fact, the new boundary condition is gauge equivalent to $\mathcal{A}$ and can be written, upon regularization, as %
\begin{equation}
A|_{\partial M_{d+1}} = A^{(e^\alpha)} = e^{-\alpha} \cA e^{\alpha} + e^{-\alpha} \dd e^{\alpha}, \quad \alpha = X^a P_a \int_{r_0 -\epsilon}^{r} \dd r' \rho(r') \,.
\end{equation}
Using the results of appendix~\ref{sec:App_finite_transform}, one can
find the appropriate diffeomorphism corresponding to the gauge transformation above. It is generated by a vector field $\xi = \xi(r)$ which depends only on the perpendicular direction. This vector field is
shown to satisfy 
\begin{equation}
\xi^\mu(r) = \begin{cases} 0 & r< r_0 -\epsilon \\ X^\mu & r> r_0 + \epsilon\end{cases}\,,
\end{equation}
with some modulation inside the shell $|r-r_0|<\epsilon$. As we remove the regulator $\epsilon \rightarrow 0^+$ one gets \eqref{Diffeoxi}.

\bibliographystyle{ytphys}
\small 
\baselineskip=.94\baselineskip
\let\bbb\bibitem\def\bibitem{\itemsep4pt\bbb}
\bibliography{ref}

\end{document}